\documentclass[longauth]{aa}

\usepackage{lscape}
\usepackage{txfonts}
\usepackage{float}
\usepackage{graphicx}
\usepackage{subfigure}
\usepackage{tabularx}
\usepackage{epsfig}
\newcommand{\dd}{deg$^{2}$}
\newcommand{\flux}{$\rm erg \, s^{-1} \, cm^{-2}$}

\begin{document}

\title{The XXL Survey XX: The 365 cluster catalogue
  ~\thanks{Based on observations obtained with XMM-Newton, an ESA science mission with instruments 
    and contributions directly funded by ESA Member States and NASA.
    Based on observations made with ESO Telescopes 
    at the La Silla and Paranal Observatories under programmes ID 191.A-0268
    and 60.A-9302. Based on observations obtained with
    MegaPrime/MegaCam, a joint project of CFHT and CEA/IRFU, at the
    Canada-France-Hawaii Telescope (CFHT) which is operated by the
    National Research Council (NRC) of Canada, the Institut National
    des Sciences de l'Univers of the Centre National de la Recherche
    Scientifique (CNRS) of France, and the University of Hawaii. 
    Based on observations collected at the German-Spanish Astronomical 
    Centre, Calar Alto, jointly operated by the Max-Planck-Institut f\"ur Astronomie Heidelberg 
    and the Instituto de Astrofísica de Andaluc\'ia (CSIC). This
    work is based in part on data products produced at Terapix
    available at the Canadian Astronomy Data Centre as part of the
    Canada-France-Hawaii Telescope Legacy Survey, a collaborative
    project of NRC and CNRS. This research has made use of the VizieR catalogue
    access tool, CDS, Strasbourg, France. This research has also made
    use of the NASA/IPAC Extragalactic Database (NED) which is
    operated by the Jet Propulsion Laboratory, California Institute of
    Technology, under contract with the National Aeronautics and Space
    Administration. }
    \thanks{Full table 5 is available in electronic form
    at the CDS via anonymous ftp to cdsarc.u-strasbg.fr (130.79.128.5)
    or via http://cdsweb.u-strasbg.fr/cgi-bin/qcat?J/A+A/.}
}

\author{
C.~Adami\inst{1} \and
P.~Giles\inst{7} \and
E.~Koulouridis\inst{3} \and
F.~Pacaud\inst{4} \and
C.A.~Caretta\inst{1,6} \and
M.~Pierre\inst{3} \and
D.~Eckert\inst{9} \and
M.E.~Ramos-Ceja\inst{4} \and
F.~Gastaldello\inst{8} \and
S.~Fotopoulou\inst{45} \and
V.~Guglielmo\inst{1,2,30} \and
C.~Lidman\inst{11} \and
T.~Sadibekova\inst{3} \and
A.~Iovino\inst{28} \and
B.~Maughan\inst{7} \and
L.~Chiappetti\inst{8} \and
S.~Alis\inst{40} \and
B.~Altieri\inst{31} \and
I.~Baldry\inst{20} \and
D.~Bottini\inst{8} \and
M.~Birkinshaw\inst{7} \and
M.~Bremer\inst{7} \and
M.~Brown\inst{21} \and
O.~Cucciati\inst{26,33} \and
S.~Driver\inst{18,19} \and
E.~Elmer\inst{15} \and
S.~Ettori\inst{26,41} \and
A.E.~Evrard\inst{25} \and
L.~Faccioli\inst{3} \and
B.~Granett\inst{28,32} \and
M.~Grootes\inst{24} \and
L.~Guzzo\inst{28,32} \and
A.~Hopkins\inst{11} \and
C.~Horellou\inst{27} \and
J.P.~Lef\`evre\inst{3} \and
J.~Liske\inst{22} \and
K.~Malek\inst{35} \and
F.~Marulli\inst{26,33,34} \and
S.~Maurogordato\inst{12} \and
M.~Owers\inst{11,17} \and
S.~Paltani\inst{9} \and
B.~Poggianti\inst{2} \and
M.~Polletta\inst{8,36,39} \and
M.~Plionis\inst{10,44} \and
A.~Pollo\inst{35,37} \and
E.~Pompei\inst{5} \and
T.~Ponman\inst{16} \and
D.~Rapetti\inst{42,43} \and
M.~Ricci\inst{12} \and
A.~Robotham\inst{18,19} \and
R.~Tuffs\inst{23} \and
L.~Tasca\inst{1} \and
I.~Valtchanov\inst{31} \and
D.~Vergani\inst{38} \and
G.~Wagner\inst{13,14} \and
J.~Willis\inst{29} \and
the XXL consortium
}

\offprints{C. Adami \email{christophe.adami@lam.fr}}

\institute{
Aix Marseille Univ, CNRS, CNES, LAM, Marseille, France
\and
INAF- Osservatorio astronomico di Padova, Vicolo Osservatorio 5, I-35122 Padova, Italy
\and
Laboratoire AIM, CEA/DSM/IRFU/SAp, CEA Saclay, F-91191, Gif-sur-Yvette, France 
\and
Argelander Institut f\"ur Astronomie, Universit\"at Bonn, Auf dem H\"ugel 71, D-53121 Bonn, Germany
\and
European Southern Observatory, Alonso de Cordova 3107, Vitacura, 19001 Casilla, Santiago 19, Chile
\and
Departamento de Astronom\'ia, DCNE-CGT, Universidad de Guanajuato;
Callej\'on de Jalisco, S/N, Col. Valenciana, 36240, Guanajuato, Gto.,
Mexico
\and
School of Physics, HH Wills Physics Laboratory, Tyndall Avenue, Bristol, BS8 1TL, UK
\and
INAF - IASF Milano, via Bassini 15, I-20133 Milano, Italy
\and
Department of Astronomy, University of Geneva, Ch. d'Écogia 16, CH-1290, Versoix, Switzerland
\and
Aristotle University of Thessaloniki, Physics Department, Thessaloniki, GR-54124, Greece
\and
Australian Astronomical Observatory, PO BOX 915, North Ryde, AU1670, Australia
\and
Laboratoire Lagrange, UMR 7293, Universit\'e de Nice Sophia Antipolis, CNRS, Observatoire de la 
C\^ote d’Azur, F-06304 Nice, France
\and
Department of Physics, University of Oxford, Oxford, OX1 3PU, United Kingdom
\and
Merton College, Oxford, OX1 4JD, United Kingdom
\and
School of Physics and Astronomy, University of Nottingham, University Park, Nottingham, NG7 2RD, United Kingdom
\and
Astrophysics and Space Research Group, School of Physics and Astronomy, University of Birmingham, Edgbaston Birmingham, B15 2TT, United Kingdom 
\and
Macquarie University, NSW, 2109, Australia 
\and
ICRAR, 1 Turner Avenue, Technology Park, Bentley, Western Australia, 6102 
\and
University of St Andrews, College Gate, St Andrews, KY16 9AJ, Fife, Scotland, UK
\and
Astrophysics Research Institute, Liverpool John Moores University, IC2, Liverpool Science Park, 146 Brownlow Hill, Liverpool L3 5RF, United Kingdom
\and
Monash University, Victoria 3800, Australia
\and
Hamburger Sternwarte, Universit{\"a}t Hamburg, Gojenbergsweg 112, D-21029 Hamburg
\and
Max-Planck-Institut f\"ur Kernphysik, PO Box 103980, D-69029 Heidelberg, Germany
\and
Max Planck Institut f\"ur Kernphysik, Saupfercheckweg 1, D-69117 Heidelberg, Germany
\and
Department of Astronomy, University of Michigan, Ann Arbor, MI 48109, USA; Department of Physics, University of Michigan, Ann Arbor, MI 48109, USA
\and
INAF, Osservatorio Astronomico di Bologna, via Pietro Gobetti 93/3, I-40129 Bologna, Italy 
\and
Chalmers University of Technology, Dept. of Space, Environment, and Earth, Onsala Space Observatory, 439 92 Onsala, Sweden
\and
INAF - Osservatorio Astronomico di Brera, Via Brera 28, 20122 Milano, via E. Bianchi 46, I-20121 Merate, Italy
\and
Department of Physics and Astronomy, University of Victoria, 3800 Finnerty Road, Victoria, BC V8P 1A1, Canada
\and
Department of Physics and Astronomy, University of Padova, Vicolo Osservatorio 3, I-35122 Padova, Italy
\and
European Space Astronomy Centre (ESA/ESAC), Operations Department, Villanueva de la Can\~ada, Madrid, Spain
\and
Universit\`a degli studi di Milano, via G. Celoria 16, I-20133 Milano, Italy
\and
Dipartimento di Fisica e Astronomia (DIFA), Università di Bologna, v.le Berti Pichat 6/2, I-40127 Bologna, Italy
\and
INFN, Sezione di Bologna, viale Berti Pichat 6/2, I-40127 Bologna, Italy
\and
National Centre for Nuclear Research, ul. Hoza 69, PL-00-681 Warszawa, Poland
\and
IRAP (Institut de Recherche en Astrophysique et Planétologie), Université de Toulouse, CNRS, UPS, Toulouse, France
\and
Astronomical Observatory of the Jagiellonian University, Orla 171, PL-30-001 Cracow, Poland
\and
INAF - IASF Bologna, via Gobetti 101, I-40129 Bologna, Italy
\and
Aix-Marseille Universit\'e - Pharo – 58 bd Charles Livon Jardin du Pharo - F-13007 Marseille
\and
Department of Astronomy and Space Sciences, Faculty of Science, Istanbul University, 34119 Istanbul, Turkey
\and
INFN, Sezione di Bologna, viale Berti Pichat 6/2, I-40127 Bologna, Italy
\and
Center for Astrophysics and Space Astronomy, Department of Astrophysical and Planetary Science, University of Colorado, Boulder, C0 80309, USA
\and
NASA Ames Research Center, Moffett Field, CA 94035, USA
\and
National Observatory of Athens, Lofos Nymfon, 11851 Athens, Greece
\and
Centre for Extragalactic Astronomy, Department of Physics, Durham University, South Road, Durham, DH1 3LE, UK
}

\date{Accepted . Received ; Draft printed: \today}

\authorrunning{Adami et al.}

\titlerunning{Second epoch XXL galaxy structure catalogue}

\abstract 
{In the currently debated context of using clusters of galaxies as cosmological probes, the need for
well-defined cluster samples is critical.}
{The XXL Survey has been specifically designed to provide a well characterised
sample of some 500 X-ray detected clusters suitable for cosmological studies. The main goal
of present article is to make public and describe the properties of the cluster catalogue in its present state, as 
well as of associated catalogues of more specific objects such as super-clusters and fossil groups.}
{We release a sample containing 365 clusters in total. 
In this paper, we give the details of the follow-up observations and explain the procedure
adopted to validate the cluster spectroscopic redshifts.
Considering the whole XXL cluster sample, we have provided two types of selection, both
complete in a particular sense: one based  on flux-morphology criteria,
and an alternative based on the [0.5-2] keV flux within
one arcmin of the cluster centre. We have also provided X-ray
temperature measurements for 80$\%$ of the clusters having a flux larger 
than 9$\times$10$^{-15}$$\rm \thinspace erg \, s^{-1} \, cm^{-2}$.}
{Our cluster sample extends from z$\sim$0 to z$\sim$1.2, with one cluster at z$\sim$2. 
Clusters were identified through a mean number of six spectroscopically confirmed cluster members. The largest number 
of confirmed spectroscopic members in a cluster is 41.
Our updated luminosity function and luminosity-temperature relation are
compatible with our previous determinations based on the 100 brightest
clusters, but show smaller uncertainties. We also present an enlarged list of
super-clusters and a sample of 18 possible fossil groups.}
{This intermediate publication is the last before the final release of
the complete XXL cluster catalogue when the ongoing C2 cluster spectroscopic follow-up is complete. It provides a unique inventory
of medium-mass clusters over a 50~\dd\ area out to z$\sim$1.}


\maketitle

\section{Introduction}

Most galaxy cluster-related cosmological probes rely on cluster number
counts and large-scale structure information. 
X-ray surveys have had a key role in this framework since the historical Einstein observatory Medium Sensitivity Survey 
(Gioia et al. 1990). Many other surveys were conducted with the ROSAT
observatory, and more recently, XMM-Newton and Chandra produced surveys such as the XMM-LSS, XMM-COSMOS, XMM-CDFS and Chandra-Ultra-Deep surveys 
(Pierre et al. 2004; Hasinger et al. 2007; Comastri et al. 2011; Ranalli et al. 2013). Following this path, it is now clear
 that cluster cosmological studies can only be
rigorously performed by simultaneously fitting a cosmological model, the
cluster selection function and the physical modelling of the cluster
evolutionary properties in whichever band the cluster selection has been
performed (e.g. Allen et al 2011). X-ray cluster cosmology is especially well suited to such an
approach, because the properties of the X-ray emitting intra-cluster
medium can be ab-initio predicted with good accuracy, either using an analytical model or
by means of hydrodynamical simulations.

The XMM-XXL project (XXL hereafter) covers two areas of 25 \dd\ each with XMM-Newton observations to a sensitivity of 
$\sim 5\times 10^{-15}$\flux ~(for point sources); the two areas are centred at:
XXL-N (02$^h$23' -04$^{\circ}$30') and XXL-S (23$^h$30' -55$^{\circ}$00'). In a first step, XXL aims at in-depth cluster evolutionary studies over the $0<z<1$
range by combining an extensive data set over the entire electromagnetic spectrum. In a second and ultimate step we aim at a 
standalone cosmological analysis (Pierre et al 2016, hereafter XXL paper I) and the X-ray cluster catalogue constitutes the core  of the whole 
project: its construction along with the determination of the cluster multiwavelength parameters follows an iterative process demanding 
special care. In this process, the spectroscopic confirmation of the X-ray cluster candidates has occupied a central place in the 
project over the last five years. In a first publication (Pacaud et al 2016, hereafter XXL paper II) we presented the hundred brightest 
galaxy clusters (XXL-100-GC) along with a set of 
preliminary scientific analyses, including the X-ray luminosity function, spatial correlation studies and a cosmological interpretation 
of the number counts. The present, and second,  release is the last before the publication of the complete cluster catalogue.
This will occur when the ongoing C2 cluster spectroscopic follow-up is completed.
The main goal of present article is to make public and describe the properties of the second release, as 
well as of associated catalogues of more specific objects such as super-clusters and fossil groups.
The present sample contains the complete subset of clusters for which the selection function is well determined (namely, the C1 selection) plus all 
X-ray clusters which are, to date, spectroscopically confirmed. The C1 and C2 classes are defined as in XXL paper II and will be described below.
Altogether, this amounts to 365 clusters and is referred to as 
the XXL-365-GC sample (cf. Table~\ref{tab:sumBen}). Along with the cluster list itself, we provide an update of the X-ray cluster properties and of their spatial 
distribution as presented in the 2016 XXL-100-GC publications. The cluster parameters derived in the present publication supersede the XXL-100-GC ones, even 
thought the consistency (see below) is very good.

In the next section, we describe the construction of the current sample. Section 3 gives a detailed account of the spectroscopic validation 
procedure. We present the cluster catalogue in Sect. 4. Section 5 provides updated determinations of the X-ray cluster luminosity 
function and of the luminosity-temperature relation. The results of spatial analyses  performed on the cluster catalogue (search for 
super-clusters and fossil groups) are presented in Sect. 6.  Notes on the newly detected structures and recent redshift measurements 
are gathered in the appendix. Throughout the paper, for consistency with the first series of XXL papers, we adopt the WMAP9 cosmology (Hinshaw et al., 2013, 
with $\Omega_m$ = 0.28, $\Omega_\Lambda$ = 0.72, and H$_0$ = 70 km s$^{-1}$ Mpc$^{-1}$), except if explicitely stated.
From the semantic point of view, we also mention that the structures called clusters in the present paper are not very massive structures, but are intermediate-mass  
concentrations in the mass range between groups of galaxies and very massive clusters of galaxies. 

\section{Selection of the X-ray cluster sample}

The X-ray pipeline and the cluster selection procedure along with the  XXL selection function are extensively described in XXL paper II. 
We recall here the main steps.

Our detection algorithm (the same version of Xamin used in XXL paper II, cf. also Faccioli et al. 2017, hereafter XXL paper XXIV)  enables the creation of 
an uncontaminated (C1) cluster sample  
by selecting all detected sources in the 2D [{\tt EXT; EXT\_STAT}] output parameter space.  The {\tt EXT} parameter is a measure of 
the cluster apparent size and the {\tt EXT\_STAT} parameter quantifies the likelihood of a source of being extended.
The {\tt EXT\_STAT} likelihood parameter  is a function of cluster size, shape and flux. This parameter depends on 
the local XMM-Newton sensitivity.

Simulations enable the definition of limits for {\tt EXT} and {\tt EXT\_STAT} above which contamination from point sources 
is negligible, providing the C1 sample. Relaxing slightly these limits, we define a second, deeper, sample (C2) to allow for 50\% contamination by
misclassified point sources; these can easily be cleaned up a posteriori using optical versus X-ray comparisons. 
Initially, the total number of such C2 cluster candidates was 195 and more than 60$\%$ are already spectroscopically confirmed 
(see below). We defined a third 
class, C3, corresponding to (optical) clusters associated with some X-ray emission, too weak to be characterised; the selection 
function of the C3 sample is therefore undefined. Initially, most of the C3 objects were not detected in the X-ray waveband
and are located within the XMM-LSS subregion. We refer the reader to Pierre et al. (2004) for a more detailed description of these 
classes.

With the present paper, we publish all C1 clusters (XXL-C1-GC hereafter, cf. Table 1) supplemented by the C2 and C3 clusters 
which are spectroscopically confirmed. C3 clusters
were not specifically targeted, but were sometimes confirmed as by-products of existing galaxy spectroscopic surveys. Table~\ref{tab:sumAng} gives 
statistics of the XXL-365-GC sample in terms of C1, C2, and C3 clusters.
This amounts to 207 C1 (among them, 183 spectroscopically confirmed to date, 4 with some spectroscopy but needing more data, 13 with a 
photometric redshift, and 7 without redshift estimation), 119 C2 and 39 C3. The C1 selection provides a complete sample in the two-parameter 
space outlined above. In order to allow straightforward comparisons with different X-ray processing methods,  we give, for information only, 
the approximate completeness flux limit of the XXL-365-GC sample computed from simulated detections. 
We performed the measurements within a radius of 1 arcmin around the cluster centre (defined from the X-ray data). 
We assume, as in XXL paper II, that the XMM-Newton count-rates are computed in the [0.5-2] keV band and converted into
fluxes assuming an Energy Conversion Factor (ECF) of $9.04\times 10^{-13}$ \flux\ /(cts/s). The completeness flux limit (the 100$\%$ completeness 
flux limit averaged across the entire survey area) is then $\sim$ 1.3 $\times$ 10$^{-14}$ $\rm \thinspace erg \, s^{-1} \, cm^{-2}$.
We emphasise that since a flux of $10^{-14}$ corresponds to $\sim$100 photons on-axis for 10ks exposures (MOS1+MOS2+PN), uncertainties are
large, which may affect the cluster ranking as a function of the flux by 10\% or more.\\ 

\begin{table*}[t!]
\caption{\label{tab:sumBen}Statistics of the XXL-365-GC, XXL-C1-GC, and XXL-100-GC samples. Numbers within parentheses are the numbers of 
spectroscopically confirmed clusters for the considered selection.}
\begin{tabular}{cccccc}
\hline
\hline
 Sample  & selection & N C1+C2+C3 & N C1 & N C2 & N C3  \\
\hline						
\hline						
XXL-365-GC    & All C1 clusters       & 365 (341)  & 207 (183)  &  119 (119) & 39 (39) \\   	      
              & + spectros. C2/C3     &            &            &            &         \\   	      
\hline						
XXL-100-GC    & 100 brightest clusters & 100 (99)   & 96 (95)    &  4 (4)     & 0       \\   	      
\hline
\end{tabular}
\end{table*}

\begin{table}[t!]
\caption{\label{tab:sumAng}Statistics of the XXL-365-GC sample in terms of C1, C2, and C3 clusters. Col.1: considered
classes. Col.2: numbers within the total XXL-365-GC sample.  Col.3: numbers of spectroscopically confirmed clusters within the 
XXL-365-GC sample. Col.4: numbers of spectroscopically confirmed clusters with at least three spectroscopic redshifts within the 
XXL-365-GC sample. Numbers within parentheses are for the northern and southern areas.}
\begin{tabular}{cccc}
\hline
\hline
Classes  & XXL-365-GC & Spect. & $\ge$3 redshifts \\
\hline						
\hline						
C1       &  207 (114/93)     &  183 (105/78)    & 160 (96/64) \\   	      
\hline						
C2       &  119 (59/60)      &  119 (59/60)     & 70 (42/28)  \\ 
\hline						
C3       &  39 (38/1)        &  39 (38/1)       & 31 (31/0)   \\ 
\hline
All      &  365 (211/154)    &  341 (202/139)   & 261 (169/92) \\ 
\hline
\end{tabular}
\end{table}

\section{Spectroscopic redshifts}

\subsection{Collecting the spectroscopic information}

The spectroscopic surveys conducted on the XXL fields are listed in XXL paper I (Table 3). We provide below a short description of this rather 
heterogeneous data set. In order to perform the spectroscopic validation and further dynamical studies of the XXL clusters, all available 
spectroscopic information on galaxies located in the XXL fields has been stored in the CEntre de donn\'eeS Astrophysiques de  Marseille 
(http://www.lam.fr/cesam/). Their astrometry was matched with the CFHTLS T0007 catalogue (http://www.cfht.hawaii.edu/Science/CFHTLS/T0007/) for 
XXL-N and with the BCS catalogue (cf. Desai et al. 2012) for XXL-S. The public and private surveys stored in CESAM and 
relevant to XXL are described in the following. All in all, the total number of redshifts present in the CESAM database are $\sim$145000 and $\sim$8500 for 
the XXL-N and XXL-S fields respectively (as of December 2016, including multiple measurements).

\subsubsection{XXL extended sources spectroscopic follow-up campaigns}

We conducted our own spectroscopic follow-up to complement the already available public spectroscopic data sets. C1 clusters were the primary targets,
but we also targeted C2 clusters when possible. The targets were chosen in order to favour the cluster confirmation by galaxies within
the X-ray contours. We note that the X-ray contours are created from a wavelet filtered photon image. The contours are run in each frame for the range between 
0.1 cts/px corresponding to the typical background level for exposition time of 10ks ($\sim$10$^{-5}$ cts/s/px) and a maximal value in the frame spaced by 15 
logarithmic levels.

a) We made extensive use of the ESO optical facilities (NTT/EFOSC2 and VLT/FORS2). We were granted three PI allocations, 
including a Large Programme (191-0268) and a pilot programme (089.A-0666). We give the details of these new PI ESO programmes in Table~\ref{tab:detailPI}.

\begin{table}[t!]
\caption{\label{tab:detailPI}Details of the three ESO PI runs.}
\begin{tabular}{lccccc}
\hline
\hline
ESO Id & Instrument & Duration & Semesters Nb \\
\hline						
191.A-0268 & FORS2	  & 132h     &	4	\\
191.A-0268 & EFOSC2 	  & 15n	     &	4	\\
089.A-0666 & FORS2 	  & 15h	     &	1	\\
60.A-9302  & MUSE         & 3h 	     &  1	\\
\hline
\end{tabular}
\end{table}

FORS2 and EFOSC2 galaxy targets were first choosen according to their strategical place inside the clusters, taking into account the already
known redshifts from other surveys, and their location regarding the X-ray contours. Then, we put as many slits as possible on other objects.
We measured the spectroscopic 
redshifts by means of the EZ code (Garilli et al. 2008) that was already used  for the VIPERS survey (Guzzo et al. 2014, Scodeggio et al. 2017). 
We adopted  the same approach: the only operation that required 
human intervention is the verification and validation of the EZ measured redshift. Each spectrum is  independently measured by two team members. 
At the end of the process, discrepant redshifts are 
discussed and homogenised. The quality of the redshift measurements is defined as in the VVDS and VIPERS surveys:

 - Flag 0: no reliable spectroscopic redshift measurement.

 - Flag 1: Tentative redshift measurement with a $\sim$50$\%$ chance that the redshift is wrong. These redshifts 
are not used.

 - Flag 2: Confidence estimated to be greater than 95$\%$.

 - Flag 3 and 4: highly secure redshift. The confidence is estimated to be higher than 99$\%$.

 - Flag 9: redshift based on a single clear feature, given the absence of other features. These redshifts are generally 
reliable.

b) We also made use of the AAOmega instrument on the AAT. A first observing campaign was published in Lidman et al. 2016 (hereafter XXL Paper XIV), while
supplementary observations done in 2016 will be included in Chiappetti et al. 2017 (hereafter XXL Paper XXVII, in prep.). 
For the first run, cluster galaxies were the prime targets and we used {\tt Runz} (Hinton et al. 2016) to measure redshifts.
X-ray AGN in the XXL-S field were the prime targets for the second run and only spare fibres were put on cluster galaxies. 
We used {\tt Marz} (Hinton et al. 2016) to measure redshifts. For each spectrum, we assign a quality flag that
varies from 1 to 6. The flags are identical to those used in the OzDES redshift survey (Yuan et al. 2015). We used AAT quality flags 3 or 4 which
are equivalent to the ESO flags 2, 3, or 4.\\
c) We also obtained Magellan spectroscopy at Las Campanas observatory from an associated survey (A. Kremin, private communication). We only used the 262 
most reliable redshifts. \\
d) We collected redshifts at the William Herschel Telescope (WHT hereafter, cf. Koulouridis et al. 2016, hereafter XXL paper XII). Redshifts were measured and 
quality flags were assigned in the same way as for the ESO data.\\

\subsubsection{Redshifts from the XMM-LSS survey}

We included all redshifts obtained for the XMM-LSS pilot survey (11 \dd\ precursor and subarea of XXL-N, Pierre et al. 2004). 
The sample is described in Adami et al. (2011).

\subsubsection{Literature data} 

The XXL-N area was defined to overlap with the VIPERS survey (VIMOS Public Extragalactic Redshift survey: Guzzo et al. 2014, 
Scodeggio et al. 2017) and to encompass the VVDS survey (Le F\`evre et al. 2013). We therefore included the redshifts from these 
VIMOS-based redshift surveys. The redshifts are measured in our own ESO spectroscopic follow-up exactly in the same way as
VIPERS and VVDS did, with 
the same quality flags. We also note that all redshifts from VIPERS, covering the redshift range 0.4$\leq$z$\leq$1.2, were made 
available for this analysis prior to the recent public release (Scodeggio et al. 2017).

GAMA, 2dF, 6dF, SDSS: These four catalogues were ingested and used without remeasuring the redshift of the galaxies. They provide
robust spectroscopic quality flags. We considered as reliable the GAMA, 2dF, and 6dF redshifts with quality flags 3 and 4 (e.g. 
Liske et al. 2015 and Baldry et al. 2014 for GAMA, and Folkes et al. 1999 for 2dF), equivalent to the ESO flags 2, 3 or 4. SDSS 
spectra with 'zWarning' between 0 and 16 were also used. We note that the GAMA spectroscopy inside the XXL area is issued from the GAMA G02 
field where fibres were also intentionally put on preliminary proposed XXL galaxy targets. G02 will be public within the GAMA DR3 data 
release (Baldry et al., in preparation).

In addition, we considered other smaller public redshift catalogues: Akiyama et al. from Subaru (2015), Simpson et al. (2006, 2012), Stalin et al. (2010),
SNLS survey (e.g. Balland et al., 2009). We remeasured and checked the redshift values for these surveys, when spectra were available, using the methods 
developed for our own spectroscopic follow-up.
We finally collected and assumed as correct all other redshifts on the XXL areas, currently available in the NED database.

\subsection{Redshift reliability and precision}

\begin{figure}[h]
\includegraphics[width=6.5cm,angle=270]{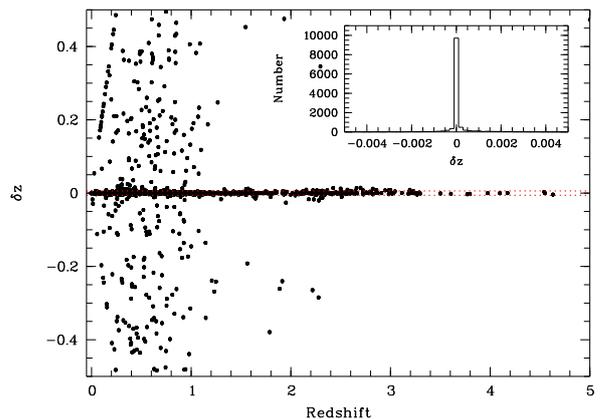}
\caption{\label{fig:deltaz2}Redshift difference versus redshift for the $\sim$12 000 objects measured twice within
the spectroscopic survey. The two red dotted lines represent the $\pm$3$\times$600 km s$^{-1}$ level (cf. section 3.2). We also give the histogram
of the redshift difference within the [-0.005,0.005] interval.}
\end{figure}

Our spectroscopic redshift catalogues come from various telescopes, with different instruments,
different setups and were obtained under different observing conditions. We thus needed to evaluate on an objective 
basis the overall reliability of the data set. Although we tried to limit multiple observations, we ended up with a non-negligible number 
of galaxies present in different surveys. We used these redundant measurements to evaluate the statistical reliability of our redshifts. 
The simplest approach consists in plotting the redshift difference versus redshift (cf. Figure~\ref{fig:deltaz2})
for the $\sim$12 000 objects measured twice in the whole spectroscopic sample. 
Out of these, 15$\%$ had a spectroscopic quality flag of 4, 61$\%$ a quality flag of 3, 24$\%$ a 
quality flag of 2, and less than 1$\%$ a quality flag of 9. We only consider flags greater than two in the following.
 
\begin{figure}[h]
\includegraphics[width=6.5cm,angle=270]{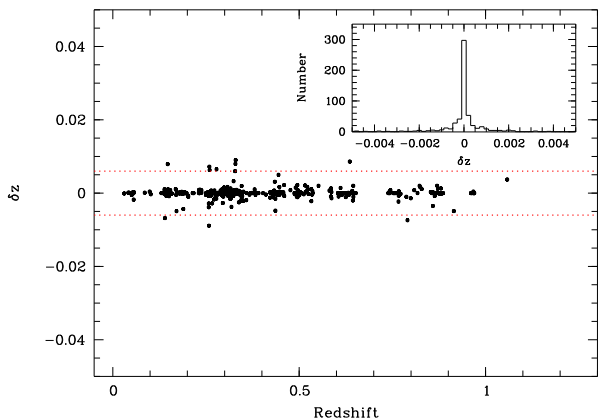}
\caption{\label{fig:deltaz2inside}Redshift difference versus redshift for the galaxies (at less than $\pm$3$\times\sigma_{v,200}$ from the cluster mean redshift
and within one Virial radius) measured two times within the spectroscopic survey. The two red dotted lines represent the $\pm$3$\times$600 km s$^{-1}$ level.
We also give the histogram of the redshift difference within the [-0.005,0.005] interval.}
\end{figure}

a) To estimate the fraction of incompatible redshifts, we selected in Figure~\ref{fig:deltaz2} all double measurements 
 differing by more than $\pm$3$\times$600 km s$^{-1}$ (600 km s$^{-1}$ is a typical value based on the VVDS and VIPERS surveys: cf. Le F\`evre et al. 
2013 and representing a good compromise between the spectrographs resolution and the possible real difference between redshifts, at the 3-$\sigma$
level). This points to strongly discrepent redshifts for 5$\%$ of the sample. A comparable percentage is expected in Guzzo et al. (2014) for 
the VIPERS survey. We therefore conclude that our sample is similar to the VIPERS survey in terms of incompatible redshifts (cf. 
Scodeggio et al. 2017). \\

b) For measurements within $\pm$3$\times$600 km s$^{-1}$, the statistical 1-$\sigma$ redshift scatter is 
$\sim$0.00049$\times$(1+z). This represents almost 150 km s$^{-1}$. We note that Figure~\ref{fig:deltaz2} may give the feeling that the
dispersion is much larger at low redshifts. However, this is mainly due to the fact that many objects are concentrated along the zero difference
level. The statistical 1-$\sigma$ uncertainty is for example $\sim$0.00049 at z$\leq$1 and $\sim$0.00057 at z$\leq$0.5. \\

c) The previous estimates pertain to the full galaxy sample. We also performed a similar analysis on the cluster galaxies alone. These galaxies 
have different types and luminosities and are therefore potentially subject to different selections.
To select these galaxies, we limited the sample to galaxies within one Virial radius and with a velocity 
within $\pm3 \times \sigma_{v,200}$, the equivalent galaxy velocity dispersion inferred from scaling laws within the Virial radius, from the cluster centre.
We could have tried to use instead the galaxy velocity dispersion computed with galaxy redshifts, but our sampling is too sparse to have precise estimations.
This will be treated in a future paper.
 Virial radius and 
$\sigma_{v,200}$ were estimated from 
X-ray data given in Table~\ref{tab:listeSL} and described in the following. Applying the same method as with the complete sample, we find an
incompatible redshift percentage of $\sim$4$\%$ (cf. Figure~\ref{fig:deltaz2inside}), even better than for the total sample. The 1-$\sigma$ redshift scatter is 
$\sim$0.00041$\times$(1+z), or 120 km s$^{-1}$ in terms of radial velocity uncertainty, also similar to the estimate for the total sample. 
Finally, we do not see any significant variation of the 1-$\sigma$ uncertainty between redshifts 0 and 0.9.\\
The last issue is to estimate the relative weight of the various telescopes in the cluster redshift compilation.  
Considering the sample of cluster galaxies only, we find that $\sim$45$\%$ 
are coming from ESO (VIMOS and FORS2 instruments), $\sim$45$\%$ from AAT  (AAOmega instrument), and 
$\sim$7$\%$ from SDSS. The remaining $\sim$3$\%$ have various origins (Subaru, WHT, LasCampanas, ..etc...).

As a remark, for a given object with multiple redshift measurements, we used the measurement coming from the highest quality spectrum. We did not notice
systematic redshift differences in the considered surveys.

\subsection{Cluster spectroscopic confirmation}

Starting from the list of extended X-ray sources (C1 or C2), the cluster spectroscopic confirmation is an iterative process. 

1) We first collected all available spectroscopic redshifts along a given line of sight towards a cluster candidate. 
We selected the spectroscopic redshifts within the X-ray contours and  searched for gaps larger than 900 km s$^{-1}$ in the resulting redshift histogram. 
This is intended to separate different concentrations in the redshift space. We searched for concentrations
of three or more redshifts between two gaps and preliminarily assigned the largest concentration to the extended source in question. This allows us to 
estimate the angular distance of the source in question. \\
2) We then repeated the process, this time within a 500$\thinspace$kpc radius. This has sometimes led us to consider larger regions than the ones defined by the X-ray contours. We
checked whether the inferred redshift was compatible with the previous one. If yes, we considered the cluster to be confirmed at the considered redshift. If not, 
we restarted the full process with another redshift concentration. In practice, this process was convergent at the first pass for the large majority
of the cases.\\
We kept open the possibility of manually assigning a redshift to a cluster when the two previous criteria did not agree (cf.
below the peculiar case of XLSSC 035). This mainly occurred when dealing with projection effects along the line of sight (cf. the eight
cases in appendix B).
Some of the lines of sight were however poorly sampled, with typically fewer than three redshifts. In this 
case, we attempted to confirm the cluster nature of the X-ray source by identifying the cluster dominant galaxy (BCG 
hereafter) in the i' band 
and close to the X-ray centroid. If the choice of such a galaxy was obvious and this galaxy had a spectroscopic redshift, we confirmed the cluster as well. This 
was the case for 30 clusters (with only the BCG), and for another 50 clusters (with the BCG plus another concordant galaxy).\\
The C3 clusters – X-ray sources too faint to be characterised as C1 or
C2 – that we present in this paper are only those resulting from the spectroscopic follow-up of X-ray sources in the
XMM-LSS pilot survey. We did not perform any systematic cluster search or follow-up for the full list of X-ray sources.

\begin{figure}[h]
\includegraphics[width=6.5cm,angle=270]{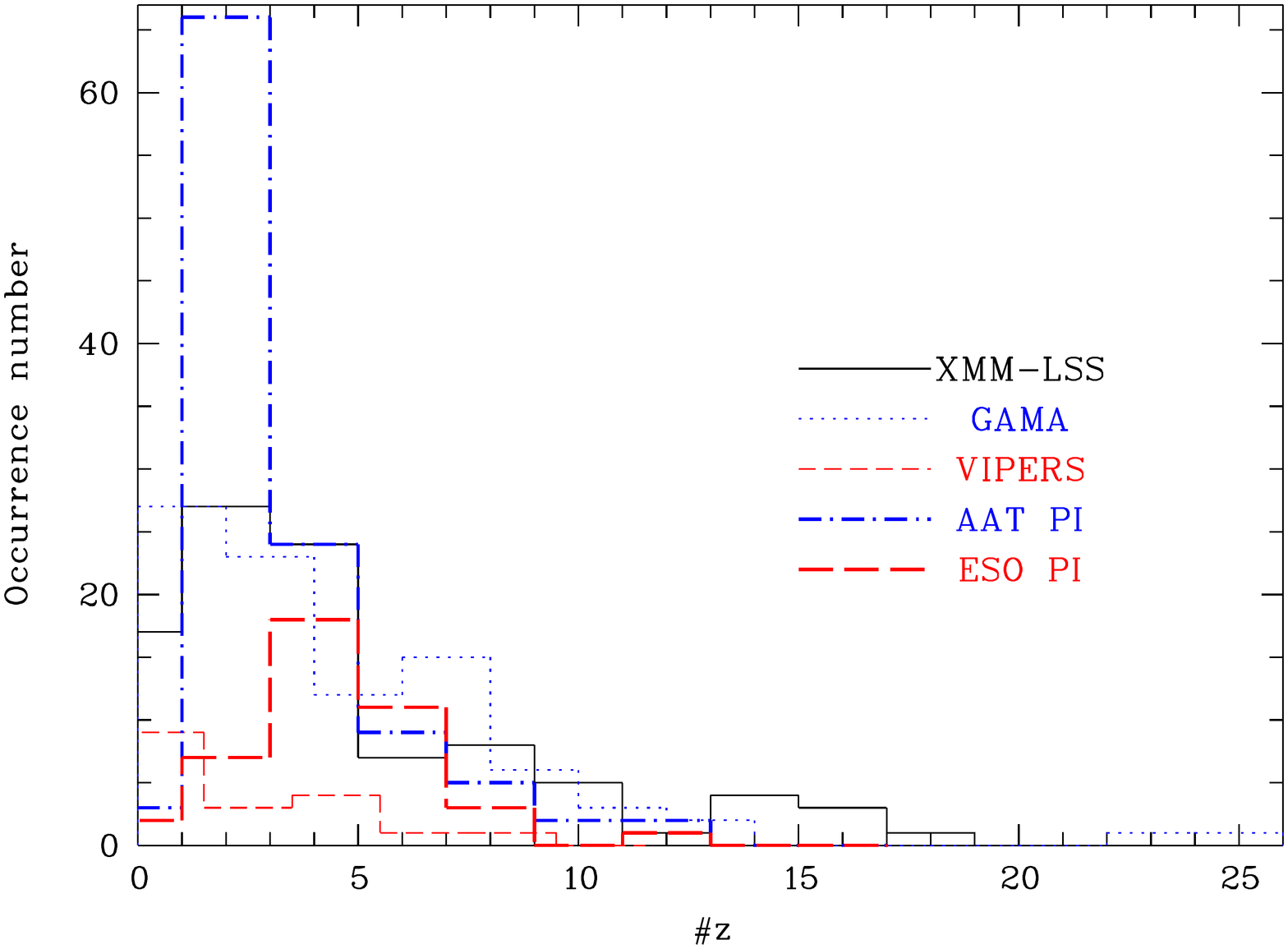}
\includegraphics[width=6.5cm,angle=270]{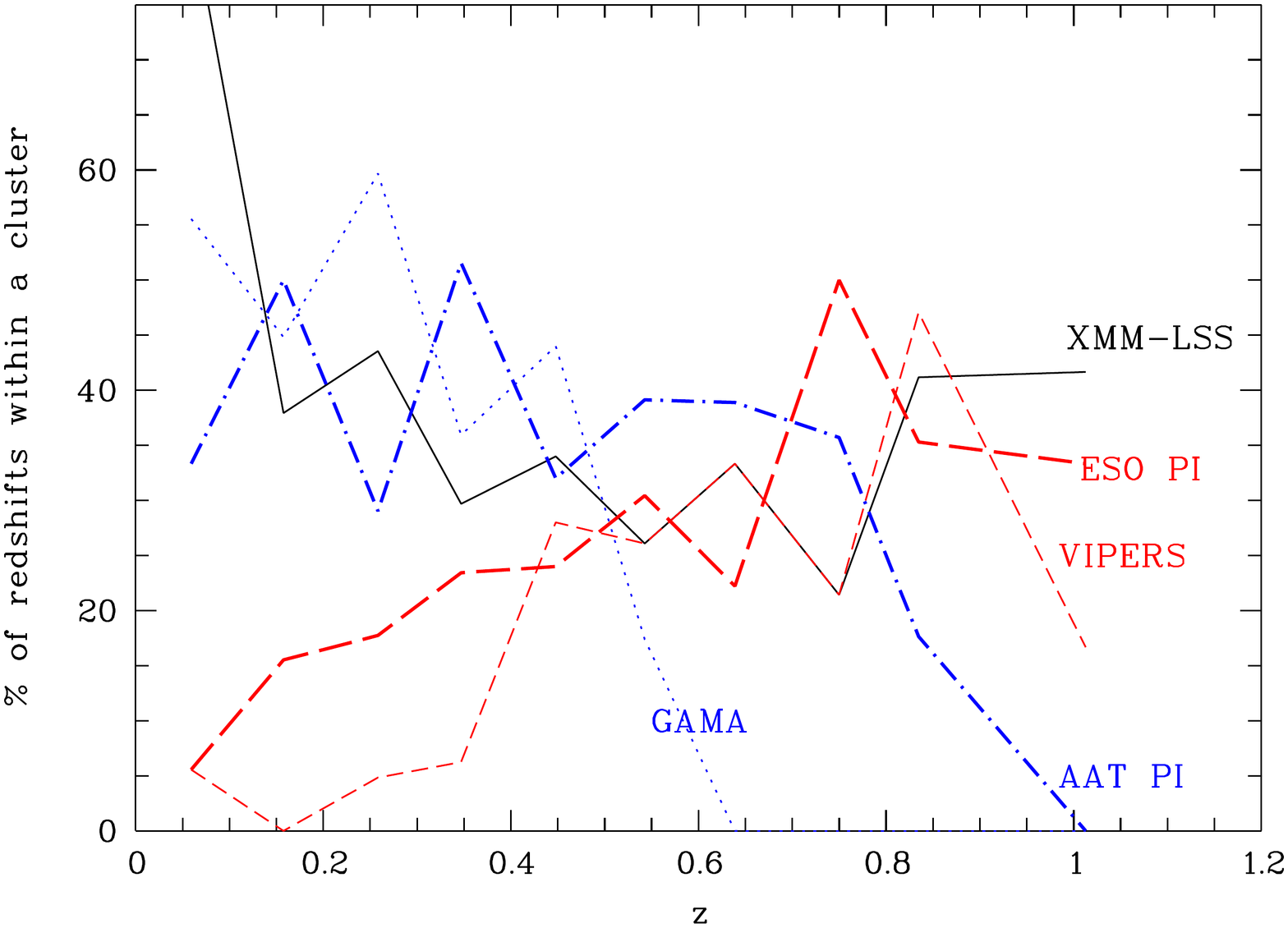}
\caption{\label{fig:contrib}Upper panel: y-axis: number of confirmed clusters. x-axis: number of galaxy redshifts sampling the confirmed clusters.
Different colours and line styles are from different spectroscopic surveys. Bottom panel: percentage of galaxy redshifts inside the confirmed clusters 
coming from a given survey and for a 
given redshift bin. Because of multiple galaxy spectroscopic measurements, the sum of the percentages for a given redshift bin is larger than 100$\%$.}
\end{figure}

In Figure~\ref{fig:contrib}, we give the contribution of the major spectroscopic surveys used in the present paper. This is showed both in terms of
the number of clusters with a given number of galaxy redshifts coming from a given spectroscopic survey, and in terms of 
number of galaxy redshifts coming from a given survey for a given redshift bin. This for example shows that the XXL ESO and XMM-LSS PI allocations were
efficient to confirm clusters in the z$\sim$[0.2-1] range while other major surveys were more specialised in terms of redshift coverage: VIPERS
at z$\geq$0.45, and AAT PI and GAMA at z$\leq$0.7 and z$\leq$0.4 respectively. 
In terms of cluster spectroscopic sampling, XXL ESO PI allocations enabled us to measure the largest number
of galaxy redshifts per cluster ($\sim$5); other surveys yielded various samplings.
The largest samplings are achieved by the XMM-LSS spectroscopic survey (most of the time for well identified peculiar or distant clusters) and
by the GAMA spectroscopic survey for nearby clusters.

Major surveys such as VIPERS or GAMA have science objectives related to field studies,
and are therefore under-represented in Figure 3 because only a small fraction
of these redshifts falls within a given cluster.
We therefore give in Table~\ref{tab:detailPI2} the mean numbers of redshifts per
line of sight (over the full redshift range of the XXL Survey, and within angular radii corresponding to 500$\thinspace$kpc at
the redshifts of the clusters). This allows us to appreciate the respective contribution of these surveys to the characterisation of
both clusters and projection effects. In such a table, intensive field surveys as VIPERS or GAMA show their great importance.

\begin{table}[t!]
\caption{\label{tab:detailPI2}Mean number of redshifts per cluster line of sight from the different surveys considered in Figure~\ref{fig:contrib} 
for the total XXL Survey, north, and south fields.}
\begin{tabular}{cccc}
\hline
\hline
Survey & XXL & XXL-N & XXL-S \\
\hline						
XMM-LSS & 9 & 15 &	1	\\
AAT PI &  /	  &  0    &	5	\\
VIPERS  &   /      &  16     &  0	\\
GAMA  &   /      &  10	     &  0	\\
ESO PI  &   2      & 2	     &  3	\\
\hline
\end{tabular}
\end{table}

\section{The cluster catalogue}

In this section, we first provide a global description of the sample. We then present the direct (spectral) measurements we made of luminosity, 
temperature, gas mass, and flux. These measurements are obviously more robust than using scaling relations, but they require higher 
quality data and therefore cannot be computed for the whole sample of clusters. Scaling relations were therefore used in order to complete the sample 
for some of the following studies.

\begin{table*}[t!]
\caption{\label{tab:C1C2cat}List of spectroscopically confirmed C1 and C2 clusters of galaxies. 
Col.1: official XLSSC name. Col.2 and 3: X-ray cluster coordinates. Col.4: cluster mean redshift. Col. 5: number of 
measured spectroscopic redshifts (X: means redshift is computed from X-ray spectroscopy directly). Col. 6: XXL class.
Col. 7: gas mass inside a physical radius of 500$\thinspace$kpc along with lower and upper uncertainties. Col. 8: r$_{500,MT}$.
Col. 9: X-ray temperature with lower and upper uncertainties. Col. 10: L$^{XXL}_{500,MT}$ X-ray luminosity and uncertainty
in the [0.5-2] keV rest-frame energy range. Col. 11: X-ray flux and uncertainty as in XXL paper II and in the [0.5-2] keV band.  
Col. 12, flags: "+" means the cluster was already published in the XMM-LSS releases, * means that we have a note on this cluster in the appendix, $l$ 
means that the considered cluster is brighter than the flux completeness limit ($\sim$ 1.3 $\times$ 10$^{-14}$ 
$\rm \thinspace erg \, s^{-1} \, cm^{-2}$), F means that the structure is a candidate fossil group. Complete table is available online (see text). 
Blank places are undetermined values (too low signal-to-noise).}
\begin{tabular}{cccccccccccc}
\hline
\hline
XLSSC  & $\alpha$ & $\delta$ & z & N$_{gal}$ & Class & M$_{gas, 500kpc}$  & r$_{500,MT}$ & T$_{300kpc}$ & L$^{XXL}_{500,MT}$  &  F$_{60}$    & flag \\
       &          &          &   &           &       & 10$^{11}$          &           &              & 10$^{42}$           &  10$^{-15}$  &  \\
       & deg      & deg      &   &           &       & M$\odot$           & kpc       & keV          & erg s$^{-1}$        &  $\rm erg \, s^{-1} \, cm^{-2}$   \\
\hline						
      199 & 	30.192	&	-6.708	&	0.339	&	2	&	1	&	73$_{-6}^{+4}$		     	 &	644	& 2.1$_{-0.3}^{+0.2}$	& 32$\pm$3        &      67$\pm$5	       	        &	   $l$    \\   
      200 &	30.331	&	-6.830	&	0.333	&	2	&	1	&	48$_{-3}^{+3}$		     	 &	653	& 2.1$_{-0.4}^{+0.3}$	& 16$\pm$2        &      31$\pm$3	       	        &	  	$l$    \\   
      114 &	30.425	&	-5.031	&	0.233	&	6	&	2	&	40$_{-3}^{+3}$		     	 &		&           	   	&         &      35$\pm$8	       	       &	  	$l$    \\   
      179 &	30.482	&	-6.574	&	0.608	&	5	&	1	&	43$_{-12}^{+11}$	     	 &	 	&           	   	&         &      14$\pm$4	       	       &	  	$l$    \\   
      113 &	30.561	&	-7.009	&	0.050	&	9	&	1	&	8$_{-1}^{+1}$		     	 &	 	&           	   	&         &      115$\pm$8	       &	  	$l$    \\   
      174 &	30.592	&	-5.899	&	0.235	&	8	&	1	&	41$_{-4}^{+3}$		     	 &	570	& 1.5$_{-0.1}^{+0.1}$	& 8$\pm$1	        &      25$\pm$4	       	               &	  	$l$    \\   
      094 &	30.648	&	-6.732	&	0.886	&	3	&	1	&	106$_{-12}^{+12}$	     	 &	581	& 3.0$_{-0.6}^{+0.5}$	& 224$\pm$32      &      48$\pm$5	       	        &	  +$l$    \\   
      196 &	30.728	&	-7.652	&	0.136	&	8	&	1	&	26$_{-3}^{+2}$		     	 &	563	& 1.3$_{-0.2}^{+0.1}$	& 4$\pm$1	        &      32$\pm$4	       	               &	  	$l$    \\   
      178 &	30.753	&	-6.285	&	0.194	&	2	&	2	&	29$_{-5}^{+3}$		     	 &	655	& 0.8$_{-0.1}^{+0.1}$	& 3$\pm$1	        &      17$\pm$3	       	               &	  	$l$    \\   
      156 &	30.766	&	-7.101	&	0.336	&	4	&	2	&	33$_{-3}^{+3}$		     	 &	 	&           	   	&         &      28$\pm$4	       	       &	  	$l$    \\   
      157 &	30.865	&	-6.929	&	0.585	&	5	&	1	&	70$_{-7}^{+7}$		     	 &	721	& 3.2$_{-0.7}^{+0.8}$	& 42$\pm$7        &      19$\pm$3	       	        &	  	$l$    \\   
      197 &	30.923	&	-7.785	&	0.439	&	2	&	1	&	107$_{-5}^{+5}$		     	 &	755	& 3.0$_{-0.5}^{+0.4}$	& 76$\pm$9        &      97$\pm$7	       	        &	  	$l$    \\   
      096 &	30.973	&	-5.027	&	0.520	&	6	&	1	&	89$_{-5}^{+5}$		     	 &	951	& 5.0$_{-0.5}^{+0.9}$	& 63$\pm$8        &      36$\pm$4	       	        &	  *+$l$   \\   
      155 &	31.134	&	-6.748	&	0.433	&	2	&	1	&	36$_{-5}^{+4}$		     	 &	576	& 1.8$_{-0.3}^{+0.3}$	& 16$\pm$3        &      23$\pm$3	       	        &	  	$l$    \\   
      173 &	31.251	&	-5.931	&	0.413	&	3	&	1	&	47$_{-4}^{+4}$		     	 &	930	& 4.3$_{-0.3}^{+0.3}$	& 17$\pm$2        &      24$\pm$3	       	        &	  	$l$    \\   
      177 &	31.290	&	-4.918	&	0.211	&	7	&	2	&	37$_{-3}^{+3}$		     	 &	 	&           	   	&         &      22$\pm$4	       	       &	  	$l$    \\   
      102 &	31.322	&	-4.652	&	0.969	&	3	&	1	&	138$_{-7}^{+7}$		     	 &	638	& 3.9$_{-0.9}^{+0.8}$	& 167$\pm$25      &      42$\pm$4	       	        &	  +$l$    \\   
      106 &	31.351	&	-5.732	&	0.300	&	14	&	1	&	83$_{-3}^{+3}$		     	 &	777	& 2.8$_{-0.3}^{+0.2}$	& 43$\pm$3        &      91$\pm$4	       	        &	  +$l$    \\   
      107 &	31.354	&	-7.594	&	0.436	&	3	&	1	&	67$_{-5}^{+4}$		     	 &	672	& 2.4$_{-0.4}^{+0.4}$	& 49$\pm$6        &      56$\pm$5	       	        &	  +$l$	  \\   
      160 &	31.521	&	-5.194	&	0.817	&	4	&	2	&	        		     	 &	 	&           	   	&         &      6$\pm$4		       &	  	          \\   
\hline																				         
\end{tabular}																			         
\end{table*}																			         

\subsection{Sample description}

The C1 + C2 clusters are listed in Table~\ref{tab:C1C2cat}. The table is sorted according to increasing RA and 
only the first twenty entries are displayed. Blank places in the Table are undetermined values. 
We note that the XLSSC 634 cluster was confirmed by Ruel et al. (2014) with Gemini/GMOS data. 
The spectroscopically confirmed C3 objects are listed in Table~\ref{tab:C3cat}. Both tables are also available in the XXL Master Catalogue browser 
at http://cosmosdb.iasf-milano.inaf.it/XXL/ and Table~\ref{tab:C1C2cat} is available at the CDS. For each source, we provide (when available):\\
- the XLSSC identifier (between 1 and 499, or 500 and 999 for XXL-N or XXL-S respectively\\
- RA and Dec\\
- the redshift and the number of galaxies used for the redshift determination \\
- the class, C1, C2 (Table~\ref{tab:C1C2cat} only) or C3 (Table~\ref{tab:C3cat} only)\\
- basic X-ray and X-ray related quantities for the clusters of the present release (X-ray fluxes, M$_{gas, 500kpc}$, r$_{500,MT}$, T$_{300kpc}$, 
and L$^{XXL}_{500,MT}$). We note that we give in the present paper the value of M$_{gas, 500kpc}$, contrary to what was given in XXL paper XIII where
M$_{gas, 500}$ was provided. \\
- a flag indicating whether there is a note on the cluster in one of the appendices, whether the 
cluster was already published in XXL paper II or in former XMM-LSS releases, and whether the cluster is a member of the flux limited sample.\\

\subsection{X-ray direct measurements}

\subsubsection{Luminosity and temperature}

Full details of the analysis of the cluster X-ray properties  can be found in XXL Paper XXVI (Giles et al., in preparation), and we 
outline the main steps of the spectral analysis here. First, we only used the single best pointings for spectral analyses when sources fell on multiple 
pointings. As a conservative approach, the extent of the cluster emission was defined as the radius beyond which no significant cluster 
emission is detected using a threshold of 0.5$\sigma$ above the background level.  Due to the low number of counts and low 
signal-to-noise of many of the clusters below the XXL-100-GC threshold, we performed a detailed modelling of the background, instead of a simple 
background subtraction.  We followed the method outlined in Eckert et al. (2014), who performed this detailed modelling 
to study a source whose emission barely exceeded the background.  We modelled the non X-ray background (NXB) using
 closed filter observations, following a phenomenological model. For observations contaminated by soft protons (where 
the count rate ratio between the in-FOV, beyond 10 arcminutes, and out-of-FOV regions of the detector was $>$1.15), we included an 
additional broken power-law component, with the slopes fixed at 0.4 and 0.8 below and above 5 keV respectively.  The sky background 
was modelled using data extracted from an offset region (outside the cluster emission determined above), using a three-component model as detailed in 
Eckert et al. (2014). Within the XSPEC environment, cluster source spectra were extracted for each of the XMM-Newton cameras and fits were performed in 
the [0.4-11.0] keV band with an absorbed APEC (Astrophysical Plasma Emission Code, Smith et al., 2001) model (v2.0.2), with a fixed 
metal abundance of Z=0.3Z$_{\odot}$.

We denote the luminosity within r$_{500,MT}$ \footnote{$r_{500,MT}$ is defined as the 
radius of the sphere inside which the mean density is 500 times the critical density $\rho_c$ of the Universe at the cluster's redshift,
$M_{500,MT}$ is then by definition equal to $4/3 \pi 500 \rho_c r_{500,MT}^3$} as L$^{XXL}_{500,MT}$, within the [0.5-2.0] keV band (cluster rest frame).
Luminosities quoted within r$_{500,MT}$ are extrapolated from 300$\thinspace$kpc (see below) out to r$_{500,MT}$ by integrating under a $\beta$-profile 
assuming a core radius $r_{\rm c}$=0.15r$_{500,MT}$ and an external slope $\beta$=0.667 (cf. XXL paper II).
Values for cluster r$_{500,MT}$ are calculated using the mass-temperature relation of Lieu et al. (2016: hereafter XXL paper IV).

Given that we are dealing with much fainter sources than in XXL paper II, it was not possible to measure X-ray temperatures for all clusters. 
In particular, several C1 clusters were located in pointings affected by flaring, had very low counts, were contaminated by point sources, or were at 
very low redshift so with a bad spatial coverage. 

\subsubsection{Gas mass}

We analytically computed gas masses for clusters with redshifts following closely the method outlined in Eckert et al. (2016; 
hereafter XXL paper XIII). Here we briefly recall the various steps of the analysis. First, we extract surface-brightness profiles in the [0.5-2] keV band starting from the 
X-ray peak 
using the {\sc Proffit} package (Eckert et al. 2011). We compute the surface-brightness profiles from mosaic images of the XXL fields instead of individual pointings, which 
allows us to improve the signal-to-noise ratio and measure the local background level more robustly compared to the analysis presented in XXL paper XIII. The surface-brightness 
profiles are then deprojected by decomposing the profile onto a basis of multiscale parametric forms. Cash (1979) statistics are used to adjust the model to the data, and the Markov 
chain Monte Carlo (MCMC) tool {\sc emcee} (Foreman-Mackey et al. 2013) is used to sample the large parameter space. 
The deprojected profiles are then converted into gas density profiles using X-ray cooling functions calculated using the APEC plasma emission code (Smith et al. 2001). Finally, 
the recovered gas density profiles are integrated over the volume within a fixed physical scale of 500$\thinspace$kpc. The gas masses measured for XXL-100-GC clusters using this procedure are 
consistent with the values published in XXL Paper XIII, with a mean value M$_{new}$/M$_{old}$ = 0.984. For more details on the analysis procedure we refer the reader to XXL Paper XIII.
In Table~\ref{tab:C1C2cat}, we give only the gas masses for clusters with an uncertainty on the flux F$_{60}$ (see below) lower than the third of the flux itself. 
We also similarly do not provide gas mass estimates for C3 clusters.

\subsubsection{X-ray flux}

To be able to directly compare our estimate of the X-ray luminosity function (see next section) with the results
of XXL paper II, we adopted for the X-ray photometry the same procedure to estimate aperture fluxes in a radius of 60\arcsec (F$_{60}$).
We performed the measurements on the pointing within which each cluster was most significantly detected - as indicated by the 
C1/C2/C3 classification. This approach was preferred compared to the other approach consisting of combining all available pointings
for a given cluster as it allowed us to keep good spatial resolution for the shape estimate.
Whenever a cluster was detected in several pointings with the same classification, we therefore retained the one
where the cluster was closest to the optical axis. The analysis then relies on a semi-interactive procedure initially developed 
for Clerc et al. (2012). It first defines a preliminary source mask based on the output of the XXL detection pipeline and allows 
the user to manually correct the mask. Then the signal in a user-defined background annulus around the source is modelled with 
a linear fit to the local exposure map (thus allowing for both a vignetted and an unvignetted background component). Finally, 
count-rates in each detector are estimated, propagating the errors in the background determination, and turned into a global 
flux using average energy conversion factors relevant to each field\footnote{Those assume an APEC v2.0.2 thermal spectrum with 
T=2\,keV and Z=0.3\,$\mathrm{Z_{\odot}}$. The difference between the two fields comes from their average absobing column 
density of $n_H=2.3{\times}10^{20}\,\mathrm{cm}^{-2}$ for XXL-N and $1.25{\times}10^{20}\,\mathrm{cm}^{-2}$  for XXL-S.}.
Of course the final estimated flux depends somewhat on the chosen background sample. In our case, the sizes of the adopted 
background annuli varied significantly, reflecting the large spread in cluster size and flux in the catalogue. They ranged 
from 90 to 300\arcsec\ for the inner radius and 180 to 500\arcsec\ for the outer bound. The shifts in the measured fluxes 
recorded when changing the background aperture were always well within the statistical errors, provided that the background
annulus was free from apparent cluster emission.

\subsection{Cluster parameters from scaling relations}

In order to allow studies of the global properties of the full sample, we also provide mean parameter estimates derived from scaling relations 
(Table~\ref{tab:listeSL}).

To estimate luminosity and temperature from scaling relations (without a spectral fit), we first extracted the XMM-Newton pn in the [0.5-2] keV band 
within 300$\thinspace$kpc from the cluster centre.
Count rates were computed starting from values and bounds for the intensity S of the source using counts and exposure data obtained in source and background 
apertures. The background-marginalised posterior probability distribution function (PDF) of the source was then
calculated, assuming Poisson likelihoods for the detected number of source counts and background counts in the given 
exposure time. The mode of this PDF was determined, and the lower and upper bounds of the confidence region were determined by
summing values of the PDF alternately above and below the mode until the desired confidence level was attained. When the
mode was at S = 0 or the calculation for the lower bound reached
the value S = 0, only the upper confidence bound was evaluated, and was considered as an upper limit. 

We converted this count rate to the corresponding X-ray luminosity by adopting an initial gas temperature, a 
metallicity set to 0.3 times the solar value (as tabulated in Anders $\&$ Grevesse 1989)
 and the cluster’s redshift (without propagating the redshift uncertainties). The same value of the temperature is used to estimate 
$r_{500,MT}$, using the mass-temperature relation for the sample XXL+COSMOS+CCCP in Table~2 
of XXL paper IV. The luminosity is then extrapolated from 300$\thinspace$kpc out to $r_{500,scal}$ (similar as $r_{500,MT}$ but computed during the 
process of the cluster parameters estimate from scaling relations) by integrating over the cluster’s emissivity 
represented by a $\beta-$model with parameters $(r_c, \beta) = (0.15 r_{500,scal}, 2/3)$. Hence, a new temperature is evaluated from the best-fit 
results for the luminosity--temperature relation quoted in Table~2 of Giles et al. (2016: hereafter XXL paper III). 
The iteration on the gas temperature is stopped when the input and output values agree within a tolerance value of 5$\%$.

\begin{figure}[h]
\includegraphics[width=8.5cm,angle=0,bb=0 365 560 760]{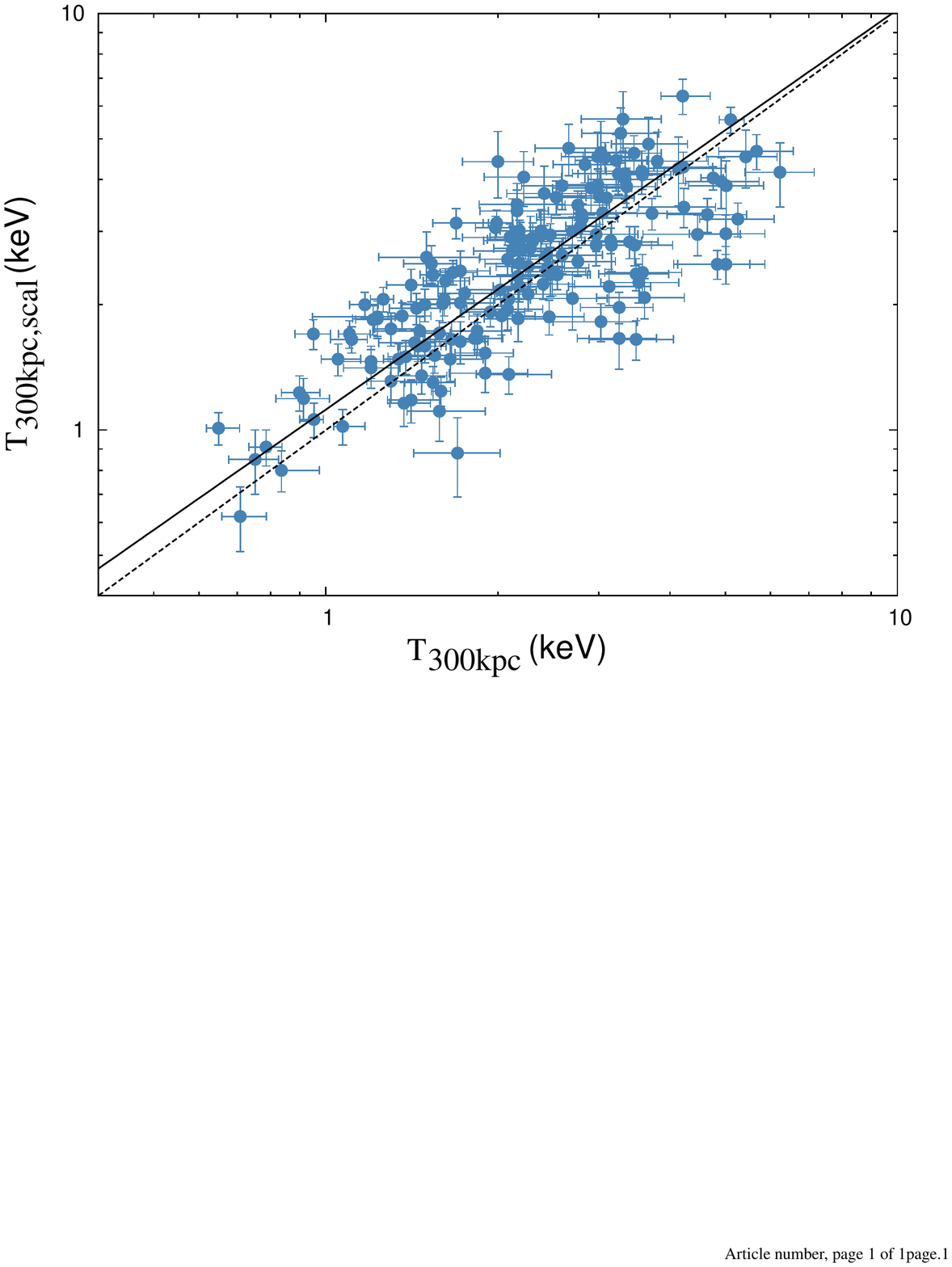}
\caption{\label{compareT} Comparison between the true temperature measurements (from Table~\ref{tab:C1C2cat}) and estimates from the scaling relations
(from Table~\ref{tab:listeSL}). The dotted and solid lines show the 1:1 relation and the actual regression to the data respectively.}
\end{figure}

Usually, this process converges in few steps (2--3 iterations). We provide estimates of the X-ray temperature, T$_{300kpc,scal}$, 
of the bolometric luminosity in the [0.5-2] keV range within $r_{500,scal}$, L$^{bol}_{500,scal}$, of the mass M$_{500,scal}$ within $r_{500,scal}$, and of 
relative errors propagated from the best-fit results of the X-ray temperature, $r_{500,scal}$, and the bolometric luminosity.
A comparison between the measured cluster temperatures and those obtained from the scaling relations is displayed in Fig.~\ref{compareT}; the observed scatter 
around the 1:1 line simply reflects the intrinsic scatter of the luminosity-temperature relation. 
In some cases (mainly for C2 clusters), this procedure converges to an M$_{500,scal}$ value that falls below the mass range of the XXL-100-GC sample
(cf. XXL paper IV), used for derivation of the scaling relations. In this case, no values are given.

\section{Updated cluster statistics}

With the current sample having twice as many C1 clusters as in XXL-100-GC (and 341 spectroscopically confirmed clusters in total), we are in a 
position to update a number of statistical results presented in the 2016 XXL release (a.k.a. DR1). Detailed analyses of these quantities in the 
current XXL-C1-GC sample will however be the subject of forthcoming papers. In this paper, we concentrate on a few basic properties of the 
XXL-C1-GC.

Regarding the 207 C1 clusters of XXL-C1-GC, only 191 are in pointings not affected by flares. All results involving the cluster selection function
are therefore based on this subsample of 191 objects. 

Five among these 191 clusters do not have a redshift determination and are therefore 
modelled using an incompletness factor in the selection function. Excluding these five, the remaining sample of 186 clusters is used to compute the 
cluster luminosity function.

Eight out of these 186 clusters have no temperature measurement and their X-ray luminosity was estimated through scaling relations.  This 
sample of 176 clusters is used to constrain the luminosity-temperature relation.

\subsection{Redshift distribution and spectroscopic redshift sampling}

\begin{figure}[h]
\includegraphics[width=6.5cm,angle=270]{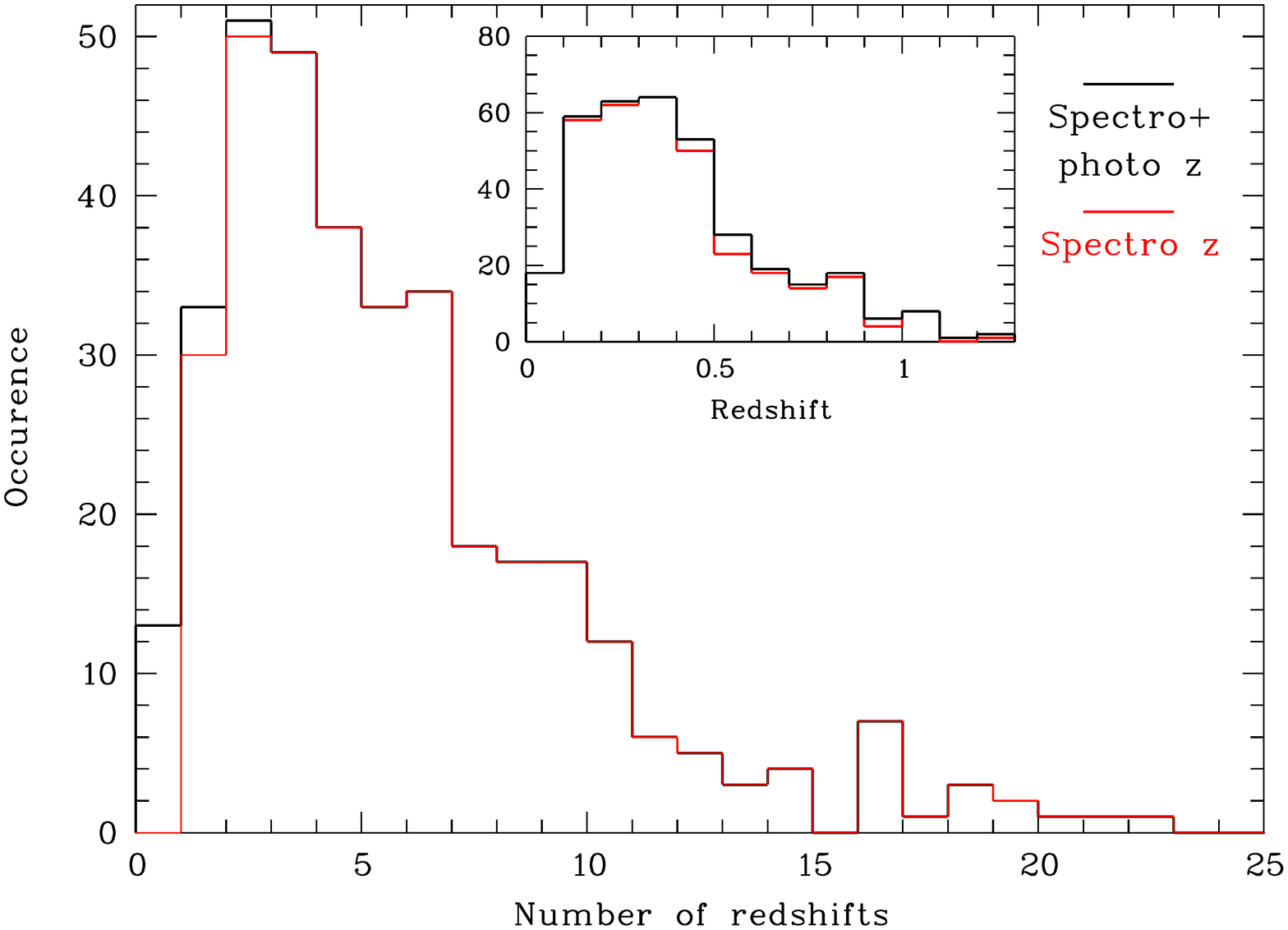}
\includegraphics[width=6.5cm,angle=270]{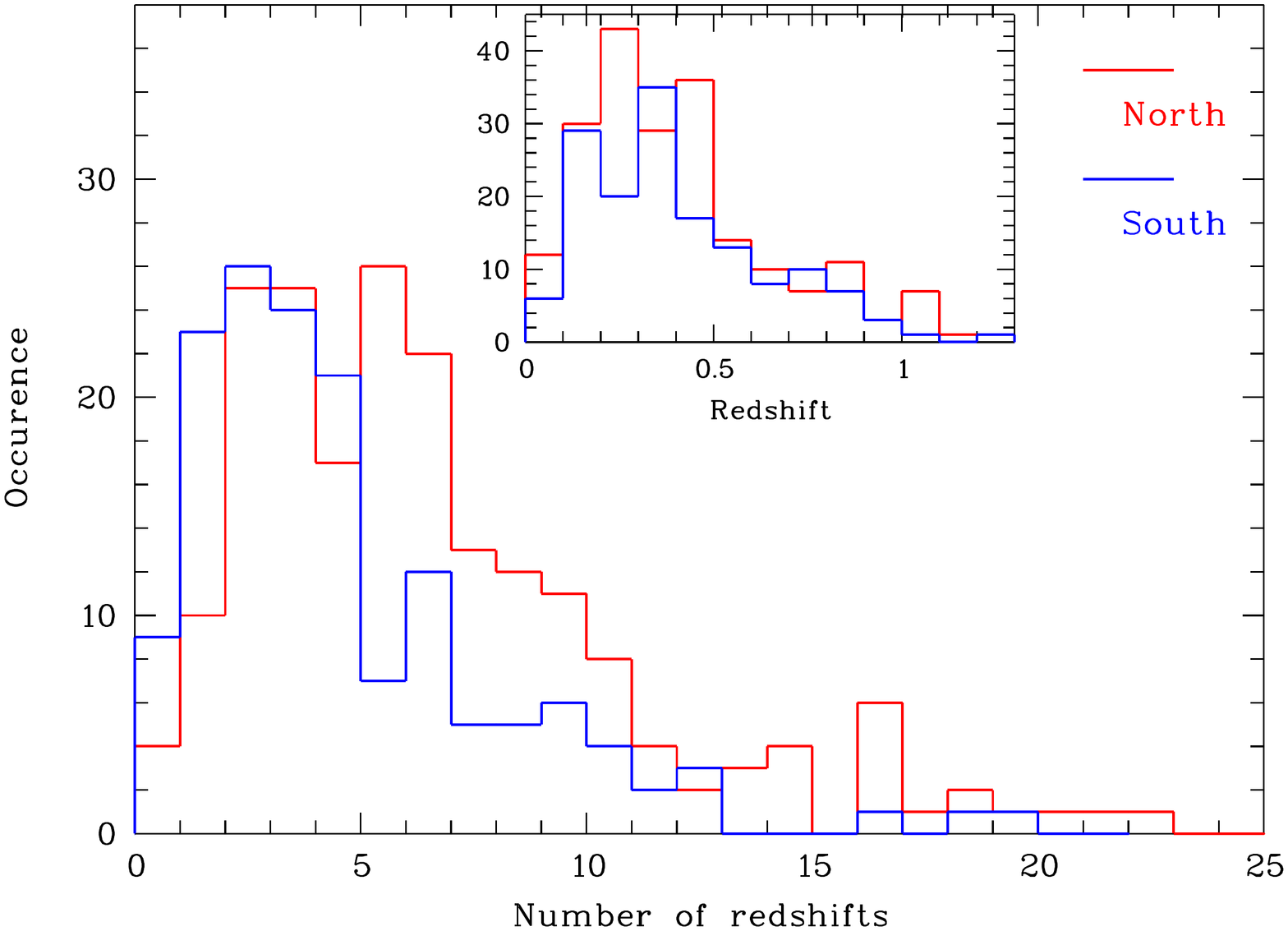}
\caption{\label{Zdist1} Distribution of the number of spectroscopic redshifts inside clusters with a redshift measurement. The insets show the 
redshift 
histograms of these samples. Top panel: Spectroscopic + photometric redshift sample (black histograms), and spectroscopic redshift sample (red 
histograms) clusters. Bottom panel: XXL-N (red histograms) and XXL-S (blue histograms) clusters. Photometric redshifts are used in replacement
of spectroscopic redshifts in these two histograms when spectroscopic redshifts are not available.}
\end{figure}

\begin{figure}[h]
\includegraphics[width=6.5cm,angle=270]{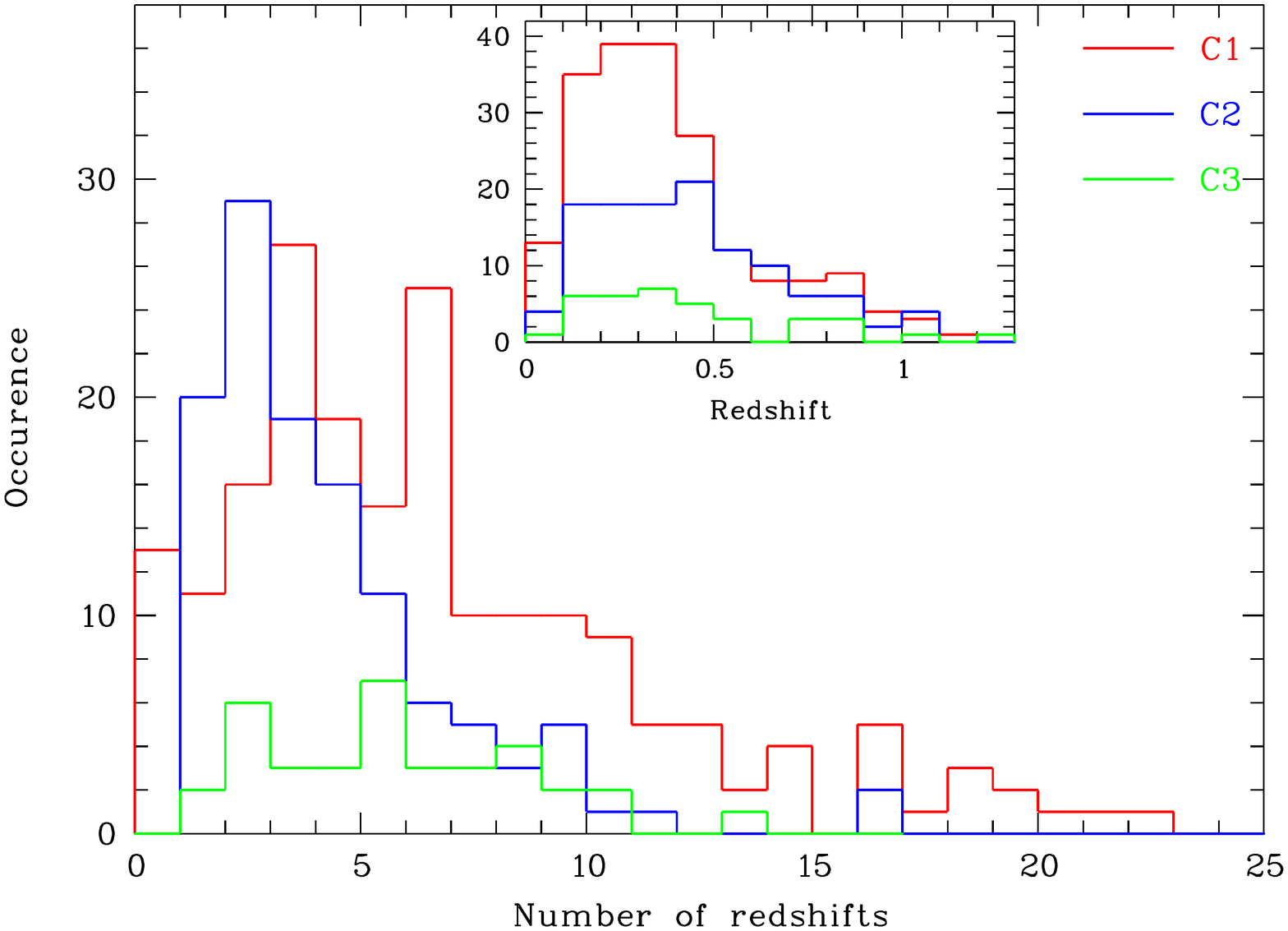}
\includegraphics[width=6.5cm,angle=270]{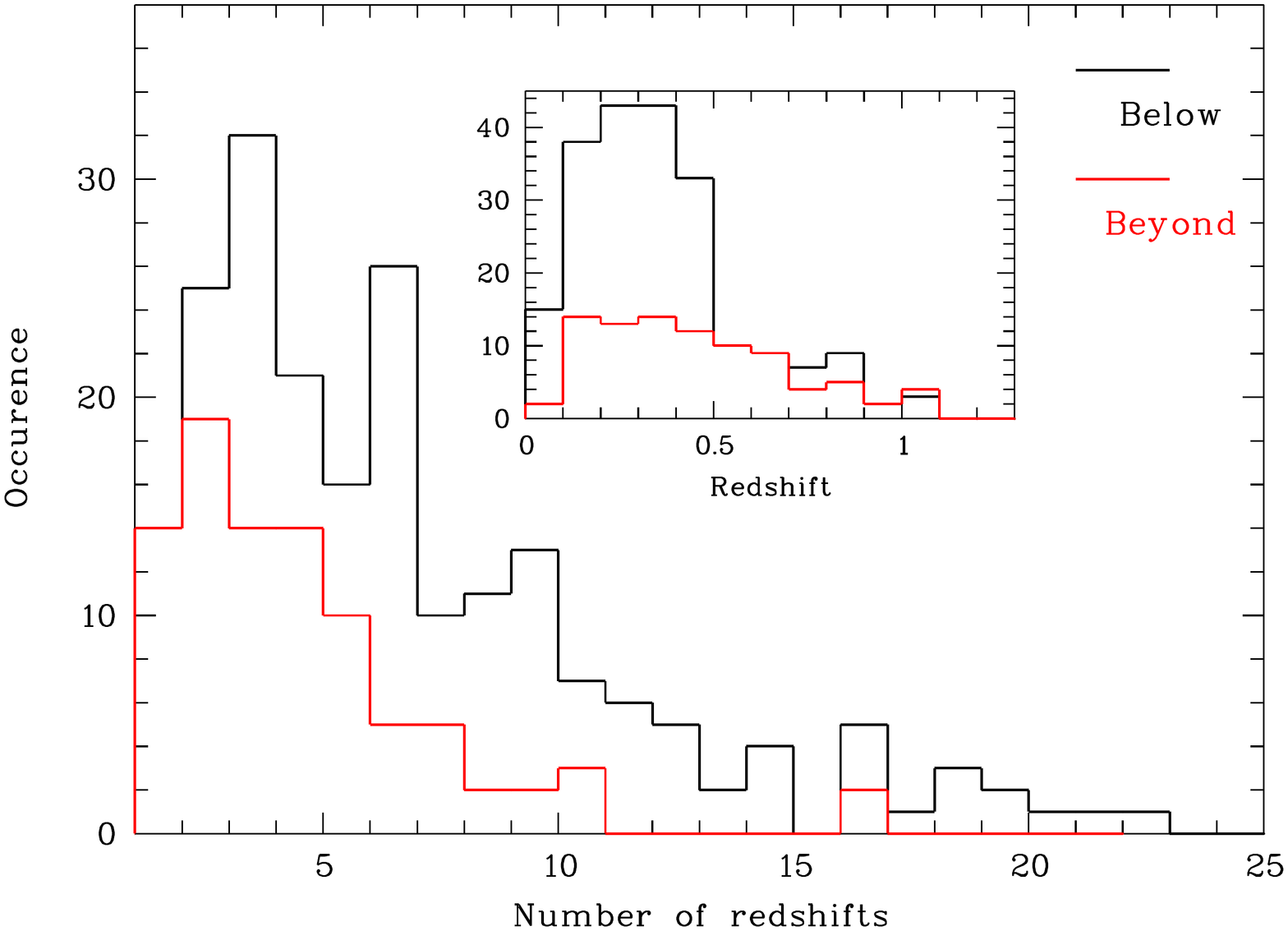}
\caption{\label{Zdist2} Distribution of the number of spectroscopic redshifts inside clusters with a redshift estimate. The insets show the 
redshift histograms of 
these samples. Top panel: C1 (red histograms), C2 (blue histograms), and C3 (green histograms) clusters. Bottom panel: clusters with also a flux 
estimate fainter (black histograms) 
and brighter (red histograms) than the reference flux completeness limit of 1.3 $\times$ 10$^{-14}$ $\rm \thinspace erg \, s^{-1} \, cm^{-2}$.}
\end{figure}

The galaxy redshift sampling of clusters and the cluster redshift distributions are displayed in Fig.~\ref{Zdist1} and Fig.~\ref{Zdist2} for various 
cluster selections. Our total sample is the full list of clusters
quoted in the present paper, including the few not yet spectroscopically confirmed clusters in Table \ref{tab:listetot7}. 

We see that the full list is very similar to the list of spectroscopically confirmed clusters, cf. top panel of Figs~\ref{Zdist1}. A 
Kolmogorov-Smirnov test shows no difference (at better than the 99.9$\%$ level) both for the redshift and the redshift sampling distributions.
This figure also shows that, among the non-spectroscopically confirmed clusters, thirteen do not have any spectroscopic redshift, three of them
have a single spectroscopic redshift (not the BCG), and one has two spectroscopic redshifts (the BCG being not 
available, spectroscopic confirmation is not validated either).

The XXL-N and XXL-S cluster samples are also similar in terms of redshift distribution (99.9$\%$level for a Kolmogorov-Smirnov
test). We however have on average more spectroscopically confirmed members
(typically more than six spectroscopic redshifts) in the northern field compared to the southern field (see below for a more quantitative analysis of the 
cluster sampling). The probability of having similar samples is only at the 28$\%$level with a Kolmogorov-Smirnov test.

C1, C2, and C3 cluster distributions are obviously different, as demonstrated by a Kolmogorov-Smirnov
test. C2 and C3 clusters have lower spectroscopic sampling than C1 as these were not our primary spectroscopic targets. 
C3 mainly appears as a subpopulation of intermediate redshift clusters, with also a few distant (z$\geq$1) structures.

Finally, clusters brighter and fainter than the reference flux completeness limit of  1.3 $\times$ 10$^{-14}$ $\rm \thinspace erg \, s^{-1} \, 
cm^{-2}$ cover almost the same redshift range. Their redshift distribution is however different (probability of having similar samples only at the 
53$\%$ level) with, not surprisingly, a lot more bright clusters at redshifts below 0.5. They also are very 
different (at the 98$\%$ level) in terms of spectroscopic sampling, the brightest clusters being better spectroscopically sampled.

\subsection{X-ray luminosities and fluxes}

\begin{figure}[h]
\includegraphics[width=6.5cm,angle=270]{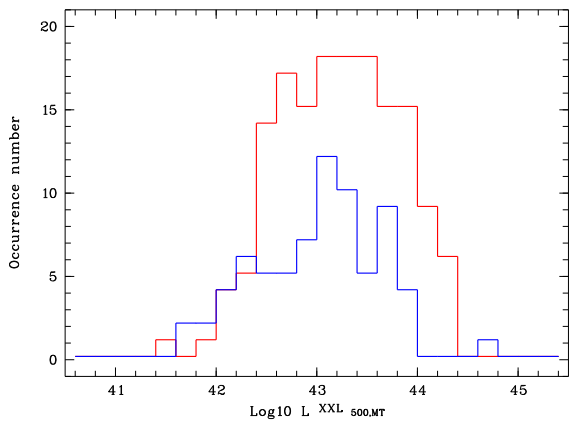}
\caption{\label{LZdist} X-ray luminosity (L$^{XXL}_{500,MT}$ in log unit of erg s$^{-1}$ in the [0.5-2] keV band) distribution of clusters having a spectroscopic redshift and 
a luminosity determination. Red histogram: the C1 sample; blue histogram: the C2 sample.}
\end{figure}

\begin{figure}[h]
\includegraphics[width=6.5cm,angle=270]{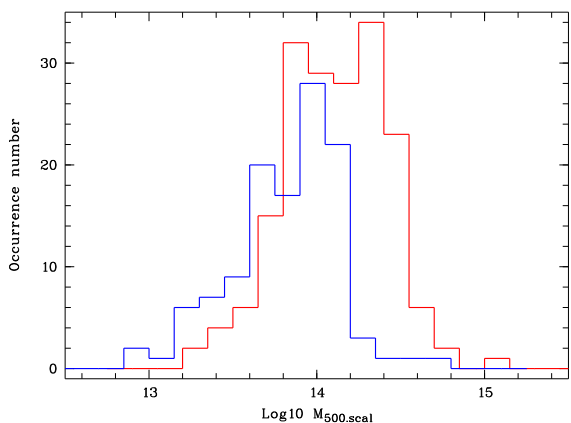}
\caption{\label{MZdist} Mass (in log units of M$\odot$) distribution of the clusters with a spectroscopic redshift estimate. Red histogram: 
the C1 sample; blue histogram: the C2 sample. The mass data points have been derived from scaling relations based on the cluster 
luminosities (cf. section 4.3 and appendix F).}
\end{figure}

We display in Fig.\ref{LZdist} the distribution of cluster luminosities L$^{XXL}_{500,MT}$ (only when available through spectral fit, so C3 clusters are excluded) for the C1 and 
the C2 samples. In addition, Fig.\ref{MZdist} shows the cluster mass M$_{500,scal}$ (derived from scaling relations) distribution for the same subsamples. We note that the cluster masses 
do not pertain here to direct spectral measurements (since temperatures are not available for the 
full sample) but were derived using scaling relations; we show these graphs to allow global comparisons with other cluster samples. In XXL paper XIII, we mentioned 
the possibility that our total CFHTLS lensing masses were overestimated. Deep Subaru-HSC observations will provide higher signal to noise information and help us
understand the contribution of non-thermal pressure in the total mass budget (Umetsu et al in prep).

\begin{figure}[h]
\includegraphics[width=6.5cm,angle=270]{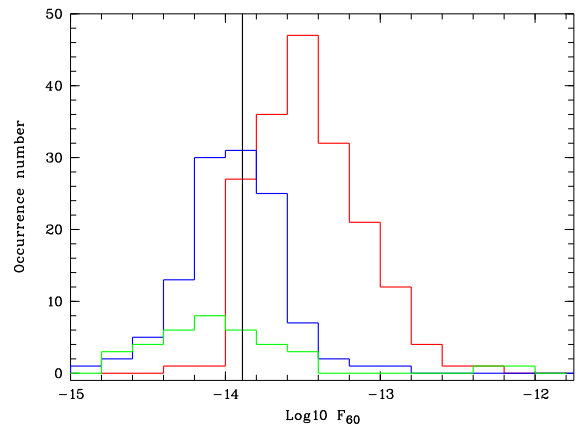}
\caption{\label{F60} X-ray flux (F$_{60}$ in log unit of $\rm erg \, s^{-1} \, cm^{-2}$, within a 60'' radius in the [0.5-2] keV band) distribution for the clusters having 
a spectroscopic redshift. Red histogram: the C1 sample; blue histogram: the C2 sample; green histogram: the C3 sample. The black vertical line is the estimated
 reference flux completeness limit of 1.3 $\times$ 10$^{-14}$ $\rm \thinspace erg \, s^{-1} \, cm^{-2}$.}
\end{figure}

Finally, in order to compare the C1 and C2 subsamples with the C3 subsample, we show in Fig.\ref{F60} the F$_{60}$ (flux within a 60'' radius in the [0.5-2] keV band) distribution of 
the three subsamples. As expected, C1 clusters are brighter than the C2 clusters. C3 clusters pertain to two distinct populations as already stated in the previous section and showed 
in Adami et al. (2011). A large part of them are structures slightly fainter than the C2 clusters, and a few are bright and distant structures.

\begin{figure}[h]
\includegraphics[width=8.5cm,angle=0]{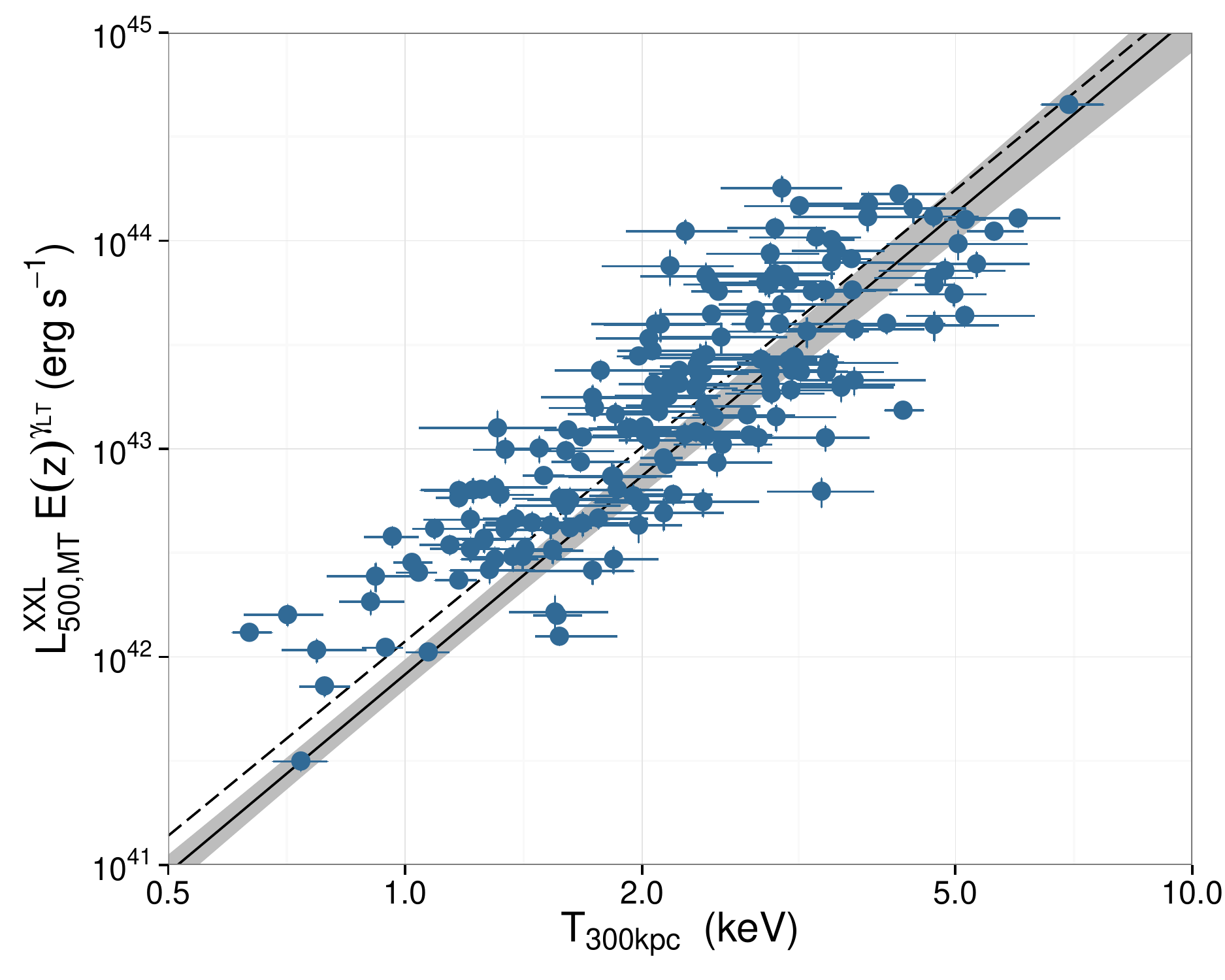}
\includegraphics[width=8.5cm,angle=0]{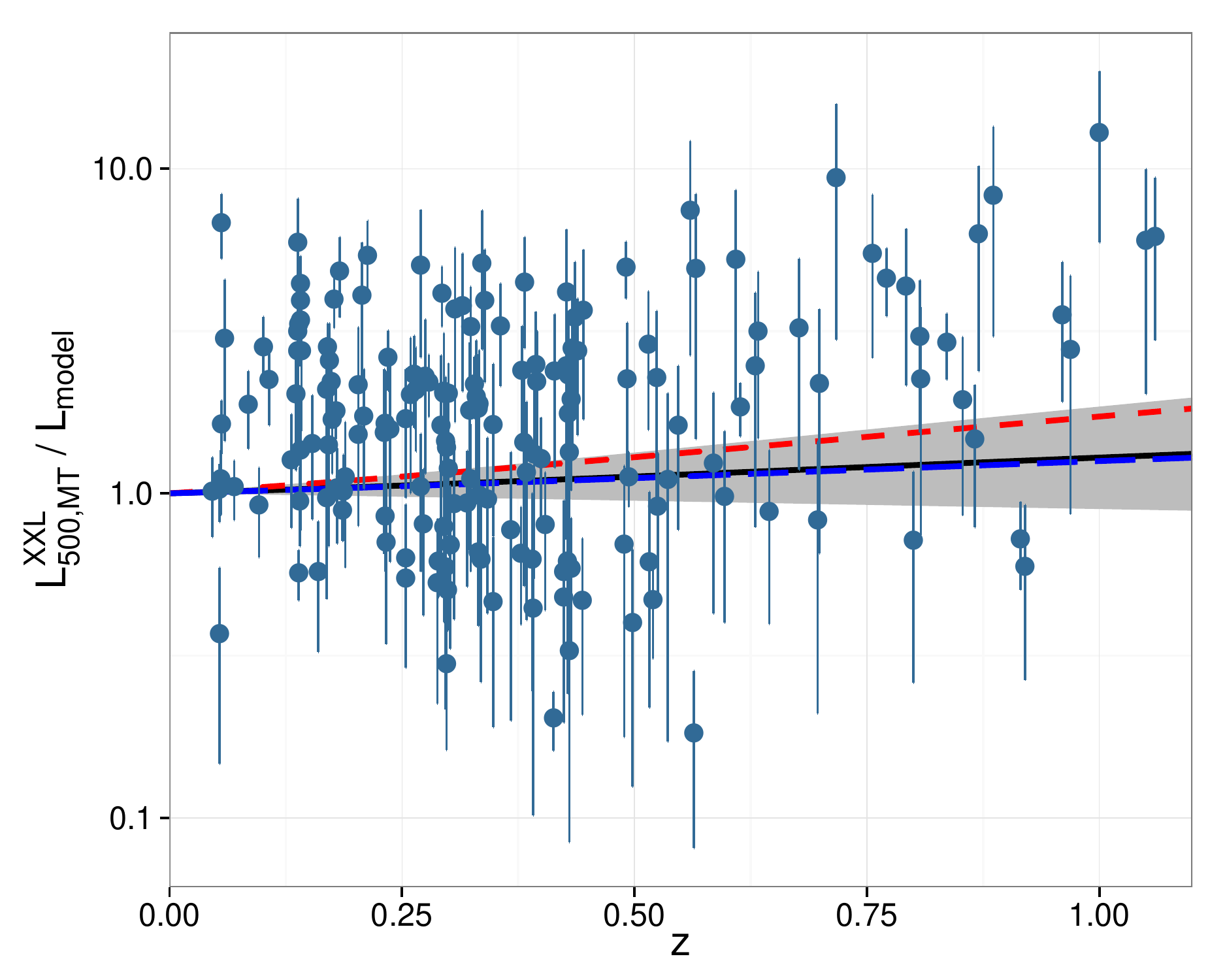}
\caption{\label{ls} Upper panel: Luminosity-temperature relation with the best-fitting
  models.  The light blue circles show the XXL-C1-GC clusters;  the
  best-fitting model (including selection effects) is shown by the solid black line, the 1$\sigma$
  uncertainty represented by the grey shaded region.  The best-fitting model
fitted to the data using the BCES regression is shown as the dashed line.
Bottom panel: Evolution
  of the luminosity-temperature relation for XXL-C1-GC.  The XXL-C1-GC clusters are
  represented by the light blue circles and the best-fitting model is
  given by the black solid line;  the grey shaded region highlights
  the 1$\sigma$ uncertainty.  The `strong' and `weak' self-similar
  expectations are given by the red dashed and blue dashed lines,
  respectively.}
\end{figure}

\subsection{Luminosity-temperature relation of the C1 sample}

Fig.~\ref{ls} shows the XXL luminosity-temperature relation for the XXL-C1-GC sample (both parameters derived from spectral measurements). A fit to the data using a power law of the form

\begin{align}
\hspace{2.5cm}\left(\frac{L}{L_{0}}\right) &= E(z)^{\gamma_{LT}}A_{\rm LT}\left(\frac{T}{T_{0}}\right)^{B_{\rm LT}}
\label{equ:ltunbias}
\end{align}

was performed, where $A_{LT}$, $B_{LT}$, and $\gamma_{LT}$ represent the normalisation, slope, and power of the evolution correction respectively.  
The power law was fit to the data, first using the BCES orthogonal regression in base ten log space (Akritas $\&$ Bershady, 1996) assuming self-similar 
evolution ($\gamma_{LT}$=1).  The best fit parameters are given in Table~\ref{tab:lttab}.  Comparing the XXL-C1-GC BCES fit to the XXL-100-GC fit, we find that the slope and
normalisation are consistent.     

We next fit the XXL-C1-GC scaling relation using the procedure outlined in XXL paper III, taking fully into account the selection effects (we refer to
Sections 4.3 and 4.4 in XXL paper III for specific details).  However, the selection function was updated to match the current sample, instead
of  the XXL-100-GC selection function previously used. Figure~\ref{ls} (upper panel) shows the XXL luminosity-temperature relation, with the 
best-fitting (bias-corrected) model given by the black solid line and the corresponding 1$\sigma$ uncertainty shown by the grey shaded region.  
The best-fitting parameter values and their uncertainties are summarised by the mean and standard deviation of the posterior chains for each parameter
from a Markov Chain Monte Carlo output.
We used four parallel chains of 50,000 iterations each.  To test for convergence, the stationary parts of the chains were compared using the Gelman and 
Rubin (1992) convergence diagnostic.  The largest value of the 95$\%$ upper bound on the potential scale reduction factor was 1.02, indicating that the chains had 
converged. 

\begin{figure*}[htbp]
\includegraphics[width=18cm,angle=0]{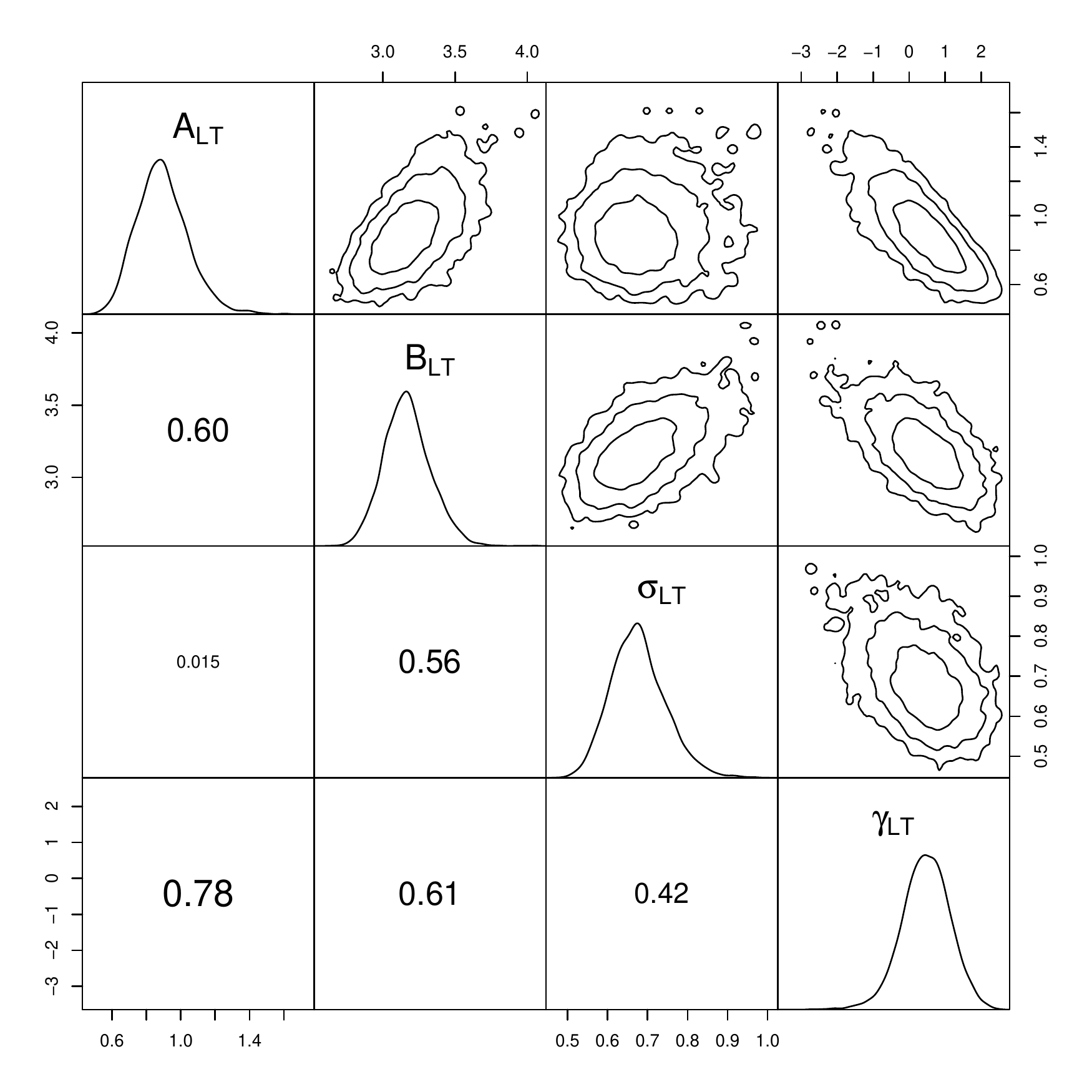}
\caption{\label{mat} Scatterplot matrix for the fit of the
    luminosity-temperature relation of the XXL-C1-GC sample. The posterior densities are
    shown along the diagonal; the $1\sigma$, $2\sigma$, and $3\sigma$
    confidence contours for the pairs of parameters are shown in the
    upper right panels.  The lower left panels show the Pearson's
    correlation coefficient for the corresponding pair of parameters
    (text size is proportional to the correlation
    strength).}
\end{figure*}

\begin{figure*}[htbp]
\includegraphics[width=18cm,angle=0]{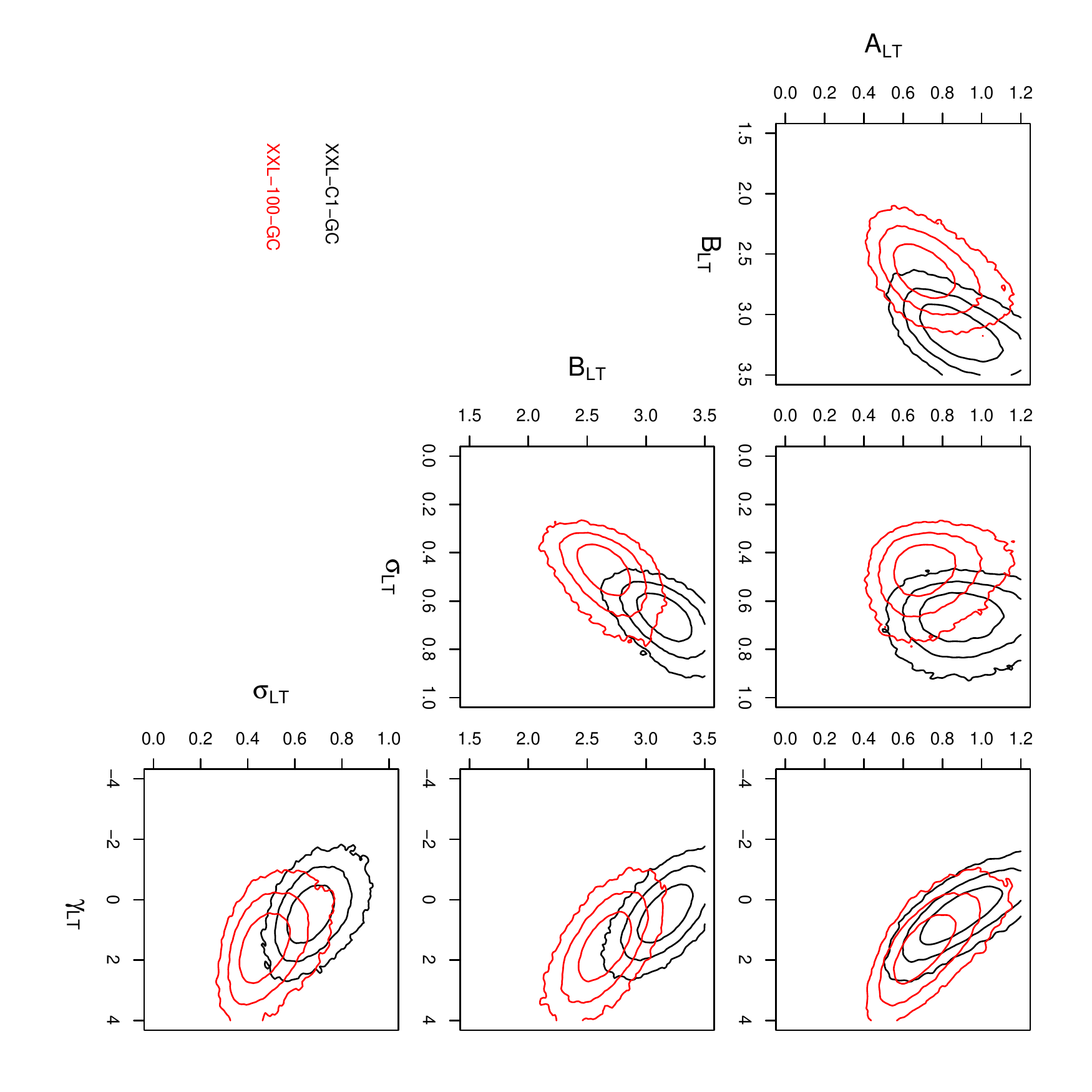}
\caption{\label{comparPaul} Matrix plot comparing the 1$\sigma$, 2$\sigma$ and 3$\sigma$ contours for pairs of parameters of the luminosity-temperature relation, with the 
XXL-C1-GC and XXL-100-GC contours given by the black and red contours respectively.}
\end{figure*}

The parameters of the luminosity-temperature scaling relation are given in Table~\ref{tab:lttab}, and illustrated with the scatterplot matrix in Fig.~\ref{mat}.  We find that, 
within errors, the normalisation, slope, evolution and scatter ($\sigma$$_{LT}$) of the XXL-C1-GC luminosity-temperature relation agree with those of the XXL-100-GC sample.
Figure~\ref{comparPaul} shows the comparison of the parameters with the XXL-C1-GC and XXL-100-GC samples. 
We find a lower normalisation than that found when using the BCES regression fit to the XXL-C1-GC sample (which did not account for selection biases), although the difference is 
minor, only weakly significant at the 1.7$\sigma$ level.

Figure~\ref{ls} (bottom panel) displays the evolution of the luminosity-temperature relation as inferred from our best-fitting model.  The best-fit 
evolution is given by the black solid line along with the 1$\sigma$ uncertainty, and the strong and weak self-similar expectations are given by the red
and blue dashed lines, respectively. The best fit evolution is consistent with that found in XXL paper III.  

\begin{table*}[t!]
\caption{\label{tab:lttab} Best-fitting parameters for the luminosity-temperature relations
  modelled in this work (with the 176 best C1 clusters, see beginning of section 5) with Eq.\ref{equ:ltunbias} where
  $L_{0}$=3$\times$10$^{43}$ \thinspace erg s$^{-1}$ and $T_{0}$=3 keV.  (1) Luminosity-temperature
  relation; (2) fit method; (3) normalisation; (4) slope; (5)
  evolution term (E(z)$^{\gamma_{LT}}$); (6) intrinsic scatter ($\sigma$$_{LT}$).}
\begin{tabular}{lccccc}
\hline
\hline
   Relation & Fit & A$_{LT}$ & B$_{LT}$  & $\gamma$$_{LT}$  & Scatter $\sigma$$_{LT}$ \\
   (1) & (2) & (3) & (4) & (5) & (6) \\
\hline						
  $L$-$T$ & BCES & $1.20\pm0.09$ & $3.10\pm0.15$ & 1 (fixed)  & $0.64\pm0.05$ \\
  $L$-$T$ & XXL & $0.89\pm0.14$ & $3.17\pm0.16$ & $0.47\pm0.68$  & $0.67\pm0.07$ \\
\hline
\end{tabular}
\end{table*}

Large outliers in the luminosity-temperature relation were also inspected for possible AGN contamination.  Initial visualisation of the X-ray images 
sometimes revealed point sources near the centre of the X-ray emission. These clusters where then removed from the sample to compute the 
luminosity-temperature relation.  At present, a systematic search for possible contamination of all clusters has yet to be performed.  However, this 
will be addressed with the release of the full XXL catalogue, where an improved pipeline will be used for joint cluster and AGN detection.  

\begin{figure*}[htbp]
\includegraphics[width=18cm,angle=0]{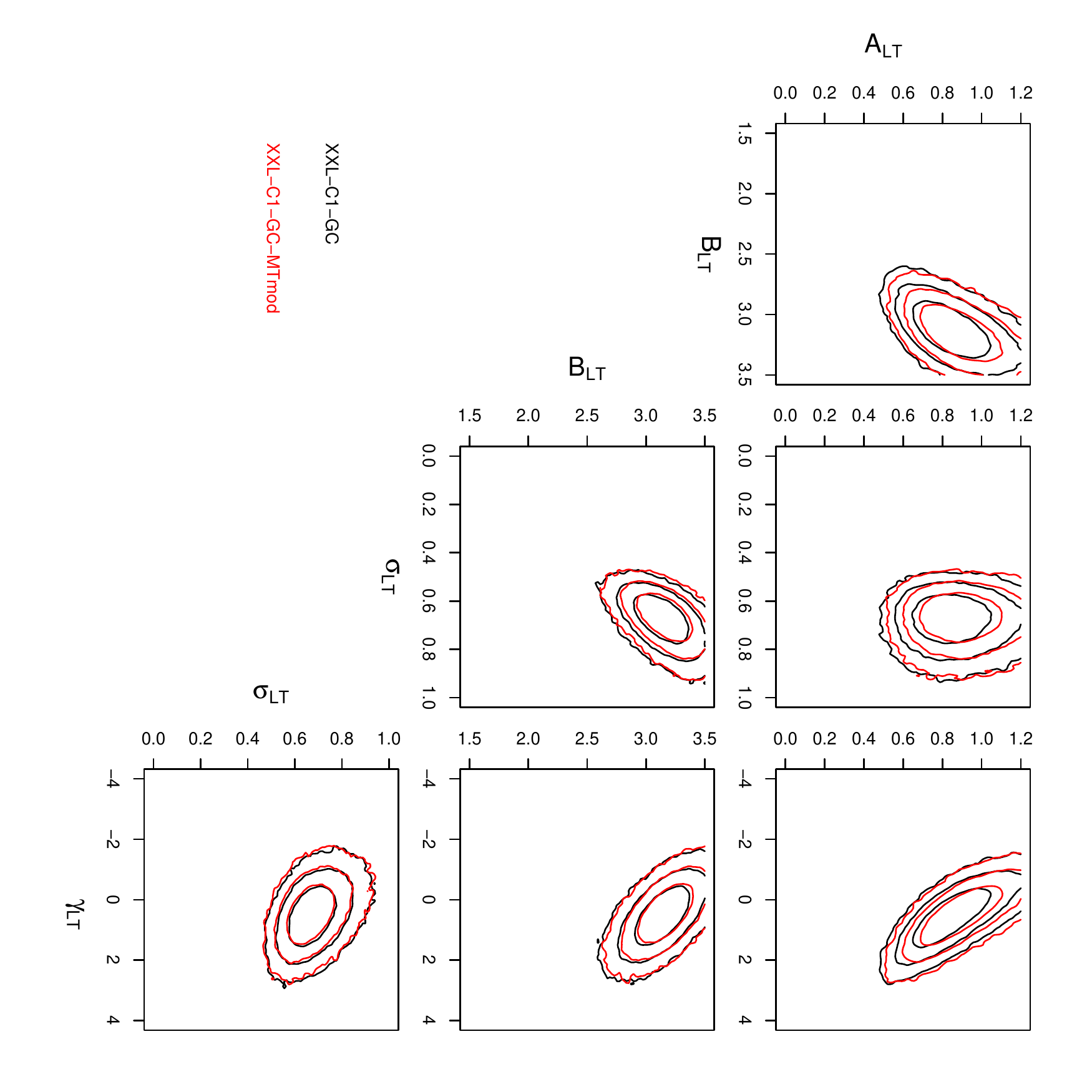}
\caption{\label{PaulBen}Fit contours of the luminosity-temperature relation parameters using both the XXL paper IV mass 
temperature relation (based on the XXL-C1-GC sample: black contours) and the scaled relation (normalisation of the 
mass-temperature decreased by 20$\%$ and using the same slope as in XXL paper IV: red contours).}
\end{figure*}

In order to test the effect of possible uncertainties on the mass temperature relation (cf. XXL paper IV), we scaled down the normalisation of the XXL paper IV 
mass temperature relation by 20$\%$. We found that the luminosity-temperature relation parameters did not change significantly, as demonstrated in 
Figure~\ref{PaulBen}, showing the parameters contours using both the XXL paper IV mass temperature relation and the scaled relation.

\subsection{X-ray luminosity function}

\begin{figure}[h!]
\begin{center}
\includegraphics[width=9.5cm]{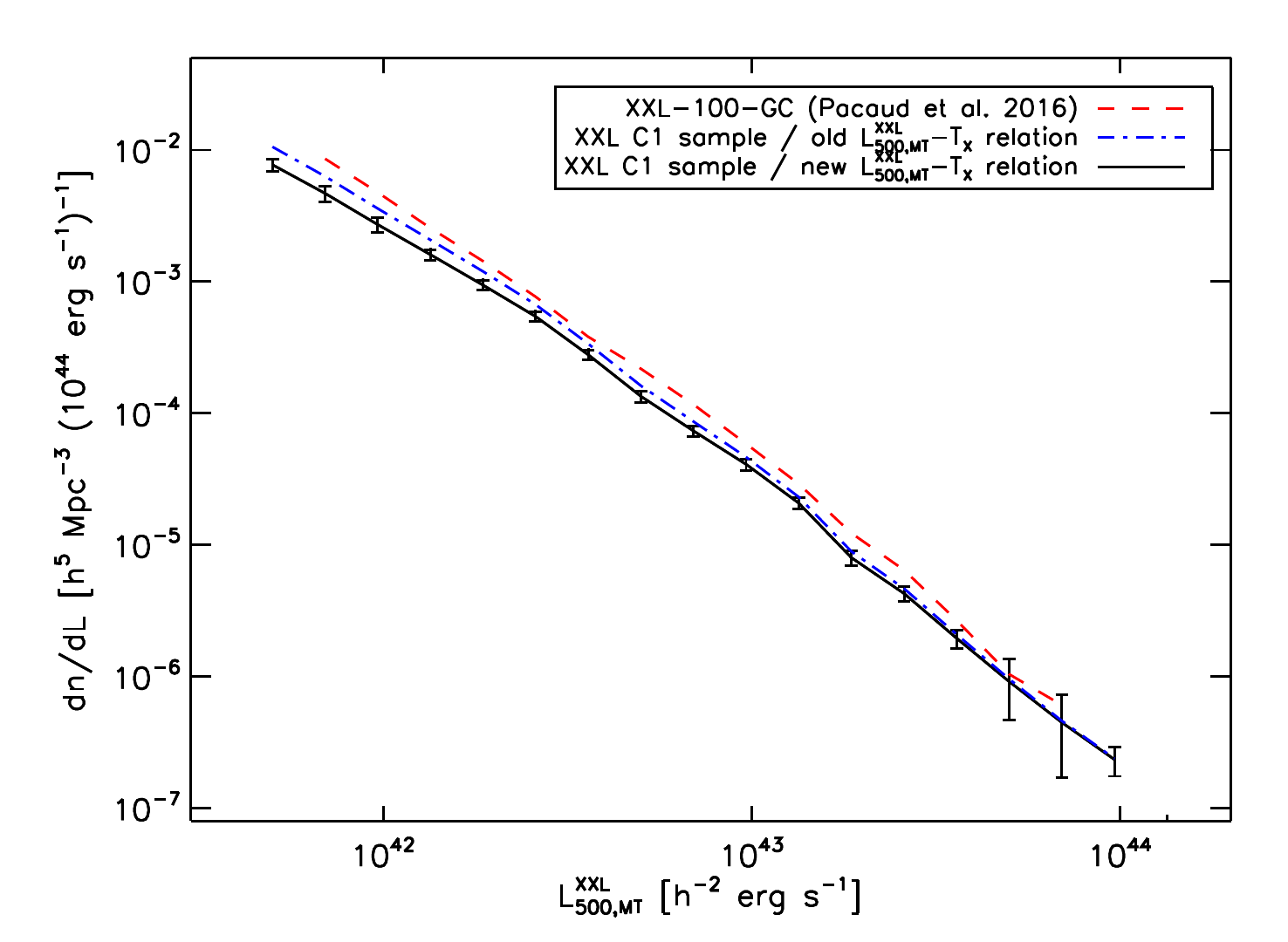}
\includegraphics[width=9.5cm]{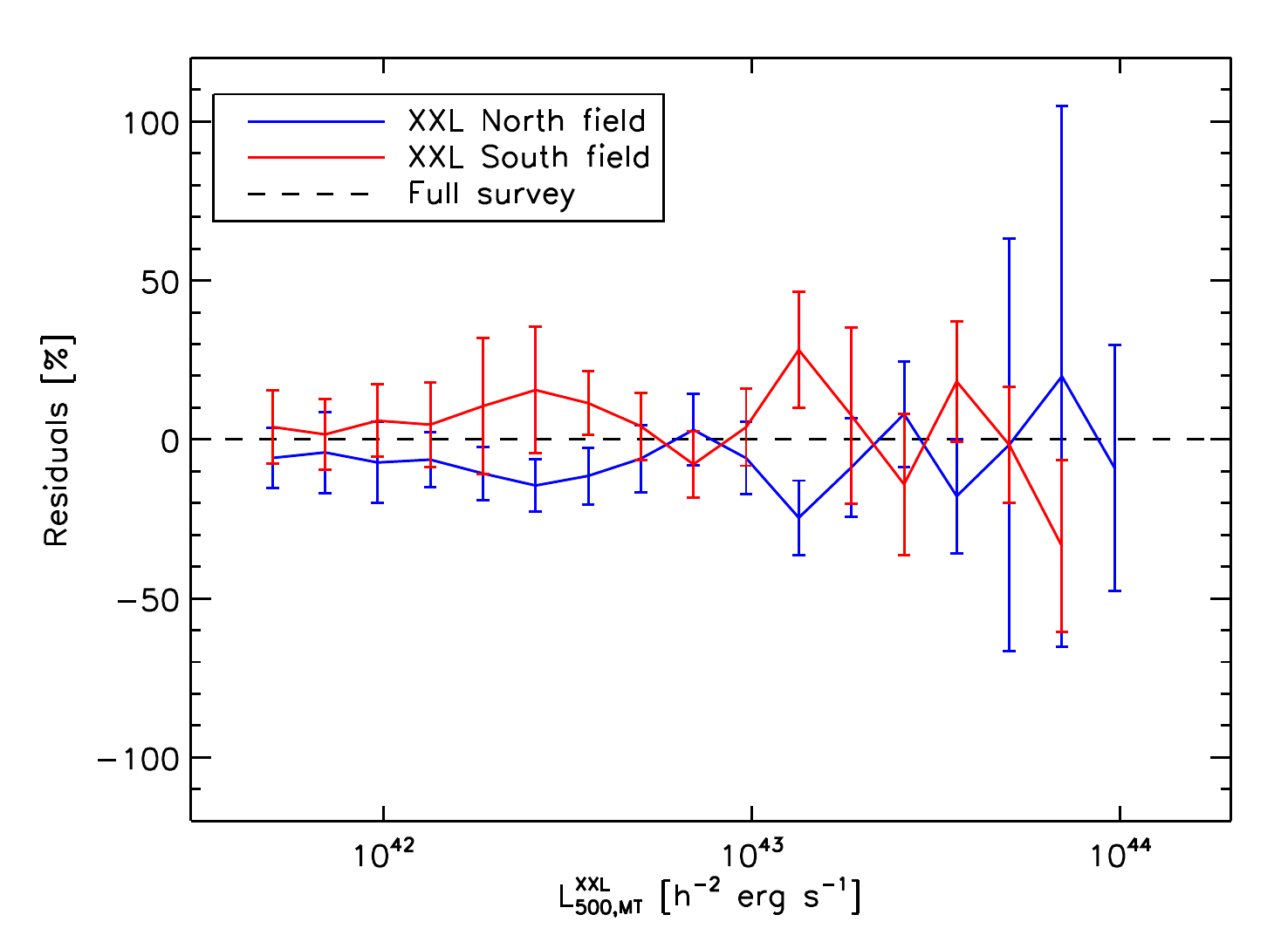}
\caption{Upper panel: X-ray luminosity function ([0.5-2] keV band) of the C1 cluster sample based on the 186 C1 clusters in good pointings and 
with redshift information. The calculation is averaged over the whole survey volume ($z$ in 0.0 - 1.3) and includes an incompletness 
factor of 2.6\% for the five C1 clusters without any redshift estimate. The method is the same as in XXL paper II. For comparison, the 
luminosity function of the XXL brightest 100 cluster sample (XXL-100-GC) is shown with the red dashed line. Finally, the dot-dashed 
blue line indicates the luminosity function of the C1 sample recomputed for with the old $L_{X}-T$ relation of XXL paper III, as 
was assumed for the XXL-100-GC sample.
Lower panel: Residuals of the C1 luminosity functions computed from only the northern or southern XXL field with respect to the 
complete luminosity function shown in the upper panel.
\label{fig:LumFunc}}
\end{center}
\end{figure}

\begin{figure}[h]
\begin{center}
\includegraphics[width=9.5cm]{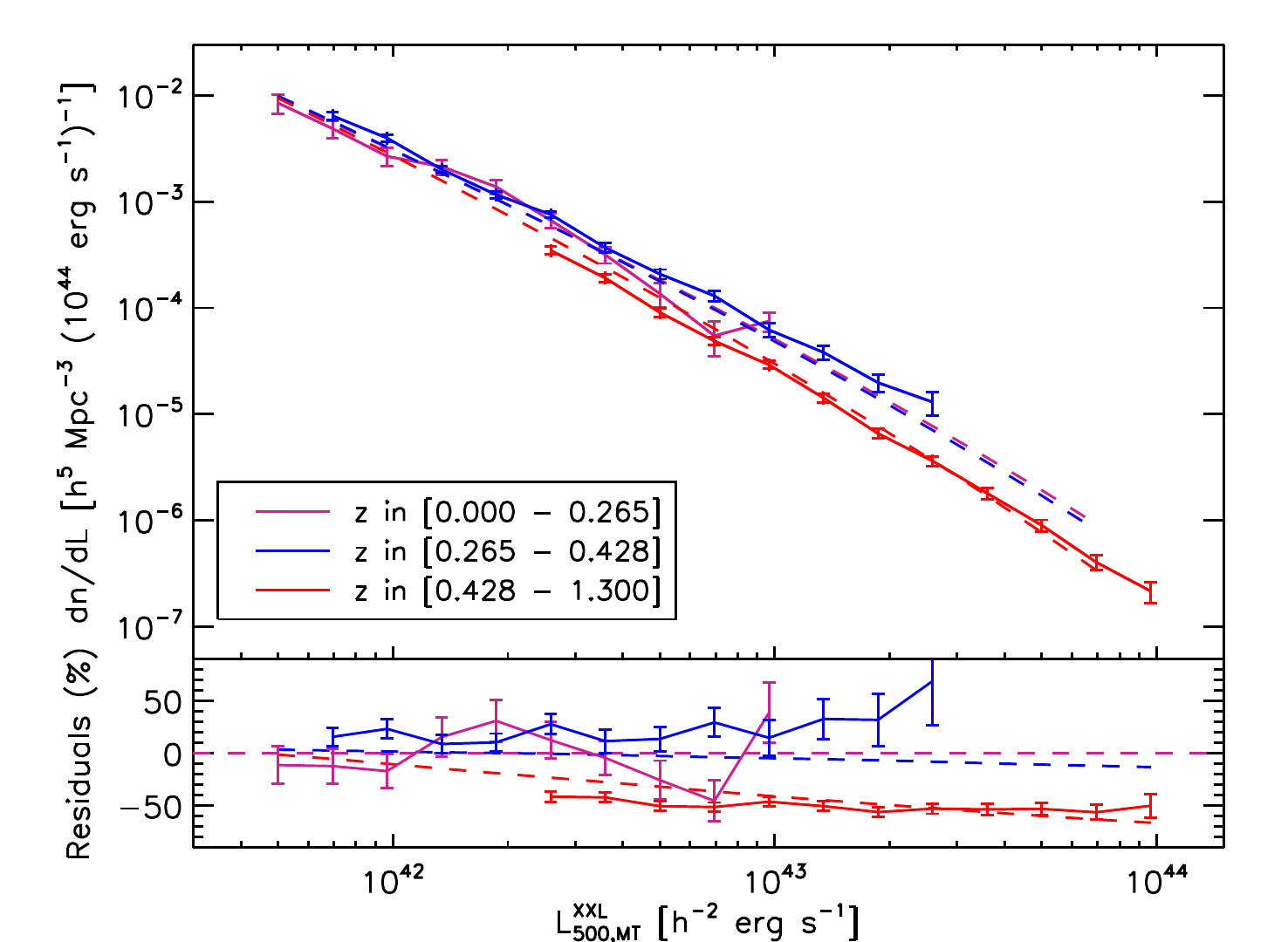}
\caption{Redshift evolution of the C1 X-ray luminosity function. The calculation relies on the same assumptions as for the full survey 
volume luminosity function of Fig.~\ref{fig:LumFunc}, but the sample is split into three redshift bins containing approximately the 
same number counts of clusters. The dashed lines show, for the same redshift bins, the luminosity function expected in the WMAP9 cosmology from 
our scaling relation model ($\mathrm{M_{500,WL}-T_{300\,kpc}}$ from XXL paper IV and $\mathrm{L^{XXL}_{500,MT}-T_{300\,kpc}}$). 
For better visualisation, the bottom panel shows the same information in the form of a residual plot with respect to the WMAP9 expectation
at low redshift. A significant negative evolution is visible at $z\geq0.43$\label{fig:LumFuncEvol}}
\end{center}
\end{figure}

Based on the new enlarged sample, we also revised our estimate of the cluster X-ray luminosity function
from XXL paper II. As for the luminosity-temperature relation, such a computation must  rely on a complete 
subsample with measured selection function and therefore we focused on the XXL-C1-GC subsample.
We relied on the available spectroscopic redshifts of Table~\ref{tab:C1C2cat} combined with the 
$\mathrm{L^{XXL}_{500,MT}}$ ([0.5-2] keV band) resulting from the X-ray spectroscopic analysis (no estimates from scaling relations).
For sixteen C1 clusters without a confirmed spectroscopic redshift, we used instead the tentative or photometric 
redshifts provided in Table~\ref{tab:listetot7}, while the five clusters without any redshift information are 
modelled using an incompleteness factor of 2.6\%. 
This incompleteness is coming from the five C1 clusters (over 191) without a spectroscopic confirmation. During computation,
we assume that these clusters are randomly selected among the full sample, and we then diminish the survey effective volume by the same factor 
of 2.6\%. The mass and redshift distribution of these 2.6\% is under-dominant compared to statistical errors.
Finally, it was not possible to obtain the luminosity of
eight clusters from X-ray spectroscopy, as the poor constraints on the temperature resulted in unphysical estimates 
of $r_{500,MT}$ and consequently unrealistically large or small extrapolation factors from the circular 300$\thinspace$kpc 
extraction region. For those eight clusters, we used instead the luminosity estimate based on 
scaling relations. This introduces a small level of inhomogeneity 
in our initial data set but we believe that the attached uncertainty is smaller than the effect of a large 
incompletness. Indeed, higher redshift (fainter) clusters are more likely to be missing from our 
spectroscopically confirmed (X-ray spectroscopic) samples, which would distort the shape of the luminosity 
function.

From this sample, we estimated the luminosity function in our reference WMAP9 cosmology using the updated
scaling relation model obtained in the previous section. The computation relied on the `cumulative 
effective volume correction' method introduced in appendix B of XXL paper II. This method is based on numerical 
derivation of a direct estimate of the cumulative luminosity, which has the advantage of reducing the Poisson
noise by effectively relying on information from several luminosity bins to derive each value. This comes at the 
cost of a large bin-to-bin correlation but the tighter constraints on each bin remain unbiased.

The redshift averaged luminosity function for the whole sample is shown in the top panel of 
Figure~\ref{fig:LumFunc}. Compared to our estimate of the luminosity function of XXL-100-GC in paper II, the 
probed luminosity range only midly increases while the errors are reduced by about 20\%. However, the new luminosity 
function appears to be lower than the previous one, particularly at the low luminosity end where the discrepancy 
exceeds 3$\sigma$. These measurements are perfectly consistent between the two XXL subfields, as illustrated by the 
bottom panel of Figure~\ref{fig:LumFunc}, effectively excluding a number of possible systematic errors in the modelling 
of the selection function like the dependence on absorption, depth or pointing layout. To further investigate the 
origin of the discrepancy, we also computed the luminosity function based on the old luminosity-temperature relation
of XXL paper III (blue dot-dashed line in Fig.~\ref{fig:LumFunc}) which revealed that the tension originates
from the change both in the number of detected sources per luminosity and redshift bin in the new sample, and in the effective 
volumes computed for different scaling relation models. With the old model, the tension between XXL-C1-GC and 
XXL-100-GC would mostly be lower than 2$\sigma$ (even at the low luminosity end where it just reaches 2$\sigma$).
In other words, when using the old model for computing the Luminosity-Temperature relation, 
all the discrepancy can be understood in terms of cosmic variance. If we compare the differences between red and blue curves
of Fig.~\ref{fig:LumFunc} (upper figure) with statistical uncertainties and north versus south variations, the observed differences
are not significant. 

We also investigated the redshift evolution of the luminosity function by splitting the sample into three redshift bins 
containing approximatively the same number of clusters. As shown in Fig.~\ref{fig:LumFuncEvol}, there is no evidence
for evolution below z$\sim$0.43 while a significant negative evolution is observed at z$>$0.4. 
This result is fully consistent with expectations calculated using the WMAP9 cosmological model 
and our preferred set of scaling relations. The absence of evolution below z$\sim$0.4 also rules out different 
redshift weights as the origin of the lower luminosity function compared to XXL-100-GC, since all the constraints at 
low luminosity come from low redshift clusters.

The measured values  (both redshift averaged and in redshift bins) are provided in Table~\ref{tab:LfuncValues}
and \ref{tab:CumLfuncValues} for the differential and cumulative luminosity functions. We however stress that our effective volume 
correction method might slightly bias the cumulative distribution at low luminosities, as it relies on the full shape of the 
modelled WMAP9 luminosity function to weight the luminosity dependent effective volume.

\begin{table*}
\begin{center}
\caption{Tabulated values of the differential luminosity ([0.5-2] keV) function for the C1 sample.\label{tab:LfuncValues}}
\begin{tabular}{ccccccccc}
\hline\hline
                                         &  \multicolumn{2}{c}{Full z range}        &  \multicolumn{2}{c}{0.0 < z < 0.265}      & \multicolumn{2}{c}{0.265 < z < 0.428}     & \multicolumn{2}{c}{0.428 < z < 1.3} \rule{0pt}{2.6ex}\\
$\mathrm{L^{XXL}_{500,MT}}$                 &  dn/dL                 &  $\Delta$(dn/dL) &  dn/dL                &  $\Delta$(dn/dL)  &  dn/dL                &  $\Delta$(dn/dL)  &  dn/dL                &  $\Delta$(dn/dL)              \\[2pt]
[$10^{42}\,\mathrm{h^{-2}erg\,s^{-1}}$]  &  [LF unit]$^\dagger$   &        \%        &  [LF unit]$^\dagger$  &        \%         &  [LF unit]$^\dagger$  &        \%         &  [LF unit]$^\dagger$  &        \%                     \\ 
\hline
 0.50 & $7.77 \times 10^{-3}$ &   10.3 & $8.49 \times 10^{-3}$ &   20.5 &           -           &     -  &           -           &     -  \rule{0pt}{2.6ex}\\
 0.69 & $4.71 \times 10^{-3}$ &   12.9 & $4.87 \times 10^{-3}$ &   19.1 & $6.43 \times 10^{-3}$ &    7.7 &           -           &     -  \\
 0.97 & $2.73 \times 10^{-3}$ &   12.4 & $2.67 \times 10^{-3}$ &   19.5 & $3.97 \times 10^{-3}$ &    7.4 &           -           &     -  \\
 1.34 & $1.62 \times 10^{-3}$ &    8.0 & $2.14 \times 10^{-3}$ &   15.9 & $2.01 \times 10^{-3}$ &    7.8 &           -           &     -  \\
 1.86 & $9.49 \times 10^{-4}$ &    7.7 & $1.38 \times 10^{-3}$ &   15.0 & $1.16 \times 10^{-3}$ &    7.6 &           -           &     -  \\
 2.59 & $5.43 \times 10^{-4}$ &    7.7 & $6.67 \times 10^{-4}$ &   15.7 & $7.58 \times 10^{-4}$ &    7.4 & $3.47 \times 10^{-4}$ &    8.4 \\
 3.60 & $2.78 \times 10^{-4}$ &    8.2 & $3.16 \times 10^{-4}$ &   17.0 & $3.69 \times 10^{-4}$ &   10.0 & $1.91 \times 10^{-4}$ &    8.3 \\
 5.00 & $1.36 \times 10^{-4}$ &    9.0 & $1.36 \times 10^{-4}$ &   25.1 & $2.08 \times 10^{-4}$ &   10.3 & $9.06 \times 10^{-5}$ &    9.1 \\
 6.95 & $7.43 \times 10^{-5}$ &    8.8 & $5.46 \times 10^{-5}$ &   36.1 & $1.30 \times 10^{-4}$ &   10.6 & $4.88 \times 10^{-5}$ &    9.0 \\
 9.65 & $4.09 \times 10^{-5}$ &    9.0 & $7.51 \times 10^{-5}$ &   20.7 & $6.22 \times 10^{-5}$ &   14.7 & $2.91 \times 10^{-5}$ &    8.5 \\
13.4 & $2.07 \times 10^{-5}$ &    9.6 &           -           &     -  & $3.82 \times 10^{-5}$ &   14.4 & $1.42 \times 10^{-5}$ &    9.4 \\
18.6 & $8.13 \times 10^{-6}$ &   12.6 &           -           &     -  & $1.98 \times 10^{-5}$ &   18.9 & $6.56 \times 10^{-6}$ &   10.1 \\
25.9 & $4.35 \times 10^{-6}$ &   13.1 &           -           &     -  & $1.30 \times 10^{-5}$ &   25.0 & $3.62 \times 10^{-6}$ &   10.5 \\
36.0 & $1.98 \times 10^{-6}$ &   15.5 &           -           &     -  &           -           &     -  & $1.80 \times 10^{-6}$ &   12.4 \\
50.0 & $9.29 \times 10^{-7}$ &   47.7 &           -           &     -  &           -           &     -  & $8.99 \times 10^{-7}$ &   12.4 \\
69.5 & $4.55 \times 10^{-7}$ &   60.7 &           -           &     -  &           -           &     -  & $4.03 \times 10^{-7}$ &   15.8 \\
96.5 & $2.38 \times 10^{-7}$ &   25.5 &           -           &     -  &           -           &     -  & $2.15 \times 10^{-7}$ &   22.3 \\
\hline\hline
\end{tabular}
\tablefoot{Because of the luminosity vs redshift degeneracy in the sample, only a limited
range of luminosities is available for each redshift slice. A graphical display of these values is provided in figures~\ref{fig:LumFunc} and \ref{fig:LumFuncEvol}.
$^\dagger$: all luminosity function values in this table are in units of [\,$\mathrm{h^5\,Mpc^{-3}\,(10^{44}\,erg\,s^{-1})^{-1}}$\,].}
\end{center}
\end{table*}

\begin{table*}
\begin{center}
\caption{Tabulated values of the cumulative luminosity ([0.5-2] keV) function for the C1 sample.\label{tab:CumLfuncValues}}
\begin{tabular}{ccccccccc}
\hline\hline
                                         &  \multicolumn{2}{c}{Full z range}        &  \multicolumn{2}{c}{0.0 $\leq$ z $\leq$ 0.265}      & \multicolumn{2}{c}{0.265 $\leq$ z $\leq$ 0.428}     & \multicolumn{2}{c}{0.428 $\leq$ z $\leq$ 1.3} \rule{0pt}{2.6ex}\\
$\mathrm{L^{XXL}_{500,MT}}$                 &  n($>$L)                 &  $\Delta[$n($>$L)$]$ &  n($>$L)                &  $\Delta[$n($>$L)$]$  &  n($>$L)                &  $\Delta[$n($>$L)$]$  &  n($>$L)                &  $\Delta[$n($>$L)$]$              \\[2pt]
[$10^{42}\,\mathrm{h^{-2}erg\,s^{-1}}$]  &  [$\mathrm{h^3\,Mpc^{-3}}$]   &        \%        &  [$\mathrm{h^3\,Mpc^{-3}}$]  &        \%         &  [$\mathrm{h^3\,Mpc^{-3}}$]  &        \%         &  [$\mathrm{h^3\,Mpc^{-3}}$]  &        \%                     \\ 
\hline
 0.50 & $5.38 \times 10^{-5}$ &    9.2 & $6.29 \times 10^{-5}$ &   22.5 &           -           &     -  &           -           &     -  \rule{0pt}{2.6ex}\\
 0.69 & $4.19 \times 10^{-5}$ &    9.0 & $4.92 \times 10^{-5}$ &   24.1 & $5.93 \times 10^{-5}$ &    7.2 &           -           &     -  \\
 0.97 & $3.23 \times 10^{-5}$ &    7.9 & $4.07 \times 10^{-5}$ &   25.0 & $4.46 \times 10^{-5}$ &    7.3 &           -           &     -  \\
 1.34 & $2.46 \times 10^{-5}$ &    7.9 & $3.22 \times 10^{-5}$ &   27.3 & $3.41 \times 10^{-5}$ &    7.3 &           -           &     -  \\
 1.86 & $1.80 \times 10^{-5}$ &    8.2 & $2.18 \times 10^{-5}$ &   35.8 & $2.68 \times 10^{-5}$ &    7.2 &           -           &     -  \\
 2.59 & $1.29 \times 10^{-5}$ &    8.7 & $1.53 \times 10^{-5}$ &   46.3 & $1.98 \times 10^{-5}$ &    7.8 & $9.00 \times 10^{-6}$ &    8.2 \\
 3.60 & $8.76 \times 10^{-6}$ &    9.5 & $1.04 \times 10^{-5}$ &   65.6 & $1.39 \times 10^{-5}$ &    8.6 & $6.25 \times 10^{-6}$ &    8.5 \\
 5.00 & $6.36 \times 10^{-6}$ &    9.9 & $7.81 \times 10^{-6}$ &   85.5 & $1.11 \times 10^{-5}$ &    8.6 & $4.48 \times 10^{-6}$ &    8.5 \\
 6.95 & $4.29 \times 10^{-6}$ &   11.2 & $5.94 \times 10^{-6}$ &  110.5 & $7.06 \times 10^{-6}$ &   11.2 & $3.28 \times 10^{-6}$ &    8.3 \\
 9.65 & $2.96 \times 10^{-6}$ &   12.6 & $5.31 \times 10^{-6}$ &  121.8 & $5.16 \times 10^{-6}$ &   12.4 & $2.25 \times 10^{-6}$ &    8.7 \\
13.4 & $1.69 \times 10^{-6}$ &   17.0 &           -           &     -  & $3.11 \times 10^{-6}$ &   16.8 & $1.43 \times 10^{-6}$ &    9.5 \\
18.6 & $1.13 \times 10^{-6}$ &   20.0 &           -           &     -  & $1.80 \times 10^{-6}$ &   21.4 & $9.92 \times 10^{-7}$ &    9.6 \\
25.9 & $6.90 \times 10^{-7}$ &   26.6 &           -           &     -  & $6.88 \times 10^{-7}$ &   41.7 & $6.21 \times 10^{-7}$ &   11.4 \\
36.0 & $3.93 \times 10^{-7}$ &   37.6 &           -           &     -  &           -           &     -  & $3.74 \times 10^{-7}$ &   11.8 \\
50.0 & $2.20 \times 10^{-7}$ &   57.9 &           -           &     -  &           -           &     -  & $1.94 \times 10^{-7}$ &   16.1 \\
69.5 & $8.72 \times 10^{-8}$ &   25.4 &           -           &     -  &           -           &     -  & $7.85 \times 10^{-8}$ &   23.0 \\
96.5 & $1.17 \times 10^{-8}$ &   90.3 &           -           &     -  &           -           &     -  & $1.01 \times 10^{-8}$ &   87.2 \\
\hline\hline
\end{tabular}
\tablefoot{Because of the luminosity vs redshift degeneracy in the sample, only a limited
range of luminosities is available for each redshift slice.}
\end{center}
\end{table*}

Clusters affected by AGNs represent less than $\sim$5$\%$ of the full C1 sample and
were not removed from the calculation of the luminosity function. This
allows a direct comparison with the preliminary results of XXL paper II. 

We also tried to estimate how many clusters in the X-ray luminosity function could be affected by cluster-cluster X-ray blending, potentially leading 
to the loss of some faint clusters and the artificial addition of bright clusters. None of the cluster pairs or super-clusters listed in 
Table~\ref{tab:PL} and Table~\ref{tab:SCL} are contributing to this bias as they are detected as independent clusters. However, the line-of-sight 
superpositions and X-ray blends, listed in appendix B, can affect the X-ray luminosity function. This is the case for the line of sight of 
XLSSC 041 where a z=0.557 cluster is missed, of XLSSC 539 including two clusters at z=0.169 and 0.184, of XLSSC 096 with two clusters at z=0.203 and 
0.520, of XLSSC 151 with two clusters at z=0.189 and 0.280, of XLSSC 044 with two clusters at z=0.263 and 0.317, and of XLSSC 079 with two clusters 
at z=0.19 and probably at $\sim$0.52. This represents however less than 5$\%$ of the sample used to compute the X-ray luminosity function and the 
effect is therefore probably negligible.

\section{Witnessing the evolution of massive structures: from super-clusters to fully collapsed fossil groups}

In order to illustrate the large variety of objects detected in the XXL Survey, we will follow in this section
the history from what could be the progenitors of very massive clusters (super-clusters), to merging
clusters in an already advanced stage (e.g. XLSSC 110), and to the possible final stage of group of galaxies
(fossil groups). 

To give a general flavour of the structures present in the XXL Survey, we also present in appendix B the notable 
cluster superpositions we detected, and the most distant cluster in our survey (XLSSC 122, cf. Mantz et al 2014, hereafter XXL paper V) 
along with additional spectroscopic follow-up of this cluster.

\subsection{Super-clusters}  

We search for a-priori physical associations between individual clusters of galaxies. We will arbitrarily call
'super-clusters' the associations of at least three clusters (whatever their separation). Cluster pairs (association of only two clusters) are not considered
as super-clusters.                                                                

\subsubsection{Friends-of-friends detected super-clusters}

We used all spectroscopically confirmed C1, C2, and C3 clusters to search for 
super-clusters in the two XXL fields.
The analysis was restricted to the [0.03-1.00] redshift range.

We first performed a classical three-dimensional friends-of-friends analysis (FoF hereafter) 
to estimate the critical linking length, $\ell_c$, 
for each field, the one that maximises the number of super-clusters
(for instance Einasto et al. 2001).
We found, respectively for XXL-N and XXL-S, 27 and 29 $h_{70}^{-1}$ 
Mpc.
While a FoF analysis with this linking length would be ideal if 
the sample was relatively homogeneously distributed in z. 
In the real world, we need a weighting function to weight $\ell_c$.

We measured the cluster space densities by dividing the 
cluster sample in ten bins of redshift and calculating the 
respective cosmological volumes. The density falls roughly 
exponentially from z$\sim$0.03 up to z$\sim$0.7, then follows a plateau and, finally, 
the last bin is very undersampled. Since this density distribution can be considered as the inverse of the selection function,
we could use it to weight $\ell_c$ with redshift. We used the pure exponential fit (cf. equation 3), to bins
between 0.22 $\leq$ z $\leq$ 0.71, which reproduces very closely the exponential plus plateau behaviour.

Thus, we applied a `tunable' FoF, as for example in Chow-Mart\'inez et al. (2014),
to the sample by using an exponential fit 
in order to weight the $\ell_c$ and compute the local linking length, 
$\ell$(z), for each targeted cluster. We have

\begin{equation}
\ell(\mathrm{z}) = \left[ \frac{3}{4 \pi \, d(\mathrm{z})} \right]^{1/3} \, \ell_c
\end{equation}

\noindent where

\begin{equation}
d(\mathrm{z}) = e^{-5.724 \, \mathrm{z}}
\end{equation}

\noindent is the normalised density (weighting) function.

We found 21 super-clusters in the XXL-N field data and 14 in XXL-S, considering only super-cluster candidates with a multiplicity 
(number of member clusters) greater than or equal to 3 (cf. 
Table~\ref{tab:SCL}). We adopted the internal denomination XLSSsC for XXL super-clusters
(replacing the preliminary notation used in XXL Papers II and XII) to
avoid any confusion with regular individual clusters.
The centres of the super-clusters were calculated as the geometrical centre of 
the member clusters. Super-clusters described in the present paper have sizes up to 60$\thinspace$Mpc, and 
this is around the median value for the largest superclusters in the local Universe (e.g.
Chow-Mart\'inez et al. 2014).

We also give (in appendix E) in Table~\ref{tab:PL} the list of cluster 
pairs (16 in the XXL-N field data and 23 in the XXL-S) detected with the same FoF approach.

\begin{table*}
\caption{\label{tab:SCL} List of detected super-cluster candidates with the FoF approach.
Columns are: Id, Id in XXL paper II and paper XII, Coordinates (J2000.0 equinox), mean redshift, multiplicity (cf. section 6.1.1), R 
reliability index from the Voronoi tessellation approach (cf. section 6.1.2), and list of the members.  }
\begin{tabular}{cccccrrr}
\hline
\hline
Name & Old & $\alpha(^{\circ})$ & $\delta(^{\circ})$ & Mean z & m & R & Members (XLSSC cluster numbers)\\
\hline
XLSSsC N18 &   &  30.430 &  -6.880  & 0.336  &  3 & 3 &  156, 199, 200 \\
XLSSsC N02 & e &  32.059 &  -6.653  & 0.430  & 11 & 4 &  082, 083, 084, 085, 086, 092, 093, 107, 155, 172, 197 \\
XLSSsC N03 &   &  32.921 &  -4.879  & 0.139  &  8 & 2 &  060, 095, 112, 118, 138, 162, 176, 201 \\
XLSSsC N06 & f &  33.148 &  -5.568  & 0.300  &  5 & 4 &  098, 111, 117, 161, 167 \\
XLSSsC N12 &   &  34.138 &  -5.003  & 0.447  &  4 & 4 &  110, 142, 144, 187 \\
XLSSsC N21 &   &  34.420 &  -5.038  & 0.651  &  3 & 3 &  059, 080, 195 \\
XLSSsC N11 &   &  34.438 &  -4.867  & 0.340  &  3 & 2 &  058, 086, 192 \\
XLSSsC N15 &   &  34.466 &  -4.608  & 0.291  &  4 & 3 &  126, 137, 180, 202 \\
XLSSsC N17 &   &  34.770 &  -4.240  & 0.203  &  3 & 3 &  077, 189, 193 \\
XLSSsC N13 &   &  35.221 &  -4.666  & 0.513  &  3 & 2 &  124, 131, 183 \\
XLSSsC N19 &   &  35.629 &  -5.146  & 0.380  &  3 & 2 &  017, 067, 132 \\
XLSSsC N04 &   &  35.813 &  -4.144  & 0.828  &  8 & 3 &  003, 015, 032, 047, 064, 069, 071, 184 \\
XLSSsC N16 &   &  36.156 &  -3.455  & 0.174  &  3 & 2 &  035, 043, 182 \\
XLSSsC N20 &   &  36.159 &  -4.239  & 0.433  &  3 & 2 &  006, 012, 026 \\
XLSSsC N10 &   &  36.290 &  -3.411  & 0.329  &  4 & 4 &  009, 010, 023, 129 \\
XLSSsC N07 &   &  36.446 &  -5.142  & 0.496  &  5 & 4 &  020, 049, 053, 143, 169 \\
XLSSsC N05 & a &  36.500 &  -4.176  & 0.055  &  6 & 2 &  011, 052, 054, 062, 125, 191 \\
XLSSsC N08 & b &  36.910 &  -4.158  & 0.141  &  4 & 1 &  041, 050, 087, 090 \\
XLSSsC N14 &   &  36.917 &  -4.405  & 0.616  &  3 & 3 &  001, 089, 145 \\
XLSSsC N01 & d &  36.954 &  -4.778  & 0.296  & 14 & 4 &  008, 013, 022, 024, 027, 028, 070, 088, 104, 140, 148, 149, 150, 168 \\
XLSSsC N09 &   &  37.392 &  -5.227  & 0.190  &  4 & 1 &  074, 091, 123, 151 \\
\hline	       
XLSSsC S14 &   & 348.858 & -54.522  & 0.202  &  3 & 2 &  530, 554, 636 \\
XLSSsC S07 &   & 349.528 & -53.353  & 0.334  &  3 & 4 &  501, 503, 593 \\
XLSSsC S06 &   & 350.399 & -53.525  & 0.275  &  4 & 3 &  526, 557, 591, 622 \\
XLSSsC S08 &   & 350.654 & -52.910  & 0.355  &  3 & 2 &  504, 545, 555 \\
XLSSsC S13 &   & 351.161 & -54.174  & 0.099  &  3 & 1 &  515, 544, 590 \\
XLSSsC S12 &   & 351.551 & -55.878  & 0.808  &  3 & 3 &  521, 575, 583 \\
XLSSsC S05 &   & 352.077 & -54.657  & 0.210  &  4 & 1 &  577, 586, 595, 608 \\
XLSSsC S03 &   & 352.610 & -55.417  & 0.273  &  5 & 4 &  519, 524, 588, 610, 612 \\
XLSSsC S01 & c & 352.878 & -54.083  & 0.171  & 12 & 3 &  514, 518, 520, 535, 536, 565, 600, 601, 623, 627, 629, 635  \\
XLSSsC S09 &   & 353.034 & -53.988  & 0.384  &  3 & 4 &  573, 574, 624 \\
XLSSsC S11 &   & 354.074 & -52.961  & 0.534  &  3 & 3 &  508, 562, 626 \\
XLSSsC S02 &   & 354.299 & -53.932  & 0.321  &  6 & 3 &  548, 563, 585, 599, 614, 632 \\
XLSSsC S10 &   & 354.760 & -56.139  & 0.469  &  3 & 2 &  551, 609, 639   \\
XLSSsC S04 &   & 357.312 & -55.137  & 0.131  &  4 & 1 &  511, 568, 569, 570  \\
\hline
\end{tabular}
\end{table*}

The use of a tunable linking length made it possible to detect
super-cluster candidates even at z$\ge$0.6, where the completeness 
of the sample becomes critically low.
The algorithm supposes that there is an `additional density' at
such redshifts that maintains a mean density more or less similar
to that of nearby clusters. Of course this virtual density may
be or not connecting the clusters to form super-clusters.
In practice, the linking length becomes larger and the possibility of
having `connected' clusters by chance is higher.
Thus, we have to take these high-redshift super-clusters with caution.
At z$\sim$0.8 (the most distant super-cluster in the present paper is detected at this redshift), the linking length is $\sim$80$\thinspace$Mpc, which is
typically of the same order as the largest known super-clusters (e.g. Horologium-Reticulum, Fleenor et al. 2005, or the BOSS Great Wall, Lietzen 
et al. 2016).

\subsubsection{Voronoi tessellation detected super-clusters}

We applied a 3D Voronoi tessellation (e.g. Icke et al. 1987 and S\"ochting et al. 2012), 
to the data in the two XXL fields in order to assess the reliability of the structures previously found.
Voronoi tesselation was not used to directly detect super-clusters. It is a partitioning of a volume according to 
the distribution of objects inside this volume.
In the first step, we divided each cone volume into a number of optimum polyhedra equal to the number of 
clusters in that volume (Voronoi cells).
If the clusters are distributed with no sampling variation with redshift, the inverse of the Voronoi
cell volume represents directly the local density at the cluster positions. In our case, the sampling is not
constant with redshift because at high redshift the linking length becomes larger than the typical cluster-cluster
separation.
The next step was then to correct the Voronoi cells volume by applying the 
weighting function already applied to the linking length in the FoF 
analysis, in order to compensate for the undersampling at the highest 
redshifts.
The condition here was to adjust the distribution of volumes maintaining 
the total volume fixed (and, so, the mean volume or, equivalently, the 
mean density).
The local density for each cluster can be obtained directly from the 
inverse of its Voronoi cell volume.

Then we applied a threshold above which the local densities of the 
clusters are at least twice the mean density (i.e., a density 
contrast of 1). 
By counting the number of `overdense' clusters over the number of
member clusters in each super-cluster (detected by FoF) we could determine 
a `reliability index' $R$ in such a way that:

\begin{itemize}
\item $R$ = 1 represents super-clusters with 25\% or less of the 
member clusters in the overdense category;
\item $R$ = 2 with a fraction between 26 and 50\%;
\item $R$ = 3 with a fraction between 51 and 75\%;
\item $R$ = 4 with more than 75\% of the clusters in the overdense 
category according to the Voronoi analysis.
\end{itemize}

We compared our super-cluster list with the one of XXL paper II
also drawn from the XXL cluster sample but with a different method and 
with a more limited individual cluster sample (only the 100 XXL brightest 
ones).
The five XXL paper II and the one paper XII super-clusters are all redetected in the present paper. 
We confirm them at very similar redshifts and we sometimes add more member clusters. The only noticeable
exception is XLSSsC N08 for which three clusters were associated with the supercluster XLSSsC N03 which was not detected 
in XXL paper II when the number of XXL spectroscopically confirmed clusters was lower.
Melnyk et al. (2017, hereafter XXL paper XXI) also found that the most populated agglomerates of AGNs are associated with 
some of the superclusters listed in Table~\ref{tab:SCL}.

\subsection{Merging process: the peculiar case of the XLSSC 110 system}  

In this section we present an example of a merging system for which we collected additional data allowing us to 
examine the structure in more depth. The XLSSC 110 system is one of the most complex compact confirmed C1 clusters (z=0.445)
we detected within the XXL Survey.
Initially confirmed with six spectroscopic redshifts, this structure shows a peculiar
behaviour, with three apparent BCGs very close in redshift. The X-ray emission coming from this 
structure is also not equally distributed over the galaxy distribution. The BCG associated
with the main X-ray peak is possibly undergoing a rather rare triple merging. 
This led us to collect more spectroscopic data for this structure and
we got PMAS (PPak mode) integral field observations for this purpose at the 3.5$\thinspace$m Calar Alto 
telescope in 2015 and 2016. We describe the data collection in appendix C.

\begin{figure}[h]
\includegraphics[width=8.5cm,angle=0,viewport=41 140 570 654]{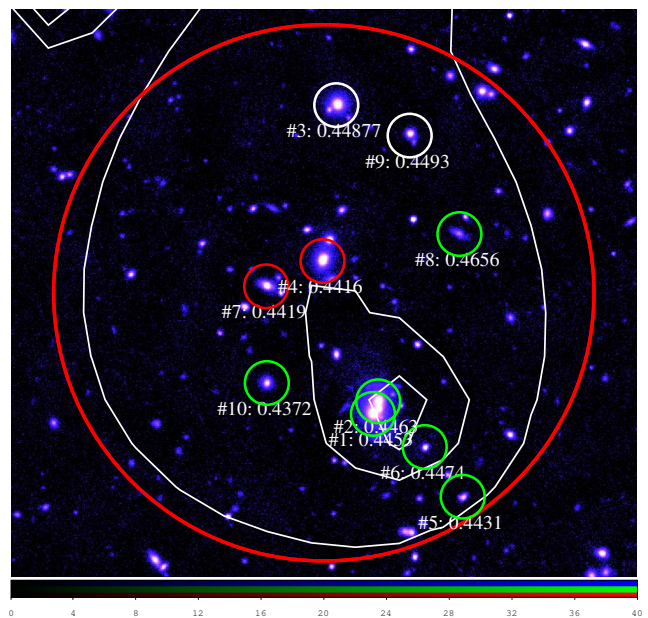}
\caption{\label{fig:test0338}CFHTLS 2.4'$\times$2.3' i' image of XLSSC 110 with known galaxy members of the
structure. Green circles are from the main dynamical structure, and 
red and white circles are from two secondary structures following 
the Serna-Gerbal technique. The large red circle represents a 500$\thinspace$kpc radius.
White contours are for the X-ray emission. The given numbers are the galaxy identification in
Table~\ref{tab:listePP} and the redshifts of the same table (giving priority to the new redshifts 
we measured ourselves).}
\end{figure}

The final list of obtained redshifts is given in Table~\ref{tab:listePP}. 
We confirm the value of three previously measured redshifts and 
successfully measure five new redshifts. 
Among these new redshifts, four are located in XLSSC 110.
This structure is clearly dominated by four bright galaxies. 
Two of them (ids 1 and 2) seem located at the bottom of the potential 
well, as traced by the X-ray contours in Figure~\ref{fig:test0338}, 
while the other two (ids 4 and 3) are located to the cluster north.

\begin{table}[t!]
\caption{\label{tab:listePP} Results of the redshift measurements on the detected galaxies
in the PMAS/PPak data for XLSSC 110 cluster. Columns are: galaxy id., known redshift, coordinates (J2000), new redshift
measurement, spectral flag. This spectral flag is the same as for other redshift measurements. 
We also recall the already known spectroscopic redshifts for this cluster (the ones with no 
z$_{new}$). Identification numbers are the ones shown in Figure~\ref{fig:test0338} if the 
galaxy is inside the structure and are arbitrary identifications if outside or if the redshift 
value is uncertain.}
\begin{tabular}{rrccrr}
\hline
\hline
Gal & z$_{prev}$ & $\alpha$ & $\delta$ & z$_{new}$ &   Flag  \\
\hline						
1   & 0.4453   & 33.5339 & -5.5927  &  0.4453  & 3 \\ 
2   & 0.4453   & 33.5335 & -5.5919  &  0.4463  & 3 \\
3   & 0.4488   & 33.5362 & -5.5730  &          &   \\
4   & 0.4420   & 33.5371 & -5.5830  &  0.4416  & 4 \\
5   & 0.4431   & 33.5282 & -5.5980  &          &   \\
6   & 0.4474   & 33.5306 & -5.5948  &          &   \\
7   & -        & 33.5407 & -5.5846  &  0.4419  & 3 \\
8   & -        & 33.5284 & -5.5812  &  0.4656  & 2 \\
9   & -        & 33.5315 & -5.5750  &  0.4493  & 2 \\
10  & -        & 33.5406 & -5.5907  &  0.4372  & 9 \\
A4  & -        & 33.5311 & -5.5883  &  0.1687  & 2 \\
\hline
B2  & -        & 33.5366 & -5.5817  &  0.4456  & 1 \\
B3  & -        & 33.5381 & -5.5811  &  0.4510  & 1 \\
A3  & -        & 33.5363 & -5.5890  &  0.3267  & 1    \\
B4  & -        & 33.5359 & -5.5798  &  0.4242  & 1    \\
B5  & -        & 33.5314 & -5.5804  &  0.4758  & 1    \\
C2  & -        & 33.5420 & -5.5859  &  0.4764  & 1    \\
\hline
\end{tabular}
\end{table}

With redshifts in Table~\ref{tab:listePP}, and only considering the 
secure spectroscopic redshifts (flags greater than or equal to 2,
considering the new measurements when available), we have the 
minimal number of redshifts to search for possible substructures 
inside XLSSC 110 with the Serna $\&$ Gerbal (1996) technique 
(SG hereafter). 
Already used in several articles (e.g. Adami et al. 2016, hereafter XXL paper VIII), this 
hierarchical method first identifies the substructures in a dynamically 
linked galaxy population, and also provides rough estimates for the mass of 
the substructures. We note that masses are estimated through a basic 
version of the Virial theorem (cf. Guennou 
et al. 2014). More precisely, the SG hierarchical method calculates 
the potential binding energy between pairs of galaxies and detects
substructures by taking positions, magnitudes, and redshifts into account. 

The SG method detects three substructures in the XLSSC 110 cluster. 
The first substructure has six galaxies and an estimated optical
dynamical mass of (1$^{+5}_{-1}$)$\times$10$^{13}$ $\thinspace$ M$_{\odot}$ (green circles in 
Figure~\ref{fig:test0338}). 
It can be considered as the cluster original structure. 
Within this structure, the more linked galaxies are $\#$1 and 2, then 5, and then 6, 10, and 8.
Galaxy $\#$8 is clearly a disk galaxy, is the one with the largest redshift, and is probably in an
infalling process onto the main structure.

Two other substructures are detected. They are smaller (2 galaxies each: red and white
circles in Figure~\ref{fig:test0338}). 
Their levels of binding energy to the main structure are different. 
The red structure of Figure~\ref{fig:test0338} is more linked to the main 
green structure than the white one. 
In physical terms, this could mean that the red structure has been in the process 
of merging with the main structure for a longer time than the white one. 
Considering the green and red structures together, the estimated optical 
dynamical mass is 3$\times$10$^{13}$ $\thinspace$ M$_{\odot}$.

This behaviour is in good agreement with our initial visual interpretation 
of the physics of this cluster.
We note, however, that our redshift catalogue remains quite sparse and more 
spectroscopic redshifts would be required to confirm this interpretation.

\subsection{Fossil groups}                                                                                

Fossil groups (FG hereafter) are peculiar structures of galaxies with an extended X-ray halo. Jones et 
al. (2003) defined them more precisely as structures with an X-ray bolometric
luminosity of more than 10$^{42}$ \thinspace erg s$^{-1}$ and a difference of two magnitudes or more 
between the first and second ranked galaxies within half the group Virial radius.
These structures mostly appear in the optical as isolated large early type galaxies. Most of the time, only X-ray
data can reveal the existing extended massive halo. Several other studies were made as in Khosroshahi, Ponman $\&$ Jones (2007) 
where additional criteria were added. 
Sometimes studied individually (e.g. Adami et al. 2007, 2012 or Ulmer et al. 2005), FGs were also the subject of statistical
studies: for example the FOGO sample (Santos et al. 2007, Girardi et al. 2014).

\subsubsection{Our selection}

Most of the studies to date, were based on optically selected FG samples (e.g. the FOGO sample). We 
propose here a first catalogue of candidates based on an pure X-ray selection. We decided not to limit 
our sample to a specific X-ray luminosity range. We therefore explored the full XXL spectroscopically confirmed cluster luminosity
range. For each of the spectroscopically confirmed galaxy structures, we examined their 
optical counterparts both in photometry (using photometric redshift techniques) and in spectroscopy (with the XXL 
spectroscopic general follow-up) to search for FG candidates.
We used a slightly different radius criterion: instead of 0.5$\times$r$_{200}$, we used 1$\times$r$_{500,scal}$ (from Table~\ref{tab:listeSL}). Following Roussel et 
al. (2000), the ratio between r$_{500,scal}$ and r$_{200}$ is 0.66. Our criterion is therefore slightly more stringent than the one of 
Jones et al. (2003).

We first selected all the spectroscopically confirmed galaxy structures in our
sample. Then we used our photometric catalogues giving the position and magnitudes of 
objects in the fields, their photometric redshift, and the associated 
redshift probability distribution function (PDF hereafter). A spectral 
star galaxy separation is also available.

We selected for each structure in a r$_{500,scal}$ radius all objects with a high 
probability to be a galaxy (probability to be a star lower than 10$\%$). When 
available, we added the spectroscopic redshift to this sample from our 
spectroscopic database (http://cesam.oamp.fr/xmm-lss/), only considering 
spectroscopic redshifts measured with more than $\sim$85$\%$ confidence 
(i.e. spectroscopic flags two or better, e.g. Le F\`evre et al. 2013).

At this step, we selected the dominant galaxy, defined as the brightest galaxy
in the r' band at less than 75$\thinspace$kpc from the X-ray centre. This distance is approximately the maximal
distance we can expect in a cluster between the BCG and the bottom of the potential
well (e.g. Adami $\&$ Ulmer 2000).

If by chance this galaxy has a
spectroscopic redshift, we check if the redshift is consistent with the 
structure redshift. The consistency criterion is defined as $\pm$3 times the
velocity dispersion of the structure estimated from X-ray luminosity (giving priority to direct spectral
measuments). If not consistent, this galaxy is removed and the next brighter galaxy is considered.

If no spectroscopic redshift is available, we consider the photometric 
redshift instead, exceptionally enlarging the consistency criterion to 
$\pm$0.1 in redshift to take into account of the larger uncertainty of the photometric redshifts ($\pm$0.056 in the south
and $\pm$0.034 in the north, Fotopoulou et al. 2016 hereafter XXL paper VI, see below) compared to the spectroscopic ones. 

The dominant galaxy being defined, we selected all galaxies along the structure 
line of sight (within r$_{500,scal}$) in the next two-magnitude interval (this requires 
obviously the magnitudes to be successfully measured). These candidate lists 
were finally scrutinised to conclude about the fossil group nature of the 
considered structures. For a considered structure and a given galaxy, we computed the probability for 
this galaxy to be outside of the previous $\pm$3 times velocity dispersion 
interval. 

- For the galaxies with a spectroscopic redshift, the galaxy probability is 1 (not FG member) or 
0 (FG member) depending on the redshift and the structure velocity dispersion.

- For the galaxies without a spectroscopic redshift, the probability is 
computed with the PDF of the galaxy, simply integrating it out 
of the structure redshift interval. 

Taking the product of these probabilities for the different galaxies in the candidate list 
(excluding the BCG) to be FG members or not, this gives finally the probability for the structure itself 
to be a fossil group.
As an example, if a structure has a single galaxy within the two-magnitude range fainter than the BCG (besides the
BCG itself) with a spectroscopic redshift within the redshift interval, the probability for this structure to be a fossil 
group will be null.

Each of our confirmed galaxy structures was then scrutinised taking into account their
probability to be a FG. After having removed obvious interlopers (e.g. structures polluted 
by bright stars, complex structure superpositions, incomplete photometric samples), we decided
to retain as FG possible candidates only structures with a probability greater than 20$\%$ to be a FG. This level was
determined a priori as the minimum percentage level below which all structures were easely classified by hand as non-FG 
structures. This level is intentionally low and it will imply a large number of false-positives. FG being however rare objects,
it is a way to not lose any of them. These potential candidates are given in Table~\ref{tab:listeFG}. 

\begin{figure}[h]
\includegraphics[width=6.5cm,angle=270]{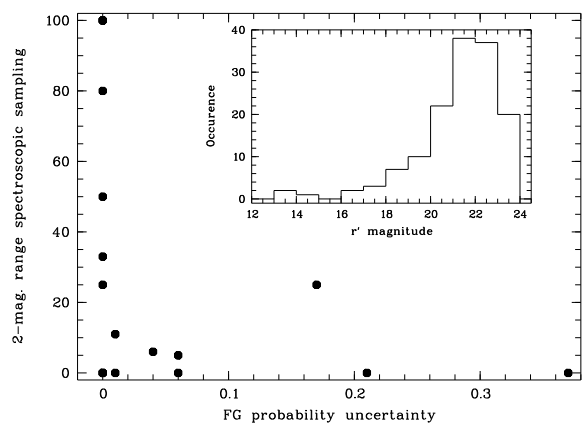}
\caption{\label{fig:figSophie} 
Percentage of spectroscopic redshifts within r$_{500,scal}$ that are found in the two-magnitude interval fainter than the BCG, as a function of the
uncertainty on the probability for a cluster to be a fossil group (given in Table~\ref{tab:listeFG}). 
Inset: r' band magnitude histogram of the galaxies without a spectroscopic redshift, and within the two-magnitude range of our FG candidates.}
\end{figure}

\begin{table}[t!]
\small
  \caption{\label{tab:listeFG}List of FG candidates. Col.1: XXL name of the confirmed galaxy structure. Col.2: probability
of the structure to be a FG. Col.3: number of galaxies affected by catastrophic errors. Col.4: typical uncertainty on the 
probability of the structure to be a FG due to catastrophic error percentages. Col.5 and 6: 
structure coordinates. Col.7: redshift of the structure. Col. 8: spectroscopic redshift sampling percentage for the considered line of sight within
the two-magnitude range (BCG excluded). }
\begin{tabular}{cccccccc}
\hline
\hline
\#    &  Prob. & N af.  & Err. & $\alpha$ & $\delta$ & z & Samp.  \\
\hline
171 &  0.49 &0&0.      & 31.986  &     -5.871    &    0.044   &  80  \\       
162 &  0.38 &0&0.      & 32.524  &     -6.093    &    0.138   &  33  \\            
128 &  0.21 &1&0.01    & 36.048  &     -3.129    &    0.480   &  0   \\             
127 &  0.38 &0&0.      & 36.850  &     -3.566    &    0.325   &  0   \\             
147 &  1.00 &0&0.      & 37.641  &     -4.625    &    0.031   &  25  \\            
554 &  0.34 &0&0.      &348.719  &    -53.626    &    0.202   &  50  \\  	   
560 &  0.85 &3&0.01    &349.420  &    -52.739    &    0.790   &  11  \\
576 &  0.24 &2&0.04    &350.542  &    -56.312    &    0.702   &  6   \\            
597 &  1.00 &0&0.      &350.765  &    -52.725    &    0.151   &  100 \\
581 &  1.00 &0&0.      &352.416  &    -54.789    &    0.138   &  100 \\           
520 &  1.00 &0&0.      &352.502  &    -54.619    &    0.175   &  0   \\      
582 &  0.53 &3&0.17    &352.610  &    -54.784    &    0.406   &  25  \\            
629 &  1.00 &0&0.      &353.928  &    -54.349    &    0.173   &  100 \\          
604 &  0.23 &1&0.37    &354.976  &    -56.254    &    0.381   &  0   \\           
566 &  0.21 &2&0.06    &357.008  &    -53.656    &    0.634   &  0   \\            
564 &  0.21 &6&0.06    &357.079  &    -53.395    &    0.981   &  5   \\         
567 &  0.56 &1&0.21    &357.222  &    -53.823    &    0.254   &  0   \\           
565 &  1.00 &0&0.      &357.339  &    -53.506    &    0.167   &  0   \\       
\hline
\end{tabular}
\end{table}

\subsubsection{Properties of our FG candidates}

We can ask the question of why 2.6 times more FG 'candidates' (5 times more if we consider only the high probability FGs) are found in the 
southern XXL field than in the northern field.

First, an obvious explanation could be the spectroscopic follow-up sampling which is much higher in the north 
thanks to the SDSS, GAMA, VIPERS, and VVDS surveys. To exclude a given galaxy structure from the FG class, we need to 
be sure that a galaxy is inside the structure with a magnitude fainter than the BCG by less than two magnitudes. Let us 
assume the existence of such a galaxy. The uncertainty on its redshift location with regard to the structure's mean 
redshift is typically the redshift measurement uncertainty. 

- If the redshift is spectroscopic, we will know quite precisely where the galaxy is and the probability 
to (wrongly) estimate that the galaxy is outside the structure will be low. The structure will therefore
be excluded from the FG class with a high probability.

- If the redshift is photometric, the probability to (wrongly) estimate that the galaxy is outside the structure will be 
much higher (photometric redshift uncertainties are typically at least twenty times larger than for spectroscopic
redshifts). The structure may therefore not be excluded from the FG class.

It is therefore much easier to exclude a structure from the FG class with spectroscopic redshifts than with only photometric 
redshifts. In our data, statistically, a northern galaxy structure line 
of sight is sampled by $\sim$17 times more redshifts than a southern line of sight. This may indeed favour the existence of a larger
number of remaining FG candidates in the southern fields. However, we note that in terms of galaxy 
structure spectroscopic members, northern structures are not significantly better sampled than southern structures. 

Second, the photometric redshifts are less precise in the south (uncertainty of 0.056 in redshift and catastrophic error percentage 
of 15$\%$) than in the north field (uncertainty of 0.034 in redshift and catastrophic error percentage 
of 3$\%$), Fotopoulou et al. (private com.). These numbers will be described in a future XXL paper, but
the catastrophic error percentage can induce non negligible uncertainties on the probability of a structure to be a FGs
(cf. Appendix A for a complete description of the computation).
 Taking these uncertainties into account (cf. Table~\ref{tab:listeFG}), basically none of the
northern FG candidates are affected, while up to four southern FG candidates may not be real FGs. 
The previously quoted ratio between northern and southern FG candidates would only be 1.8 with just this explanation.
This does not, however, explain the 1:5 ratio between the number of high probability FG candidates in the two fields, as none
of these objects are affected.

Another way to test if we have significantly more FG candidates in the southern than in the northern area is simply to sum the probability of all massive
structures (not only the FG candidates listed in Table~\ref{tab:listeFG}) in both fields of being such FGs. If the two fields are similar from
a cosmic variance point of view, these sums should be identical. We find a ratio of 1.1 between the two sums, speaking in favour of the northern and 
southern fields being indeed similar.

We also compared (using a bi-dimensional Kolmogorov-Smirnov test) the distribution of the northern and southern FG candidates within the X-ray 
luminosity vs X-ray luminosity uncertainty space, and within the X-ray luminosity vs redshift space. X-ray luminosity is here estimated from scaling 
laws. In the first case, the probability to have similar distributions in the north and in the south is more than 65$\%$. In the second case, the 
probability to have similar distributions is greater than 99$\%$.

\begin{figure}[h]
\includegraphics[width=6.5cm,angle=270]{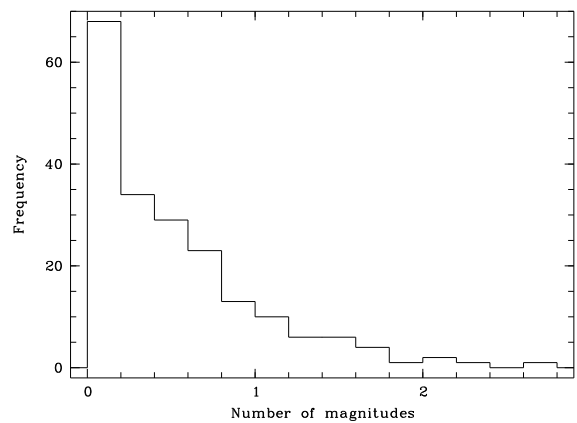}
\caption{\label{fig:gapmag} Histogram of the number of magnitudes we can add 
to the brightest cluster member r' magnitude in order to achieve a probability larger than 20$\%$ to have no galaxy within the defined magnitude 
interval (for all the spectroscopically confirmed clusters of the XXL northern area).}
\end{figure}

We however stress that only a spectroscopic follow-up of the photometric redshift classified galaxies will definitely tell us whether we have
a significant difference between the FG density in the northern and southern XXL fields, in particular when considering high probability
FG candidates. This is also the only way to assess nominatively the FG nature
of our candidates. We show in Table~\ref{tab:listeFG} and in  Figure~\ref{fig:figSophie} that the uncertainty of a candidate to be a FG is
obviously related to the spectroscopic redshift sampling percentage for the considered line of sight (within the two-magnitude range and BCG excluded). 
The fewer
spectroscopic redshifts we have along the line of sight, the more photometric redshifts we need and they are potentially affected by catastrophic errors.
Reaching a sampling percentage of better than 30$\%$ would lead to a negligible uncertainty on the FG nature due to photometric redshift catastrophic errors.
The inset in the same figure also shows that the magnitude range of the lacking spectroscopic redshifts (within the two-magnitude range) is easely reachable with an
integral field spectrograph such as MUSE/VLT.

\subsubsection{Are our FG candidates different from the general XXL cluster population?}

We cannot reproduce the same tests as in the previously quoted literature studies without performing 
additional spectroscopical follow-up of our FG candidates. We can however perform a basic test: are our FG candidates 
simply extreme cases
of the general galaxy structure population, obeying to the same formation process, evolution path, ..etc...? Or
alternatively, are FGs an independent population of galaxy structures? Several studies as Girardi et al. (2014),
Zarattini et al. (2014) and Kundert et al. (2015) seem to show that FGs behave very similarly to normal
galaxy structures. 
Taking into account the full spectroscopically confirmed cluster sample in the XXL northern area (the one which has 
the best spectroscopic sampling), we computed the number of magnitudes we can add to the brightest cluster member
r' magnitude (as defined previously for the FGs) in order to achieve a probability greater than 20$\%$ to have no galaxy
within the defined magnitude interval.
Good FG candidates therefore appear in this plot as the galaxy structures with the largest magnitude gaps, greater or 
equal to 2. In Figure~\ref{fig:gapmag}, there appears to be a continuous variation of this gap, the good FG 
candidates being only the extreme cases. The same exercise in the XXL southern area gives very similar results.
These histograms may be however polluted by interlopers (galaxies with photometric redshifts, but outside of the 
cluster if we had a spectroscopic redshift). Once again, this shows that a more complete spectroscopic redshift 
follow-up is needed for these FG candidates. 
This also speaks in favour of a common origin between regular groups of galaxies and FGs.  

FG candidates are however significantly fainter in terms of X-ray luminosity compared to the global cluster sample. A Kolmogorov-Smirnov test
shows that there are less than 4$\%$ of chances that the global sample of clusters has a similar X-ray luminosity (estimated from scaling laws) 
distribution compared to the luminosity distribution of our FG candidates.

\section{Conclusion}

In the present paper we released several catalogues based on a sample containing 365 clusters in total.
We described the follow-up observations, the precision of the measured galaxy redshifts, and explained the procedure
adopted to validate the cluster's spectroscopic redshifts.

We provided X-ray flux, luminosity, temperature, and direct gas mass measurements for a large part of the sample
extending from z$\sim$0 to z$\sim$1.2 (with a cluster at z$\sim$2). We also estimated from scaling relations luminosities, 
temperatures, and total masses.
Using this 365 cluster sample, we updated the previous XXL luminosity function and luminosity-temperature relations only
based on the 100 brightest clusters.
We presented an enlarged catalogue of super-clusters and a sample of 18 fossil group candidates.

This intermediate publication is the last before the final release of the complete XXL cluster catalogue. It provides a unique 
inventory of medium-mass clusters over a 50~\dd\ in a $0<z<1.2$ cone and gives a flavour of the general properties of the 
cluster sample.

\begin{acknowledgements}
The authors thank the referee for useful comments.
XXL is an international project based around an XMM-Newton Very Large Programme 
    surveying two 25 deg2 extragalactic fields at a depth of $\sim$5 $\times$ 10$^{-15}$ \thinspace erg 
    cm$^{-2}$ s$^{-1}$ in the [0.5-2] keV band for point-like sources. The XXL website is 
    http://irfu.cea.fr/xxl. Multiband information and spectroscopic follow-up of the X-ray 
    sources are obtained through a number of survey programmes, summarised at 
    http://xxlmultiwave.pbworks.com/. 
We gratefully acknowledge financial support from the Centre National d'Etudes Spatiales 
throughout many years. We also acknowledge financial support from 'Programme National de 
Cosmologie et Galaxies' (PNCG) of CNRS/INSU, France. We also thank Calar Alto Observatory 
for allocation of director's discretionary time to this programme. C.A.C. 
acknowledges financial support from CONACyT (Mexico) for sabbatical leave and 
Laboratoire d'Astrophysique de Marseille for hospitality and support during 
stay. Also based in part on data acquired through the Australian Astronomical
Observatory, under programme A/2016B/107.
We also thank the whole GAMA and VIPERS teams for sharing unpublished data.
DR is supported by a NASA Postdoctoral Program Senior Fellowship at the NASA Ames Research Center, administered by the 
Universities Space Research Association under contract with NASA.
MP acknowledges the financial support from Labex OCEVU (ANR-11-LABX-0060).
FP and MERC acknowledge support by the German Aerospace Agency (DLR) with
funds from the Ministry of Economy and Technology (BMWi) through grant 50 OR
1514 and grant 50 OR 1608.
SA acknowledges a post-doctoral fellowship from TUBITAK-BIDEB through 2219 program.
SF acknowledges support from  the  Swiss  National  Science  Foundation.

\end{acknowledgements}

\clearpage

\appendix

\section{Uncertainty on the probability of an XXL structure to be a FG}     

In the process of computing the probability of an XXL structure to be a FG, galaxy photometric redshift uncertainties are probably not crucial 
in our case because we are directly using the spectral energy distributions.
A high catastrophic error percentage can however have a non negligible effect as it can be the sign of a failed computation of the spectral energy 
distribution itself. Assuming this is the case, we computed for each of our FG candidates the number N of galaxies potentially affected. This number
N is the total number of galaxies along the considered FG line of sight within the two-magnitude range, minus the number of galaxies with a
spectroscopic redshift (these ones are not affected by the catastrophic errors). For
each of these galaxies, the redshift is therefore unknown. The probability to be outside of the considered structure is then equal to the ratio
of the explored redshift range (z=[0-6]) over the typical redshift range covered by a massive structure (less than 0.024). This gives
a statistical probability to be outside of the considered structure of 99.6$\%$ if we assume that no redshift selection effects are at play. 
For each of the FG candidates, we therefore simulated
100 times the replacement by 99.6$\%$ of the initially computed structure membership probability of the N galaxies. This gave us an estimate of the uncertainty
of the considered massive structures of being FGs (cf. Table~\ref{tab:listeFG}).

\section{Notable superpositions}                                                                                      

In this section we present noticeable line of sight cluster - cluster associations (most with
overlapping X-ray emissions). This is not a systematic search in the XXL Survey, but just
a summary of the evident associations we eyeballed in the survey. A list of potential cluster pairs is given in Table~\ref{tab:PL}.
All images are CFHTLS i' band images, with north to the top and east to the left. 
The white contours are the XMM X-ray contours, which were created using local minimum and maximum and a logarithmic scale 
with ten levels.

\begin{figure}[h]
\includegraphics[width=\linewidth,viewport=41 140 570 654]{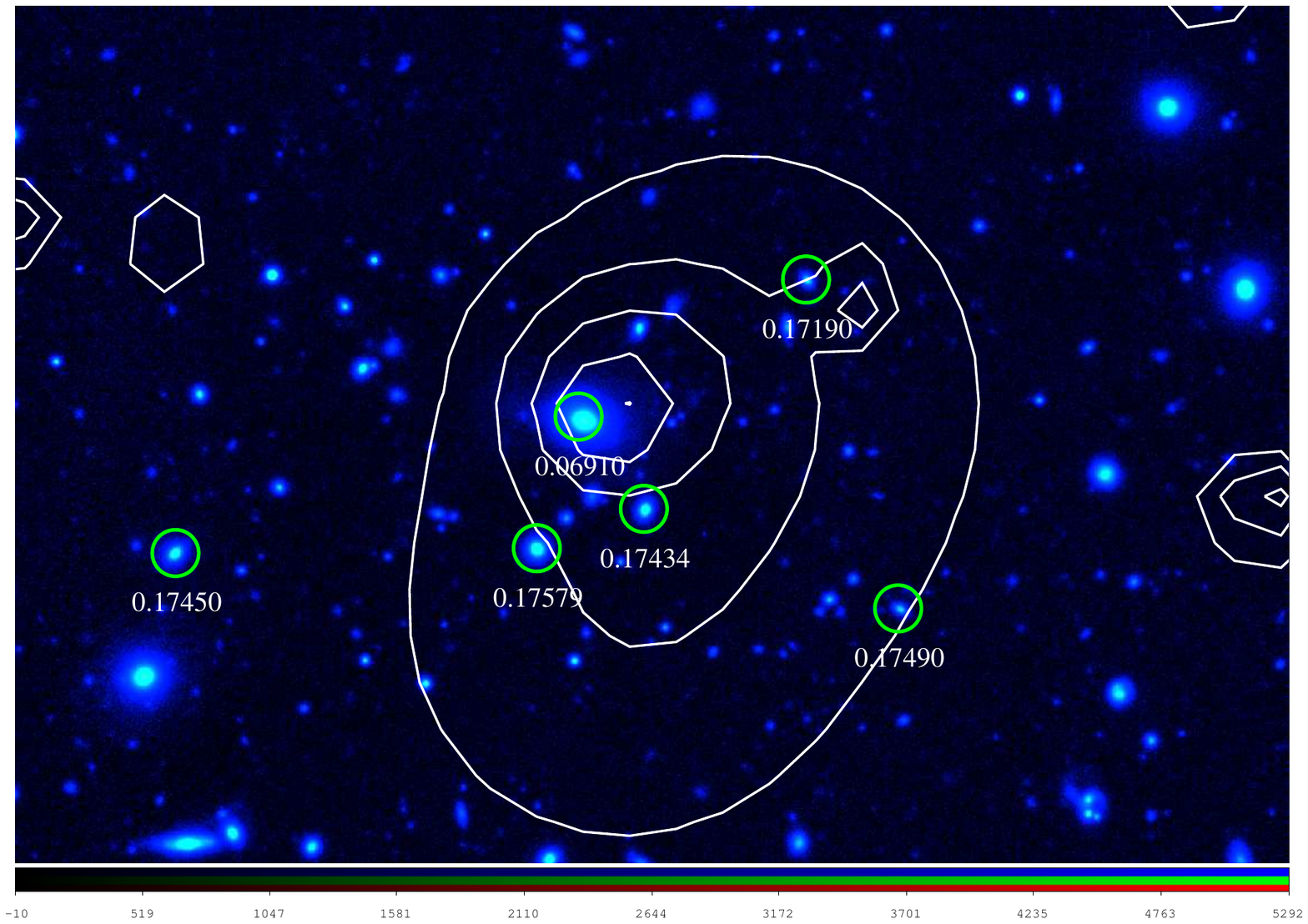}
\caption{\label{fig:n0034}CFHTLS i'band 4.5'$\times$3' image around the XLSSC 035 confirmed z=0.174 cluster. Green circles
represent the member galaxies plus the bright foreground galaxy discussed in the text. White contours are for the X-ray emission.}
\end{figure}

XLSSC 035: This structure is a C1 confirmed cluster. It presents a nice superposition of a z$\sim$0.174
cluster and of a bright galaxy at z=0.0691 (cf. Figure~\ref{fig:n0034}). Initially, this structure was classified 
at the redshift of this bright galaxy (assumed to be the BCG of the cluster). Later redshift measurements showed 
that the structure thought likely to be associated with the X-ray extended emission is in fact more 
distant. A second measure of the bright galaxy was made, confirming its z=0.0691 redshift.

\begin{figure}[h]
\includegraphics[width=\linewidth,viewport=41 140 570 654]{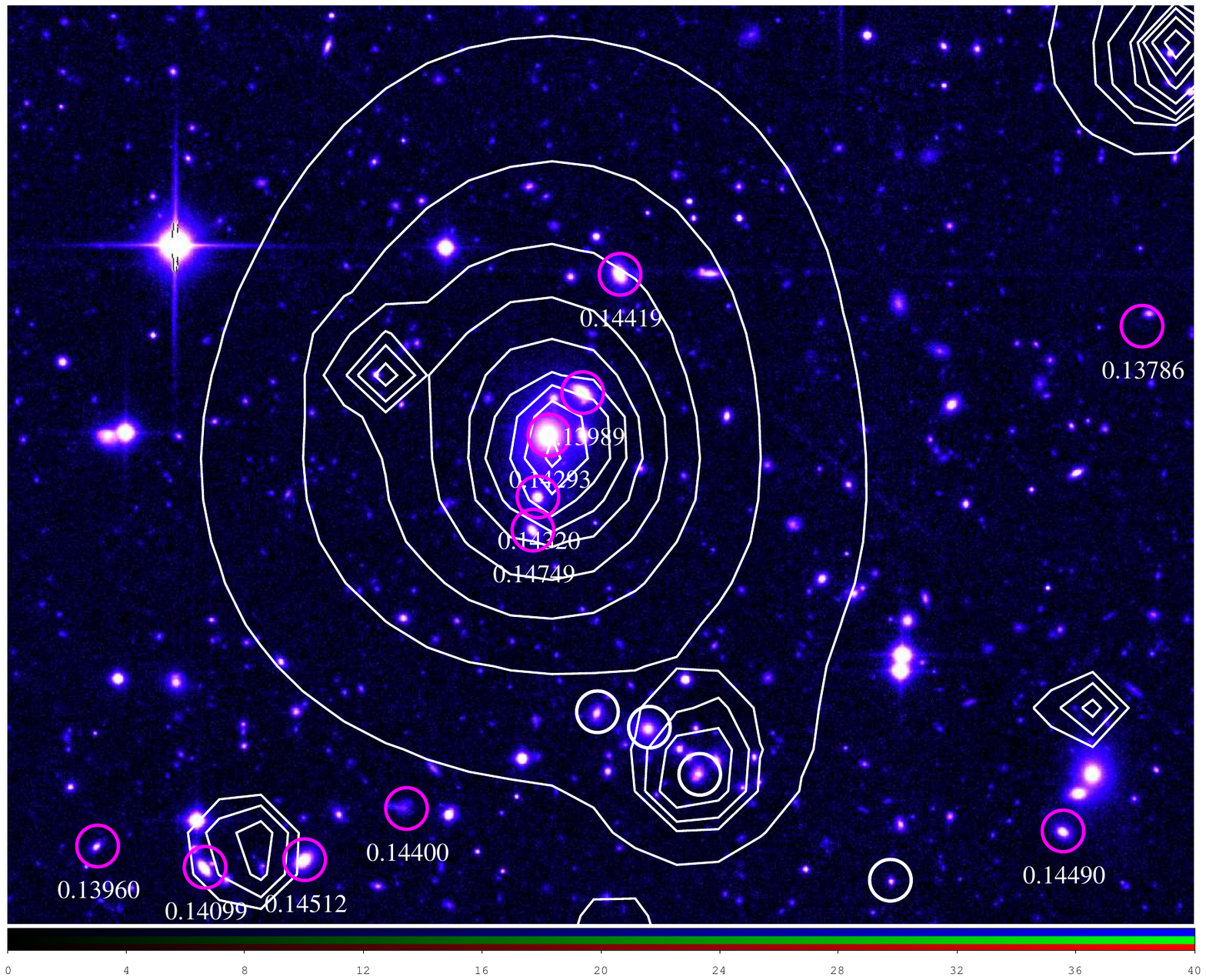}
\caption{\label{fig:n0016}CFHTLS i'band 5.5'$\times$3.5' image around the XLSSC 041 confirmed z=0.142 cluster. Magenta circles
represent the member galaxies. White circles are the galaxy members of the background cluster at z=0.557.
White contours are for the X-ray emission.}
\end{figure}

XLSSC 041: This structure is a very regular C1 cluster at z=0.142. Figure~\ref{fig:n0016} however shows another 
X-ray concentration at the (36.3682,-4.2602) coordinates towards the south-west. This secondary peak nicely corresponds to
a concentration of four galaxies at a mean redshift of 0.557. XLSSC 041 therefore seems to be a regular cluster polluted by another
line of sight z=0.557 cluster. 

\begin{figure}[h]
\includegraphics[width=\linewidth,viewport=41 140 570 654]{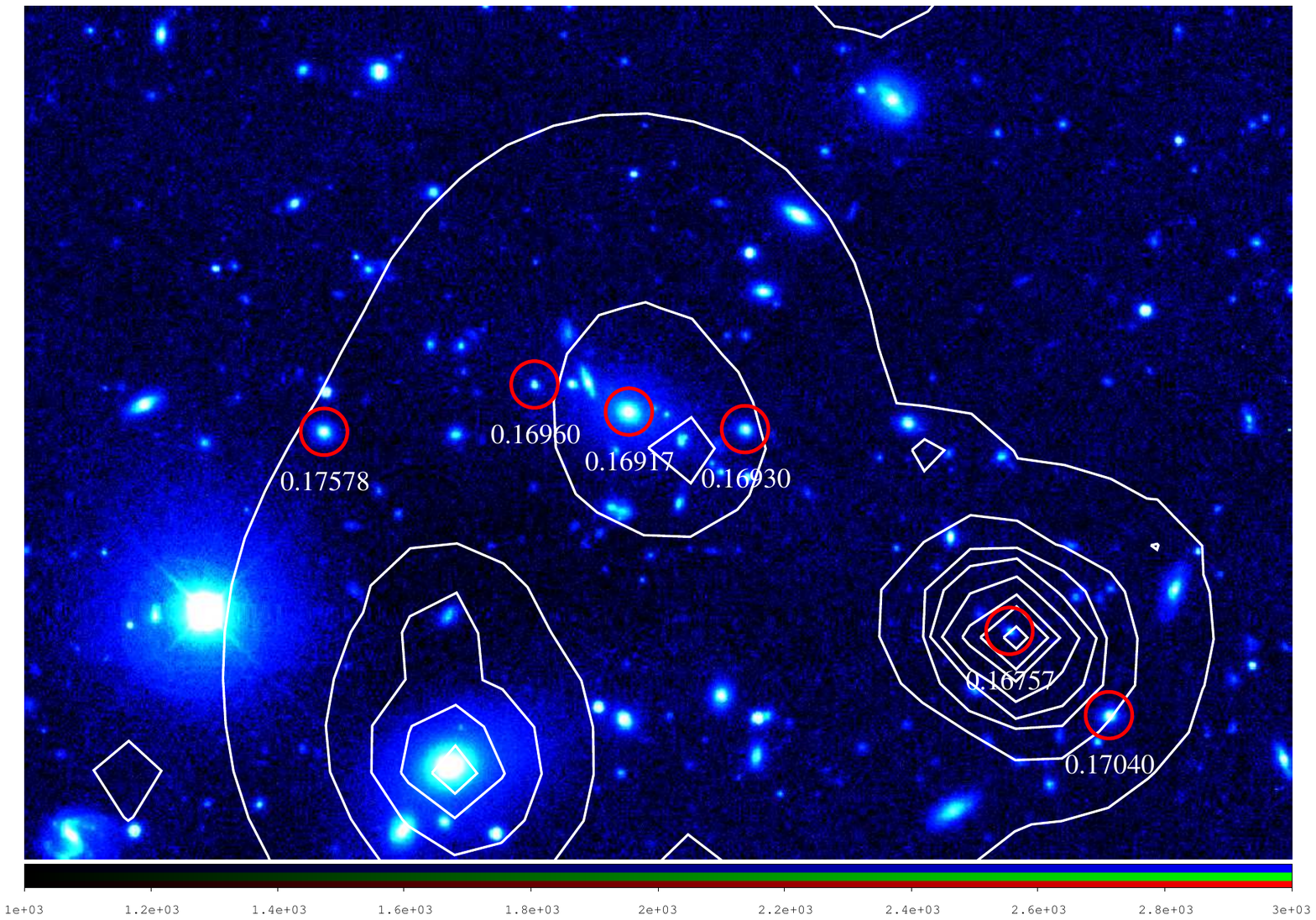}
\includegraphics[width=\linewidth,viewport=41 140 570 654]{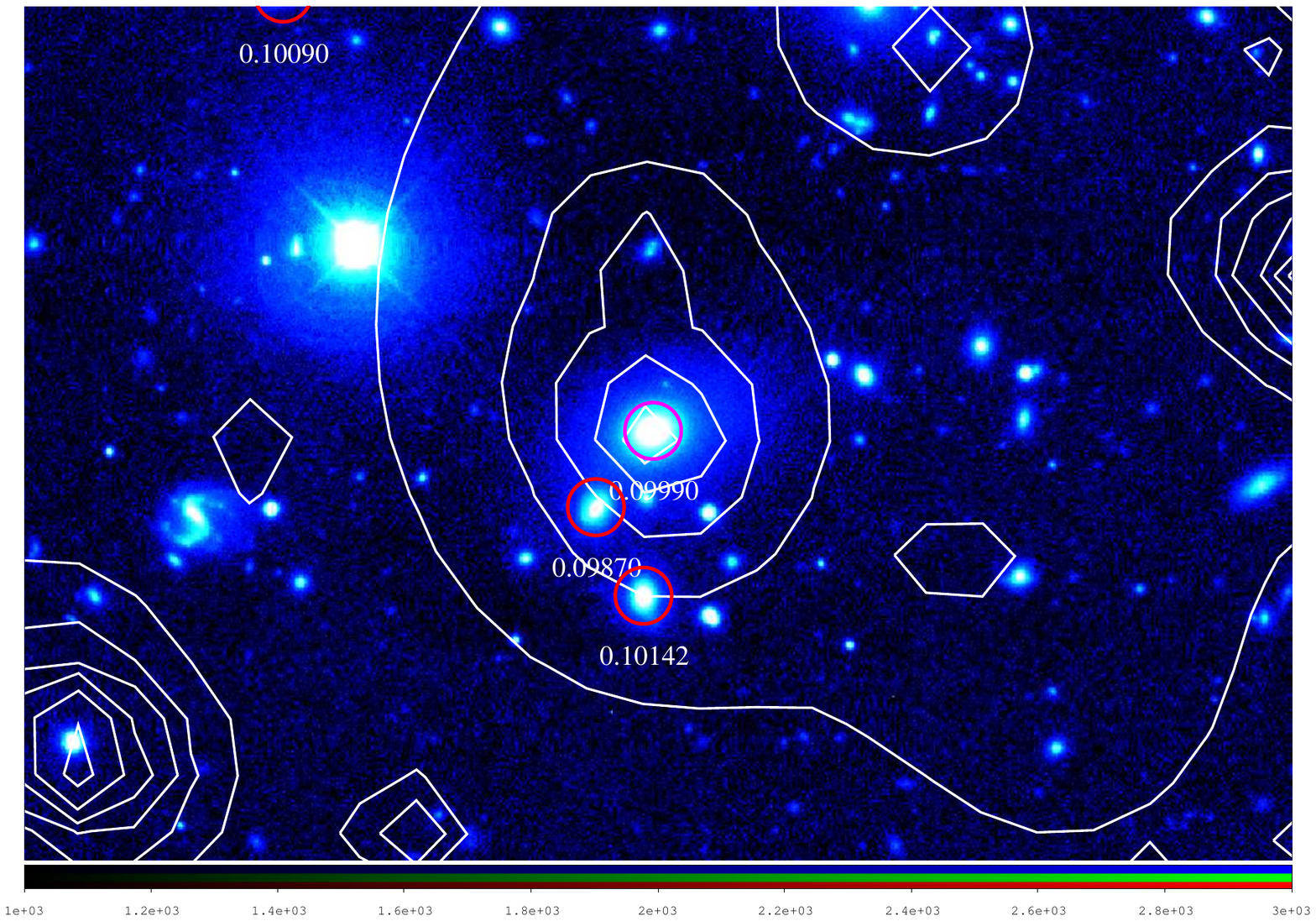}
\caption{\label{fig:s00790080}CFHTLS i'band 4.5'$\times$3' image around XLSSC 514 and XLSSC 515 confirmed z=0.101 and 0.169 
clusters. Magenta circles represent the member galaxies of the two clusters. Figures are showing the central areas
of these structures. White contours are for the X-ray emission.}
\end{figure}

XLSSC 514 and XLSSC 515: These two clusters are confirmed C1 structures and are very close in projection. Their X-ray emissions
are merged on the sky, the first one being at z=0.101 and the second one at z=0.169 (cf. Figure~\ref{fig:s00790080}).

\begin{figure}[h]
\includegraphics[width=\linewidth,viewport=41 140 570 654]{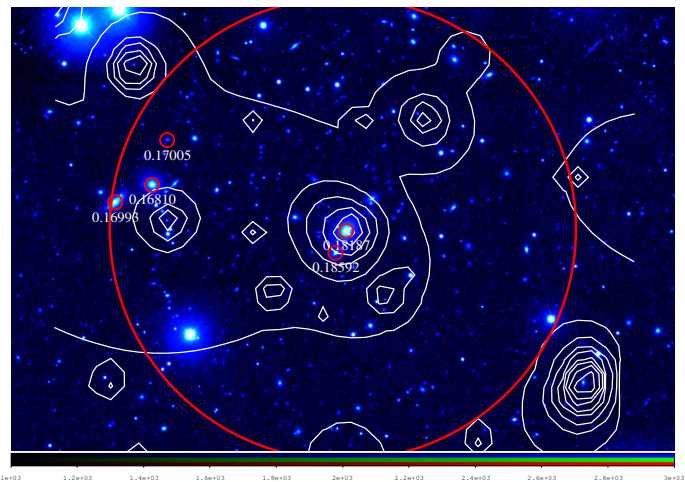}
\caption{\label{fig:s0182}CFHTLS i'band 7.5'$\times$5' image around XLSSC 539 confirmed z=0.184 and 0.169 double
structure. Red circles represent the member galaxies of the two components. We clearly see the central z=0.184
structure and the other component at z=0.169 towards the east. The large red circle represents a 500$\thinspace$kpc radius area.
White contours are for the X-ray emission.}
\end{figure}

XLSSC 539: This C1 cluster has two components as shown in Figure~\ref{fig:s0182}. The first one at z=0.184 was assumed
as the main cluster redshift because very well correlated with the X-ray peak. Another cluster at z=0.169 is also present 
towards the east and is probably an infalling structure entering a future merging state.

\begin{figure}[h]
\includegraphics[width=\linewidth,viewport=41 140 570 654]{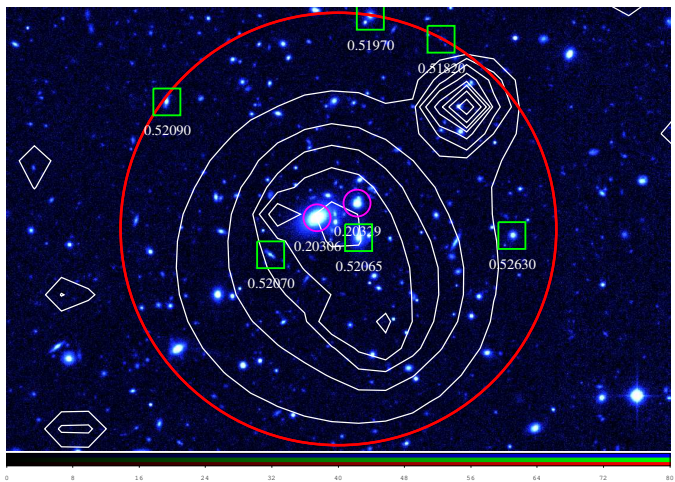}
\caption{\label{fig:n0260}CFHTLS i'band 4.'$\times$2.5' image around XLSSC 096. Green squares represent the member galaxies 
of the main structure at z=0.520 and magenta circles are the foreground structure at z=0.203. The large red circle 
represents a 500$\thinspace$kpc radius area. White contours are for the X-ray emission.}
\end{figure}

XLSSC 096: This X-ray source is a very nice example of close superposition on the sky of two different galaxy structures
(cf. Figure~\ref{fig:n0260}). The first one is sampled with six redshifts at z=0.520 (including a BCG-like galaxy at z=0.5206).
The second one is exactly on the same line of sight, at z=0.203 (sampled by two spectroscopic redshifts, including also a
BCG-like galaxy). We choose to adopt z=0.520 because of the greater richness of this component.

\begin{figure}[h]
\includegraphics[width=\linewidth,viewport=41 140 570 654]{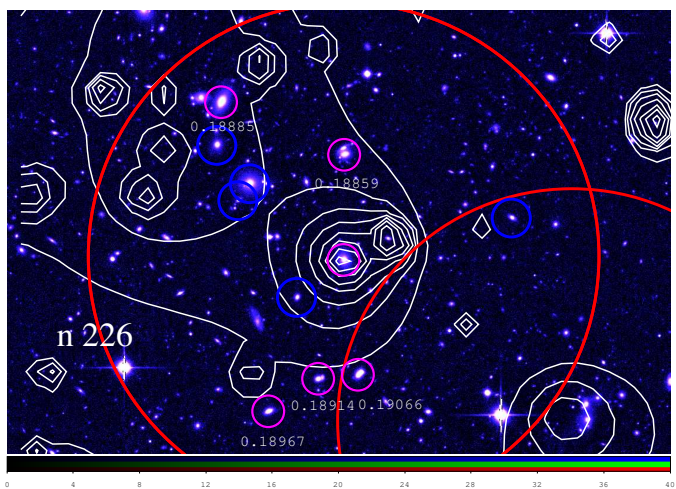}
\caption{\label{fig:n0226}CFHTLS i'band 6.5'$\times$4.5' image around XLSSC 151. Magenta circles represent the member galaxies 
of the main structure at z=0.189 and blue circles those of the background structure at z=0.280. The large red circle 
represents a 500$\thinspace$kpc radius area. White contours are for the X-ray emission.}
\end{figure}

XLSSC 151: This cluster is also a noticeable superposition. The main structure is at z=0.189, clearly located on the main X-ray 
peak (cf. Figure~\ref{fig:n0226}). However, another structure also appears at z=0.280, dominated by a BCG-like galaxy which
also may be correlated with a secondary X-ray peak.

\begin{figure}[h]
\includegraphics[width=\linewidth,viewport=41 140 570 654]{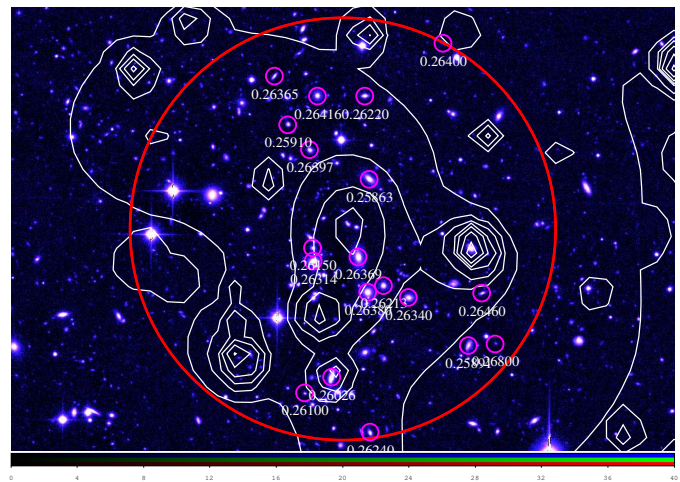}
\includegraphics[width=\linewidth,viewport=41 140 570 654]{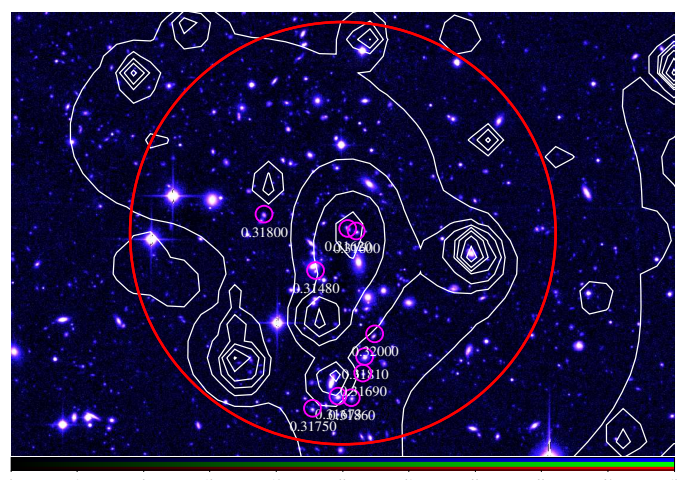}
\caption{\label{fig:n0057}CFHTLS i'band 6.5'$\times$4.5' images of two superposed structures. Magenta circles represent 
the member galaxies of the two structures at z=0.263 (upper panel) and at z=0.317 (bottom panel). The large red circle 
represents a 500$\thinspace$kpc radius area. White contours are for the X-ray emission.}
\end{figure}

XLSSC 044: This line of sight is complex with two richly sampled structures on it (cf. 
Figure~\ref{fig:n0057}). The main one at z=0.263 has nineteen known members and the other one at z=0.317 has ten 
known members.

\begin{figure}[h]
\includegraphics[width=\linewidth,viewport=41 140 570 654]{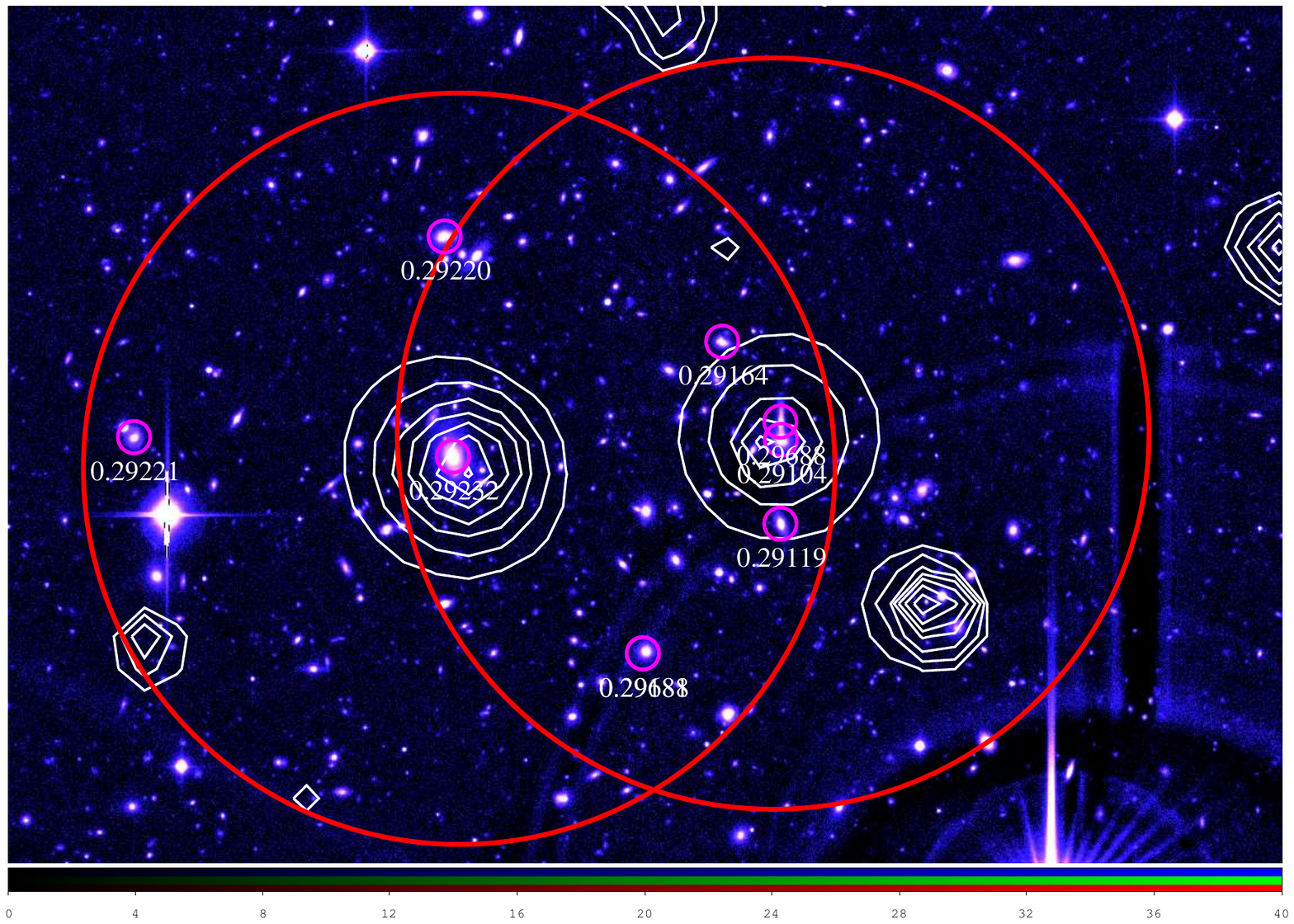}
\caption{\label{fig:n00220224}CFHTLS i'band 6.5'$\times$4.5' image of the XLSSC 149 and 150 structures. Magenta circles represent 
the member galaxies of the two structures at z=0.292. The large red circles 
represent 500$\thinspace$kpc radius areas. White contours are for the X-ray emission.}
\end{figure}

We also note than XLSSC 149 and 150 (cf. Figure~\ref{fig:n00220224}) are two clusters with non overlapping X-ray emission (at the depth of the XXL 
observations). They have however exactly the same redshift (z=0.292) and are separated by less than 500$\thinspace$kpc. This means that we deal
with two low mass structures (each at temperature of $\sim$2keV) with the potential to merge in the future. This is supported by the fact that they 
are both part of the N01 super-cluster (cf. Table~\ref{tab:SCL}).

XLSSC 079: This cluster may suffer from a superposition effect. The main structure is clearly detected at z$\sim$0.19. However, another
structure may be present on the same line of sight at z$\sim$0.52. An SDSS galaxy spectroscopic redshift is available at z=0.5171 
($\alpha$=34.49248, $\delta$=-4.86538) in the DR12, very close to the main structure. This is not enough to officially confirm this superposition,
but this value may support the detection of two literature clusters detected at the same place on a photometric redshift basis: CFHT-W CL J021757.8-045142 
(Wen et al., 2012) at z=0.537 and SXDF35XGG (Finoguenov et al. 2010) at z=0.46.

\section{IFU observations of XLSSC 110}  

The observed field was centred on the cluster position. 
The PPak-IFU of PMAS at the 3.5$\thinspace$m Calar Alto telescope is an hexagonal packed fibre-bundle instrument 
with 331 object, 36 sky and 15 calibration fibres (2.68'' diameter each,
separated by 3.57'' and 3.12'', respectively in the $\mathtt{x}$ and 
$\mathtt{y}$ coordinates, resulting in a 60\% filling factor), 
covering a projected FOV in the sky of 74'' $\times$ 65''. \\

Twelve 1200s exposures were obtained during two observing runs (2015,
December 31 and 2016, February 01-02), with two pointings to each of the six 
positions (Table~\ref{tab:PMASPPAK}) in a dithering mode, using the V300 
grating and 4k$\times$4k (15 $\mu$m pixels, 2.57-2.88 read out noise and 1.14-1.29 
$\mathtt{e}^{-1}/\mathtt{ADU}$) CCD. \\

\begin{table}[t!]
\caption{\label{tab:PMASPPAK} Details of the PMAS/PPak observations of XLSSC 110: exposure number,
run date, coordinates (J2000), air mass range.}
\begin{tabular}{cccc}
\hline
\hline
Exposure & Run & $\alpha$,$\delta$  & Air Mass Range  \\ 
\hline						
1        & 2015 Dec & 33 7941,  -5.4209 & 1.36-1-36     \\ 
2        & 2015 Dec & 33.7932,  -5.4220 & 1.38-1.41     \\
3        & 2015 Dec & 33.7936,  -5.4193 & 1.46-1.52     \\
4        & 2016 Feb & 33.7960,  -5.4242 & 1.46-1.53     \\
5        & 2016 Feb & 33.7945,  -5.4251 & 1.62-1.73     \\
6        & 2016 Feb & 33.7945,  -5.4226 & 1.89-2.09     \\
\hline
\end{tabular}
\end{table}

Calibration images consisted of one set of ten bias (0s exposures) 
for each observing run, five and twelve twilight sky flats (1-30s), respectively, 
plus continuum dome-flat fields (5s) and HgHe calibration arc lamps 
(60-120s) via fibres for every science image. Dark frame exposures 
were not taken since the instrument is regularly checked and there
is no dark current currently.
Standard stars (BD +25 4655, G191B2B and BD +33 2642) and a comparison 
elliptical (NGC 499) were also observed during the observing runs. \\

The spectral images are read separately in four CCD blocks (a, b, c and d 
quadrants, from bottom-left anti-clockwise), reduced 
together with the P3D software (cf. Sandin et al. 2010).
Data reduction followed standard steps for IFS: bias combination and 
subtraction, detection of spectra along the cross-dispersion axis, 
their tracing along dispersion axis, extraction, transmission correction,
cosmic ray events removal and wavelength calibration.
The resulted calibrated spectra cover approximately 3862 \AA \
(ranging from 3749 to 7610 \AA) with a linear dispersion of 
about 1.92 \AA/pixel (or $\sim$ 100 km s$^{-1}$ per pixel at 5700 \AA).
Differential atmospheric refraction correction was not applied.
Sky fibres were averaged for each image, excluding some (1 spaxel in 
the first run and 3 in the second) with inconsistent signal, to prepare 
master sky spectra which were subtracted from all the spaxels of that 
image. \\

Thus, the brightest objects in the field of each position of 
the dithering pattern (6) were first identified individually to have the 
spectra of the respective spaxels clipped and, after, summed.
One of the dithering positions (the middle one of the second run) was 
not used due to very low S/N.
Fifteen objects were identified, some of them not present in all the
dithering positions. The summed spectrum of each galaxy was 
searched for absorption and emission lines with different algorithms
(cross-correlation with galaxy templates and EZ code) and for redshift 
estimation. 

\section{The XLSSC 122 line of sight} 

\begin{figure}[h]
\includegraphics[width=\linewidth,viewport=41 140 570 654]{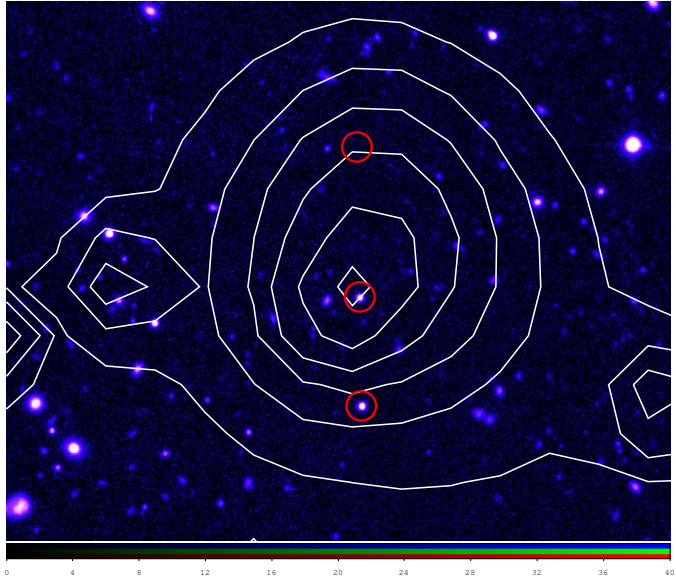}
\includegraphics[width=6.5cm,angle=270]{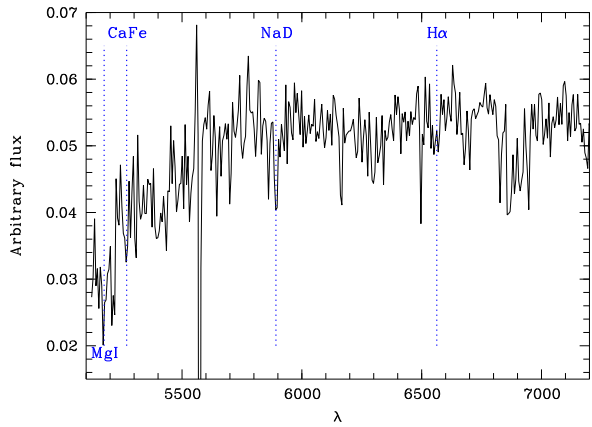}
\includegraphics[width=6.5cm,angle=270]{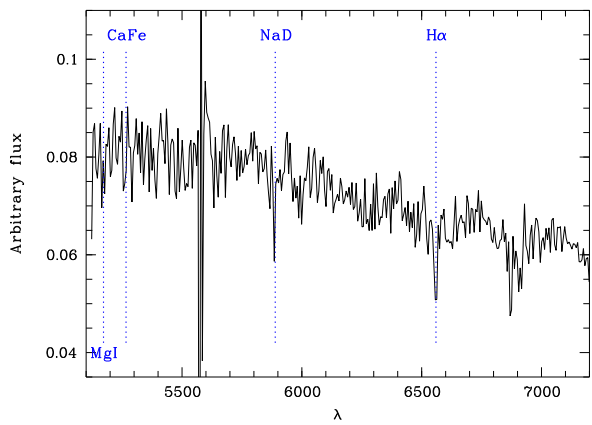}
\caption{\label{fig:n0095}Upper panel: CFHTLS i'band $\sim$2'$\times$2' image of the XLSSC 122 structure. Red circles represent 
the three detected objects along the line of sight. White contours are for the X-ray emission. Middle panel: spectrum of the southern
star. Bottom panel: spectrum of the star close to the X-ray centre. }
\end{figure}

This line of sight hosts a massive very distant (z=1.99) cluster of galaxies (XXL paper V). This is the most distant cluster detected with the
Sunyaev-Zeldovich effect to date.
A subsequent deep XMM-Newton observation allowed us to confirm its redshift (z=1.99), via X-ray spectroscopy (Mantz et al 2017, hereafter XXL paper XVII).
As part of the 191.A-0268
ESO LP, we also spectroscopically observed the line sight with VLT/FORS2 in single slit mode. The sky region of this cluster being very poorly 
sampled, this allowed us to search for possible contamination by X-ray point sources. We show in Figure~\ref{fig:n0095} the location
of the three objects we spectroscopically observed (the three red circles in upper panel of Figure~\ref{fig:n0095}). The two brightest objects (in the middle and south of
the cluster line of sight) are intermediate and cold stars (cf. the middle and bottom panels of Figure~\ref{fig:n0095}), quite unlikely to produce significant X-ray emission.

We also placed in the slit another object, at $\sim$50 arcsec from the X-ray centre towards the north. Nearly invisible on the i' band image, it shows a single isolated 
emission line in the FORS2 spectrum at $\sim$5436\AA. It may be an object at z$\sim$2.51 with CIV in emission. It also may be Lyman$\alpha$ at 
z$\sim$3.47. In this case, we may have expected to also detect CIV at $\sim$6911\AA, but this position is heavily polluted by sky lines. We could also consider
a CIII emission at z$\sim$1.85 (Lyman$\alpha$ and CIV would be outside of our spectral range in this case, and MgII would also be heavily
polluted by sky lines). Finally, it is unlikely that this line is MgII at z$\sim$0.94 because we do not detect [OII].

\clearpage

\section{Cluster pairs}                                                                                      

To publish the full results of our super-cluster detection process, in Table~\ref{tab:PL} we give the detected cluster pairs. We 
only list a sequence number as these structures are not used at all in the present paper.

\begin{table}[t!]
\caption{\label{tab:PL}List of detected cluster pairs with the FoF approach. Columns are: cluster
pair id., coordinates (J2000), mean redshift, members (XLSSC numbers).}
\begin{tabular}{rrrrr}
\hline
\hline
Id & $\alpha$ & $\delta$ & Mean Redshift & XLSSC \\ 
\hline						
 1 &  30.508 &  -5.465 & 0.234 &  114, 174 \\
 2 &  30.968 &  -5.122 & 0.814 &  160, 164 \\
 3 &  31.098 &  -6.462 & 0.901 &  094, 100 \\
 4 &  32.333 &  -6.229 & 0.043 &  115, 171 \\
 5 &  33.631 &  -4.421 & 0.156 &  057, 166 \\
 6 &  34.241 &  -3.786 & 0.758 &  076, 136 \\
 7 &  34.425 &  -4.763 & 0.195 &  079, 141 \\
 8 &  35.408 &  -3.561 & 0.230 &  039, 120 \\
 9 &  35.765 &  -4.819 & 0.322 &  018, 040 \\
10 &  35.787 &  -3.091 & 0.486 &  036, 128 \\
11 &  36.247 &  -4.458 & 0.264 &  025, 044 \\
12 &  36.336 &  -4.236 & 0.769 &  002, 037 \\
13 &  36.641 &  -4.301 & 0.584 &  038, 068 \\
14 &  36.670 &  -5.928 & 0.232 &  055, 103 \\
15 &  36.679 &  -4.115 & 0.345 &  014, 033 \\
16 &  36.705 &  -3.131 & 0.278 &  031, 051 \\
\hline
17 & 349.280 & -54.432 & 0.378 &  513, 525 \\
18 & 349.374 & -56.034 & 0.234 &  592, 594 \\
19 & 349.561 & -52.808 & 0.455 &  558, 559 \\
20 & 349.665 & -55.686 & 0.076 &  527, 579 \\
21 & 349.776 & -56.243 & 0.302 &  528, 617 \\
22 & 350.495 & -56.141 & 0.700 &  517, 576 \\
23 & 350.522 & -55.084 & 0.345 &  523, 584 \\
24 & 350.697 & -53.497 & 0.151 &  552, 597 \\
25 & 351.125 & -53.590 & 0.861 &  534, 621 \\
26 & 351.236 & -55.202 & 0.607 &  580, 611 \\
27 & 351.959 & -52.635 & 0.108 &  533, 550 \\
28 & 352.254 & -56.493 & 0.171 &  618, 620 \\
29 & 352.861 & -54.380 & 0.403 &  542, 582 \\
30 & 352.965 & -53.195 & 0.800 &  546, 549 \\
31 & 353.320 & -52.459 & 0.455 &  561, 641 \\
32 & 353.483 & -55.712 & 0.727 &  571, 572 \\
33 & 353.988 & -53.876 & 0.515 &  537, 628 \\
34 & 354.919 & -56.048 & 0.382 &  543, 604 \\
35 & 355.582 & -56.340 & 0.414 &  540, 605 \\
36 & 355.614 & -55.923 & 0.185 &  539, 541 \\
37 & 356.179 & -56.043 & 0.426 &  603, 606 \\
38 & 356.734 & -53.850 & 0.633 &  509, 566 \\
39 & 357.155 & -55.481 & 0.392 &  510, 602 \\
\hline
\end{tabular}
\end{table}

\section{Alternative measurement of X-ray parameters}  

In Table~\ref{tab:listeSL} we provide the parameter estimates derived from scaling relations (including the value of r$_{500,scal}$ which is different 
from the other estimate of Table~\ref{tab:C1C2cat}). XLSSC 603 is not included in this table because the flux in the pn detector was equal to zero, 
despite the 142 counts in the MOS.

\begin{table}[t!]
\caption{\label{tab:listeSL} List of X-ray parameters from scaling relations (cf. section 4.3 and appendix F) for the confirmed C1 and C2 
clusters of galaxies. The full table is available only in the XXL Master Catalogue browser at http://cosmosdb.iasf-milano.inaf.it/XXL/ 
juxtaposed at the side of the XXL-365-GC table. Col.1: XXL name of the galaxy structure. Col. 2:  X-ray temperature and 
uncertainty (in the [0.5;2]keV band).
Col. 3: radius corresponding to the 500 matter density contrast along with its uncertainty. Col. 4: total mass at the 500 matter 
density contrast along with its uncertainty. Col. 5: Bolometric X-ray luminosity and uncertainty. }
\begin{tabular}{ccccc}
\hline
\hline
XLSSC		&  T$_{300kpc,scal}$ & r$_{500,scal}$  & M$_{500,scal}$          & L$^{bol}_{500,scal}$     \\
		&   keV         & kpc           &  10$^{13}$ $\thinspace$ M$\odot$   & 10$^{42}$ \thinspace erg s$^{-1}$  \\
\hline	         				
001	&	4.2$\pm$0.5	&	819$\pm$94	&	30$\pm$10	&	250$\pm$20	\\
002	&	3.7$\pm$0.5	&	692$\pm$84	&	21$\pm$8	&	200$\pm$25	\\
003	&	4.4$\pm$0.7	&	745$\pm$96	&	29$\pm$11	&	360$\pm$44	\\
005	&	2.7$\pm$0.5	&	499$\pm$68	&	11$\pm$5	&	120$\pm$19	\\
006	&	6.3$\pm$0.6	&	1151$\pm$137	&	66$\pm$24	&	650$\pm$29	\\
008	&	1.6$\pm$0.2	&	579$\pm$53	&	7$\pm$2	&	17$\pm$3	\\
009	&	1.8$\pm$0.2	&	605$\pm$59	&	9$\pm$3	&	23$\pm$5	\\
010	&	2.8$\pm$0.2	&	773$\pm$73	&	18$\pm$5	&	72$\pm$6	\\
011	&	0.8$\pm$0.1	&	435$\pm$40	&	2$\pm$1	&	2$\pm$1	\\
013	&	2.0$\pm$0.2	&	635$\pm$57	&	10$\pm$3	&	26$\pm$3	\\
018	&	1.5$\pm$0.2	&	548$\pm$49	&	6$\pm$2	&	14$\pm$2	\\
020	&	2.3$\pm$0.3	&	625$\pm$64	&	11$\pm$4	&	47$\pm$8	\\
021	&	0.9$\pm$0.1	&	460$\pm$41	&	3$\pm$1	&	3$\pm$1	\\
022	&	3.1$\pm$0.2	&	835$\pm$79	&	22$\pm$6	&	91$\pm$4	\\
023	&	2.5$\pm$0.2	&	716$\pm$66	&	14$\pm$4	&	50$\pm$5	\\
025	&	2.9$\pm$0.2	&	812$\pm$75	&	20$\pm$5	&	73$\pm$4	\\
027	&	2.4$\pm$0.2	&	710$\pm$64	&	13$\pm$4	&	43$\pm$3	\\
028	&	1.5$\pm$0.2	&	545$\pm$52	&	6$\pm$2	&	12$\pm$3	\\
029	&	4.6$\pm$0.9	&	675$\pm$96	&	27$\pm$12	&	480$\pm$45	\\
030	&	1.8$\pm$0.2	&	496$\pm$53	&	7$\pm$2	&	25$\pm$5	\\
\hline									        			      		  
\end{tabular}								        
\end{table}								        

\section{C3 clusters and not yet spectroscopically confirmed C1 clusters}  

The spectroscopically confirmed C3 objects are listed in Table~\ref{tab:C3cat}.
In Table~\ref{tab:listetot7} we also give the list of C1 candidate clusters not yet spectroscopically 
confirmed (too few redshifts, and BCG identification not clear). XLSSC identifications are not available most
of the time because these clusters are not yet spectroscopically confirmed. Both tables are also available in the 
XXL Master Catalogue browser at http://cosmosdb.iasf-milano.inaf.it/XXL/, appended at the end of the
XXL-365-GC table.

\begin{table*}[t!]
\caption{\label{tab:C3cat}Parameters for C3 clusters. Col.1: official XLSSC name. Col.2 and 3: structure coordinates. 
Col.4: redshift. Col. 5: number of measured spectroscopic redshifts in the clusters. Col. 6: XXL class. 
Col. 7: X-ray flux and uncertainty as in Tab.~\ref{tab:C1C2cat}. Col. 8, flags: '+' means the cluster was already published in the XMM-LSS releases, * means that we 
have a note on this cluster in the appendix A, $^{l}$ 
means that the considered cluster is brighter than the  reference flux completeness limit, F means that the structure is a candidate fossil group.}
\begin{tabular}{cccccccc}
\hline
\hline
 XLSSC  & $\alpha$ & $\delta$ & z & N$_{gal}$ & Class &    F$_{60}$  & f.   \\
        &       &       &   &    &        &  10$^{-15}$   &  \\
        & deg      & deg      &   &    &        & $\rm erg \, s^{-1} \, cm^{-2}$ &  \\
\hline						
  164   & 30.415   & -5.050    & 0.811  &  5 &   3  	&  28$\pm$9 &  $^{l}$  \\   	      
  118   & 33.692   & -3.941    & 0.140  &  1 &   3  	& 776$\pm$15 &  $^{l}$  \\                  
  066   & 34.476   & -5.450    & 0.250  &  8 &   3  	&   9$\pm$2 & +	      \\                  
  063   & 34.654   & -5.674    & 0.276  &  3 &   3  	&  31$\pm$5 & +$^{l}$  \\                  
  136   & 34.800   & -3.749    & 0.766  &  7 &   3  	&  28$\pm$4 &  $^{l}$  \\                  
  119   & 35.366   & -4.570    & 0.158  &  9 &   3  	&   5$\pm$2 &  	      \\                  
  034   & 35.372   & -4.099    & 1.036  &  2 &   3  	&  21$\pm$9 &  $^{l}$  \\                  
  126   & 35.424   & -4.454    & 0.290  &  2 &   3  	&   7$\pm$2 &  	      \\                  
  134   & 35.515   & -5.737    & 0.744  &  2 &   3  	&  25$\pm$7 &  $^{l}$  \\                  
  132   & 35.593   & -4.888    & 0.377  &  2 &   3  	&   4$\pm$2 &  	      \\                  
  120   & 35.718   & -4.280    & 0.229  &  3 &   3  	&   6$\pm$2 &  	      \\                  
  024   & 35.744   & -4.121    & 0.291  & 10 &   3  	&   9$\pm$2 & +	      \\                  
  046   & 35.763   & -4.606    & 1.217  & 10 &   3  	&   7$\pm$2 & +	      \\                  
  026   & 35.925   & -4.514    & 0.435  &  5 &   3  	&  12$\pm$2 & +        \\ 
  015   & 35.926   & -5.034    & 0.858  &  6 &   3  	&   3$\pm$6 &  	      \\                  
  143   & 35.960   & -5.610    & 0.498  &  1 &   3  	&  14$\pm$3 &  $^{l}$  \\                  
  007   & 36.025   & -3.921    & 0.559  &  5 &   3  	&  11$\pm$3 & +        \\                  
  019   & 36.049   & -5.380    & 0.496  &  5 &   3  	&   1$\pm$3 &    \\    
  133   & 36.069   & -5.058    & 0.152  &  5 &   3  	&   7$\pm$2 &  	      \\                  
  053   & 36.112   & -4.832    & 0.495  &  5 &   3  	&  12$\pm$3 &          \\                  
  131   & 36.173   & -4.219    & 0.513  &  3 &   3  	&   3$\pm$2 &  	      \\                  
  037   & 36.288   & -4.552    & 0.767  &  4 &   3  	&   2$\pm$2 & +	      \\                  
  043   & 36.293   & -4.030    & 0.172  & 13 &   3  	&  10$\pm$3 & +        \\                  
  042   & 36.345   & -4.447    & 0.463  &  6 &   3  	&   6$\pm$2 & +	      \\            
  045   & 36.369   & -4.261    & 0.556  &  4 &   3  	&  17$\pm$5 &  +$^{l}$ \\    
  004   & 36.376   & -5.120    & 0.291  & 11 &   3  	&   2$\pm$3 &  + \\    
  068   & 36.426   & -4.411    & 0.585  &  4 &   3  	&   2$\pm$2 & +	      \\                  
  129   & 36.446   & -3.167    & 0.329  &  4 &   3  	&  10$\pm$44 &          \\                  
  069   & 36.542   & -4.522    & 0.824  &  8 &   3  	&   4$\pm$2 & +	      \\                  
  017   & 36.614   & -5.000    & 0.383  &  5 &   3  	&  10$\pm$3 & +        \\                  
  014   & 36.641   & -4.063    & 0.344  &  7 &   3  	&   9$\pm$4 & +	      \\                  
  033   & 36.717   & -4.166    & 0.345  &  8 &   3  	&   8$\pm$2 &  	      \\  
  070   & 36.863   & -4.903    & 0.301  &  9 &   3  	&   2$\pm$2 & +	      \\                  
  031   & 36.912   & -3.436    & 0.277  &  2 &   3  	&   4$\pm$3 &  	      \\                  
  125   & 36.942   & -3.736    & 0.054  &  5 &   3  	&   5$\pm$2 &  	      \\                  
  074   & 37.034   & -5.595    & 0.192  &  7 &   3  	&  22$\pm$5 &  $^{l}$  \\                  
  012   & 37.116   & -4.435    & 0.435  &  6 &   3  	&  22$\pm$2 & +$^{l}$  \\                  
  016   & 37.119   & -4.995    & 0.332  &  8 &   3  	&   6$\pm$3 & +	      \\                  
  552   & 350.629  & -54.269   & 0.150  &  2 &   3  	& 580$\pm$10 &  $^{l}$  \\    
\hline
\end{tabular}
\end{table*}

\begin{table*}[t!]
\caption{\label{tab:listetot7}C1 candidate clusters not yet spectroscopically confirmed with official name, coordinates (J2000),
guessed redshift, number of available spectroscopic redshifts along the line of sight, and X-ray flux and uncertainty as in XXL paper II and in the 
[0.5-2] keV band. The first list is for cluster candidates with not enough spectroscopic 
redshifts to be confirmed, the second list for cluster candidates with no spectroscopic redshift but a photometric redshift estimate, and
the third list for cluster candidates with neither spectroscopic nor photometric redshift estimate. P in the fifth column means that we only have a 
photometric redshift estimate. An empty fourth column means that we have no estimate at all of the redshift. $^{1}$: Photometric
redshift for 3XLSS J232713.5-560337 is given in Suhada et al. (2012). $^{2}$: 3XLSS J232624.8-524210 is heavily polluted by a very bright star so 
its C1 classification is uncertain and the determination of a photometric redshift was impossible. In the last column (flag), $l$ 
means that the considered cluster is brighter than the  reference flux completeness limit ($\sim$ 1.3 $\times$ 10$^{-14}$ $\rm \thinspace erg \, s^{-1} \, cm^{-2}$).}
\begin{tabular}{ccccccc}
\hline
\hline
IAU Name  & $\alpha$ & $\delta$ & z & N$_{gal}$ &   F$_{60}$                & flag  \\
          &          &          &   &           &   10$^{-15}$ &   \\
          & deg      & deg      &   &           &   $\rm erg \, s^{-1} \, cm^{-2}$ &   \\
\hline						
3XLSS J021210.6-061235&  33.044   & -6.210   & 0.426      &  2 &        28$\pm$5 & $l$\\        
3XLSS J021825.9-045947&  34.608   & -4.997   & 1.132      &  1 &         4$\pm$1 & \\		  
3XLSS J232704.6-525831&  351.769  & -52.975  & 0.583      &  1 &        26$\pm$4 & $l$\\	  
3XLSS J233116.6-550737&  352.819  & -55.127  & 1.296      &  1 &         5$\pm$2 & \\		  
\hline		                 	                                      		         
3XLSS J020604.1-072432&  31.517   & -7.409   & 0.563      &  P &        11$\pm$3 & \\ 		  
3XLSS J020720.0-060936&  31.833   & -6.160   & 0.460      &  P &         5$\pm$3 & \\ 		 
3XLSS J021803.5-055524&  34.514   & -5.923   & 0.450      &  P &        29$\pm$5 & $l$\\ 	  
3XLSS J022043.7-030106&  35.182   & -3.019   & 0.160      &  P &        18$\pm$5 & $l$\\  	  
3XLSS J231609.8-541617&  349.041  & -54.272  & 0.288      &  P &         7$\pm$2 & \\ 		  
3XLSS J232713.5-560337&  351.806  & -56.061  & 0.920$^{1}$&  P &        22$\pm$3 & $l$\\	  
3XLSS J232801.9-545545&  352.008  & -54.929  & 0.960      &  P &        21$\pm$3 & $l$\\ 	  
3XLSS J233407.0-523709&  353.529  & -52.619  & 0.560      &  P &        14$\pm$3 & $l$\\ 	  
3XLSS J233531.3-543511&  353.881  & -54.586  & 0.866      &  P &        31$\pm$4 & $l$\\ 	 
3XLSS J233706.8-541910&  354.279  & -54.320  & 0.524      &  P &        14$\pm$3 & $l$\\ 	 
3XLSS J233948.0-541126&  354.950  & -54.191  & 0.738      &  P &        10$\pm$3 & \\ 		  
3XLSS J234137.0-545208&  355.404  & -54.869  & 0.597      &  P &        13$\pm$3 & $l$\\ 	 
3XLSS J234154.7-550746&  355.478  & -55.129  & 0.630      &  P &        12$\pm$3 & \\ 		 
\hline		                 	                                      		         
3XLSS J021604.6-032625&  34.021   & -3.440   &            &  0 &        16$\pm$3 & $l$\\ 	  
3XLSS J022157.6-034002&  35.490   & -3.666   &            &  0 &         9$\pm$1 & \\ 		  
3XLSS J022732.0-031456&  36.883   & -3.248   &            &  0 &         2$\pm$1 & \\ 		  
3XLSS J231731.6-551424&  349.382  & -55.240  &            &  0 &        27$\pm$4 & $l$\\	  
3XLSS J231639.5-553418&  349.165  & -55.572  &            &  0 &        20$\pm$3 & $l$\\	  
3XLSS J232624.8-524209&  351.603  & -52.703  &  $^{2}$    &  0 &        21$\pm$3 & $l$\\	  
3XLSS J234550.2-535247&  356.459  & -53.880  &            &  0 &         8$\pm$3 & \\ 		  
\hline                                                                                                                                                                   
\end{tabular}
\end{table*}


\begin{thebibliography}{}

\bibitem{} Adami C., Pompei E., Sadibekova T., et al., 2016, A$\&$A 592, 7: XXL paper VIII

\bibitem{} Adami C., Jouvel S., Guennou L.,  et al., 2012, A$\&$A 540, 105

\bibitem{} Adami C., Mazure A., Pierre M., et al., 2011, A\&A 526, 18

\bibitem{} Adami C., Russeil D., Durret F., 2007, A$\&$A 467, 459

\bibitem{} Adami C., Ulmer M.P., 2000, A\&A 361, 13

\bibitem{} Akiyama M., Ueda Y., Watson M., et al., 2015, PASJ 67, 82

\bibitem{} Akritas M.G., Bershady M.A., 1996, ApJ 470, 706

\bibitem{} Allen S.W., Evrard A.E, Mantz A.B., 2011, ARA\&A 49, 409

\bibitem{} Anders E., Grevesse N., 1989, Geochimica et Cosmochimica Acta 53, 197

\bibitem{} Baldry I.K., Alpaslan M., Bauer A.E., et al., 2014, MNRAS 441, 2440

\bibitem{} Balland C., Baumont S., Basa S., et al., 2009,  A$\&$A 507, 85

\bibitem{} Cash W., 1979, ApJ 228, 939

\bibitem{} Chiappetti L., et al., 2017,  A$\&$A in prep.: XXL paper XXVII

\bibitem{} Clerc N., Sadibekova T., Pierre M., et al., 2012, MNRAS 423, 3561

\bibitem{} Chow-Mart\'inez M., Andernach H., Caretta C.A., Trejo-Alonso J.J., 2014, MNRAS 445, 4073

\bibitem{} Comastri A., Ranalli P., Iwasawa K., et al., 2011, A\&A 526, L9

\bibitem{} Desai S., Armstrong R., Mohr J. J., et al., 2012, ApJ 757, 83

\bibitem{} Eckert D., Molendi S., Paltani S., 2011,  A$\&$A 526, 79

\bibitem{} Eckert D., Molendi S., Owers M., et al., 2014,  A$\&$A 570, 119

\bibitem{} Eckert D., Ettori S., Coupon J., et al., 2016, A$\&$A 592, 12: XXL paper XIII

\bibitem{} Einasto M., Einasto J., Tago E., M\"uller V., Andernach H., 2001, AJ 122, 2222

\bibitem{} Faccioli L., et al., 2017, A$\&$A in prep.:  XXL paper XXIV

\bibitem{} Finoguenov A., Watson M. G., Tanaka M., et al.,  2010, MNRAS 403, 2063

\bibitem{} Fleenor M.C., Rose J.A., Christiansen W.A, et al., 2005, AJ 130, 957

\bibitem{} Folkes S., Ronen S., Price I., et al., 1999, MNRAS 308, 459

\bibitem{} Foreman-Mackey D., Hogg D.W., Lang D., Goodman J., 2013, PASP 125, 306

\bibitem{} Fotopoulou S., Pacaud F., Paltani S., et al. 2016, A$\&$A, 592, A5 

\bibitem{} Garilli B., Le F\`evre O., Guzzo L., et al., 2008, A$\&$A 486, 683

\bibitem{} Gelman A., Rubin D.B., 1992, StaSc 7, 457

\bibitem{} Giles P., Maughan B.J., Pacaud F., et al., 2016, A$\&$A 592, A3: XXL paper III

\bibitem{} Gioia I.M., Maccacaro T., Schild R.E., et al., 1990, ApJSS 72, 567

\bibitem{} Girardi M., Aguerri J.A.L., De Grandi S., et al., 2014, A$\&$A 565, 115

\bibitem{} Guennou L., Adami C., Durret F., et al., 2014, A$\&$A 561, 112

\bibitem{} Guzzo L., Scodeggio M., Garilli B., et al., 2014, A\&A 566, 108

\bibitem{} Hasinger G., Cappelluti N., Brunner H., et al., 2007, ApJS 172, 29

\bibitem{} Hinshaw G., Larson D., Komatsu E., et al., 2013, ApJS 208, 19

\bibitem{} Hinton S.R., Davis T.M., Lidman C., et al, 2016, A$\&$C 15, 61

\bibitem{} Icke V., van de Weygaert R., 1987, A\&A 184, 16

\bibitem{} Jones L.R., Ponman T.J., Horton A., et al., 2003, MNRAS 343, 627 

\bibitem{} Khosroshahi H.G., Ponman T.J., Jones L.R., 2007, MNRAS 377, 595

\bibitem{} Koulouridis E., Poggianti B., Altieri B., et al., 2016, A\&A 592, 11: XXL paper XII

\bibitem{} Kundert A., Gastaldello F., D'Onghia E., et al., 2015, MNRAS 454, 161

\bibitem{} Le F\`evre O., Cassata P., Cucciati O., et al., 2013, A\&A 559, A14

\bibitem{} Lidman C., Ardila F., Owers M., et al., 2016, PASA 33, 1: XXL paper XIV

\bibitem{} Lietzen H., Tempel E., Liivamagi L.J., et al., 2016, A\&A 588, L4

\bibitem{} Lieu M., Smith G.P., Giles P.A., et al., 2016, A\&A 592, 4: XXL paper IV

\bibitem{} Liske J., Baldry I.K., Driver S.P., et al., 2015, MNRAS 452, 2087

\bibitem{} Mantz A.B., Abdulla Z., Carlstrom J.E., et al., 2014, ApJ 794, 157: XXL paper V

\bibitem{} Mantz A.B., Abdulla Z., Allen S.W., et al., 2017, A\&A in press: XXL paper XVII

\bibitem{} Melnyk, O., et al., 2017,  A$\&$A in prep..: XXL paper XXI

\bibitem{} Pacaud F., Clerc N., Giles P.A., et al., 2016, A\&A 592, 2: XXL paper II

\bibitem{} Pierre M., Pacaud F., Adami C., et al., 2016, A\&A 592, 1: XXL paper I

\bibitem{} Pierre M., Valtchanov I., Altieri B., et al., 2004, JCAP 09, 011 

\bibitem{} Ranalli P., Comastri A., Vignali C., et al., 2013, A\&A 555, A42

\bibitem{} Roussel H., Sadat R., Blanchard A., 2000, A$\&$A 361, 429

\bibitem{} Ruel J., Bazin G., Bayliss M., et al., 2014, ApJ 792, 45 

\bibitem{} Sandin C., Becker T., Roth M.M., et al., 2010, A$\&$A 515, 35

\bibitem{} Santos W.A., Mendes de Oliveira C., Sodr\'e L.Jr., 2007, AJ 134, 1551

\bibitem{} Scodeggio M., Guzzo L., Garilli B., et al., 2017, A$\&$A in press, astroph: 161107048

\bibitem{} Serna A., Gerbal D., 1996, A$\&$A 309, 65

\bibitem{} Simpson C., Mart\'inez-Sansigre A., Rawlings S., et al., 2006, MNRAS 372, 741

\bibitem{} Simpson C., Rawlings S., Ivison R., et al., 2012, MNRAS 421, 3060
 
\bibitem{} Smith R.K., Brickhouse N.S.; Liedahl D.A.; Raymond J.C., 2001, ApJ 556, L91
 
\bibitem{} S\"ochting I.K., Coldwell G.V., Clowes R.G., Campusano L.E., Graham M.J., 2012, MNRAS 423, 2436
 
\bibitem{} Stalin C.S., Petitjean P., Srianand R., et al., 2010,  MNRAS 401, 294

\bibitem{} Suhada R., Song J., B\"ohringer H., et al., 2012, A$\&$A 537, 39

\bibitem{} Ulmer M.P., Adami C., Covone G., et al.,  2005, ApJ 624, 124

\bibitem{} Wen Z. L., Han J. L., Liu F. S., 2012, ApJS 199, 34

\bibitem{} Yuan F., Lidman C., Davis T.M., et al., 2015, MNRAS 452, 3047

\bibitem{} Zarattini S., Barrena R., Girardi M., et al., 2014, A$\&$A  565, 116 


\end{thebibliography}
\end{document}